\documentclass[iop, apj, iopart]{emulateapj}
\usepackage{color}
\usepackage{adjustbox}
\usepackage[encapsulated]{CJK}
\usepackage{ucs}
\usepackage[utf8x]{inputenc}
\usepackage[flushleft]{threeparttable}
\usepackage{array}
\usepackage{natbib}
\usepackage{multirow}
\newcolumntype{x}[1]{>{\centering\arraybackslash\hspace{0pt}}p{#1}}

\newenvironment{Contfigure*}{%
\renewcommand{\thefigure}{\arabic{figure} (Cont.)}%
\addtocounter{figure}{-1}%
\begin{figure*}}{%
\renewcommand{\thefigure}{\arabic{figure}}%
\end{figure*}}

\newcommand{\msun}{$\mathrm{M_{\sun}}$}
\newcommand{ \kms}{$\mathrm{km\,s^{-1}}$}

\begin{document}
\begin{CJK}{UTF8}{gbsn}

\title{Searching For $\lowercase{z}>6.5$ Analogs Near The Peak Of Cosmic Star Formation}

\author{Xinnan Du (杜辛楠)\altaffilmark{1}, Alice E. Shapley\altaffilmark{2}, Mengtao Tang\altaffilmark{3}, Daniel P. Stark\altaffilmark{3}, Crystal L. Martin\altaffilmark{4}, Bahram Mobasher\altaffilmark{1}, Michael W. Topping\altaffilmark{2}, Jacopo Chevallard\altaffilmark{5} 
}

\altaffiltext{1}{Department of Physics and Astronomy, University of California, Riverside, 900 University Avenue, Riverside, CA 92521}
\altaffiltext{2}{Department of Physics and Astronomy, University of California, Los Angeles, 430 Portola Plaza, Los Angeles, CA 90095}
\altaffiltext{3}{Department of Astronomy / Steward Observatory, University of Arizona, 933 N Cherry Ave, Tucson, AZ 85721}
\altaffiltext{4}{Department of Physics, University of California, Santa Barbara, CA 93106}
\altaffiltext{5}{Scientific Support Office, Directorate of Science and Robotic Exploration,SA/ESTEC, Keplerlaan 1, 2201 AZ Noordwijk, The Netherlands}

\slugcomment{Draft Version \today}

\shorttitle{Searching For $z>6.5$ Analogs}
\shortauthors{Du et al}

\begin{abstract}
Strong $[\textrm{O}~\textsc{iii}]\lambda\lambda4959,5007+\mbox{H}\beta$ emission appears to be typical in star-forming galaxies at $z>6.5$. As likely contributors to cosmic reionization, these galaxies and the physical conditions within them are of great interest.
At $z>6.5$, where Ly$\alpha$ is greatly attenuated by the intergalactic medium, rest-UV metal emission lines provide an alternative measure of redshift and also constraints on the physical properties  of star-forming regions and massive stars. We present the first statistical sample of rest-UV line measurements in $z\sim 2$ galaxies selected as analogs of those in the reionization era based on $[\textrm{O}~\textsc{iii}]\lambda\lambda4959,5007$ EW or rest-frame $\ub$ color. Our sample is drawn from the 3D-HST Survey and spans the redshift range $1.36\leqslant z \leqslant 2.49$. We find that the median Ly$\alpha$ and \textrm{C}~\textsc{iii}]$\lambda\lambda1907,1909$ EWs of our sample are significantly greater than those of $z\sim2$ UV-continuum-selected star-forming galaxies. Measurements from both individual and composite spectra indicate a monotonic, positive correlation between \textrm{C}~\textsc{iii}] and [\textrm{O}~\textsc{iii}], while a lack of trend is observed between Ly$\alpha$ and [\textrm{O}~\textsc{iii}] at EW$_{[\textrm{O}~\textsc{iii}]}\lesssim1000 \mbox{\AA}$. At higher EW$_{[\textrm{O}~\textsc{iii}]}$, extreme Ly$\alpha$ emission starts to emerge. Using stacked spectra, we find that Ly$\alpha$ and \textrm{C}~\textsc{iii}] are significantly enhanced in galaxies with lower metallicity. Two objects in our sample appear comparable to $z>6.5$ galaxies with exceptionally strong rest-UV metal line emission. These objects have significant \textrm{C}~\textsc{iv}$\lambda\lambda1548,1550$, \textrm{He}~\textsc{ii}$\lambda1640$, and \textrm{O}~\textsc{iii}]$\lambda\lambda1661,1665$ emission in addition to intense Ly$\alpha$ or \textrm{C}~\textsc{iii}]. Detailed characterization of these lower-redshift analogs provides unique insights into the physical conditions in $z>6.5$ star-forming regions, motivating future observations of reionization-era analogs at lower redshifts.
\end{abstract}


\section{Introduction}
\label{sec:Intro}

Characterizing galaxies prior to and during the reionization epoch remains a forefront goal for galaxy formation studies. At $z>6.5$, where Ly$\alpha$ is significantly attenuated by the increasingly neutral intergalactic medium \citep[IGM;][]{Stark2010,Vanzella2011,Treu2012,Pentericci2014,Schenker2014,Itoh2018}, spectroscopic confirmation of these early galaxies has proven to be extremely challenging. Thus far, the vast majority of information obtained for $z>6.5$ galaxies is through deep photometry \citep{Labbe2013,Bouwens2015,Finkelstein2015,Stark2016}. Integrated galaxy properties, such as stellar mass, age, and star-formation history can be inferred from modeling galaxy spectral energy
distributions \citep[SEDs;][]{Finkelstein2013,Bouwens2014,Schmidt2017}. In addition, {\it Spitzer}/IRAC infrared broadband colors have been used to estimate the rest-optical emission properties of $z>6.5$ galaxies \citep{Ono2012,Smit2014,Smit2015,RB2016,Laporte2017}.

In light of the difficulty of using Ly$\alpha$ as a redshift probe, the \textrm{C}~\textsc{iii}]$\lambda\lambda1907,1909$ emission doublet -- typically the second strongest rest-UV emission feature after Ly$\alpha$ -- has been used to measure the spectroscopic redshifts of a small sample of $z>6.5$ galaxies \citep{Stark2015b,Stark2017,Hutchison2019}. In addition, \textrm{C}~\textsc{iii}], in combination with other rest-UV emission lines, such as \textrm{C}~\textsc{iv}]$\lambda\lambda1548,1550$, \textrm{He}~\textsc{ii}, and \textrm{O}~\textsc{iii}]$\lambda\lambda1661,1665$, provides valuable constraints on various galaxy physical properties. Key quantities include the C/O abundance ratio, ionization parameter, gas-phase oxygen abundance (i.e., metallicity\footnote{In this work, we use the terms ``metallicity" and ``gas-phase oxygen abundance" interchangeably.}), and ionizing photon production efficiency. Stellar population parameters can also be inferred using models that attempt to describe broadband photometry and nebular emission lines in concert \citep{Stark2015b,Sobral2015,Stark2017,Schmidt2017}.

Based on {\it Spitzer/IRAC} photometry and a small number of near-IR spectroscopic measurements, the sample of $z>6.5$ star-forming galaxies studied thus far appears to have significantly stronger nebular and recombination emission compared to typical ``main-sequence" star-forming galaxies at lower redshifts \citep[e.g. $z\sim2-3$;][]{Shapley2003,Steidel2014,Kriek2015,Reddy2018}. Specifically, intense [\textrm{O}~\textsc{iii}] or [\textrm{O}~\textsc{iii}]$+$H$\beta$ is ubiquitously inferred at $z>6.5$ \citep{Ono2012,Finkelstein2013,Smit2014,RB2016,Stark2017,Labbe2013}. In addition, among the small number of $z>6.5$ sources confirmed spectroscopically, several rest-UV emission-line measurements have been obtained, from both metals and \textrm{He}~\textsc{ii}
\citep{Sobral2015,Stark2015b,Laporte2017}. Through photoionization modeling and SED fitting, these galaxies are found to have low stellar masses ($\lesssim10^9$\msun), high specific star-formation rates (sSFRs) ( $\gtrsim10$ Gyr$^{-1}$), low gas-phase metallicities ($\lesssim0.2Z_{\sun}$), hard radiation fields, large ionzing photon production rates (log($\xi_{ion}$/erg$^{-1}$ Hz)$\gtrsim$ 25.6), and low dust content ( $V$-band optical depth $\tau_{V}\lesssim0.05$) \citep{Ono2012,Finkelstein2013,Stark2015b,Stark2017}. Although such extreme systems may have been detected due to observational biases, the typical star-forming galaxy at $z>6.5$ is also found to have an [\textrm{O}~\textsc{iii}]$\lambda\lambda4959,5007+\mbox{H}\beta$ EW of at least several hundred angstroms \citep{Labbe2013,Smit2015}. All evidence combined points to the distinct physical properties of $z>6.5$ galaxies compared to their lower-redshift counterparts. Robust characterization of the massive stars and interstellar medium in star-forming galaxies at $z>6.5$ is crucial to our understanding of reionization, which will require a statistical sample of at least moderate S/N spectra for such galaxies.

However, the possibility of obtaining such a sample is limited by the low S/N of data attainable with 8--10-meter-class ground-based telescopes and state-of-the-art near-IR spectrographs, and the only small to moderate equivalent widths (EWs) of rest-UV metal lines. An alternative approach to learn about $z>6.5$ star-forming galaxies is to search for and study lower-redshift analogs, which show similar emission-line properties. With lower-redshift observations, for example at $z\sim 2$ during the peak epoch of star formation,
we can obtain detailed information, with higher S/N, on the physical properties (e.g., gas-phase metallicity, ionization parameter, dust extinction) from rest-UV and optical spectroscopy, and infer key stellar population parameters from SED fits using multiwavelength photometry. 

In fact, studies of extreme emission-line galaxies (EELGs) at lower redshifts have already been carried out. \citet{Stark2014} studied a sample of 17 gravitationally-lensed galaxies at $z\sim2$. These galaxies are faint, low-mass systems with large \textrm{C}~\textsc{iii}] EWs (median EW$_{\textrm{C}~\textsc{iii}]}=7.1\mbox{\AA}$) and numerous other rest-UV metal line detections. While these galaxies were not originally selected based on the presence of strong emission lines, their inferred properties (e.g., metallicity, stellar age, ionization parameters) are comparable to those of the $z>6.5$ galaxies. \citet{Mainali2019} found similar results for a new sample of four low-mass, gravitationally-lensed galaxies at $1.6 \leq z \leq 1.7$. Unlensed high-redshift analogs have also been discovered. For example, \citet{Amorin2017} identified a sample of 10 $z\sim 3$ galaxies within the VUDS survey showing simultaneous detections of Ly$\alpha$, \textrm{C}~\textsc{iii}], and \textrm{O}~\textsc{iii}]$\lambda\lambda1661,1665$. These galaxies are characterized in the median by low stellar mass ($<10^9 M_{\odot}$), significantly subsolar oxygen abundance, and high sSFR. 

There have also been attempts to select high-redshift analogs according to rest-optical emission line properties. \citet{Erb2016} studied a sample of 14 low-metallicity, high-ionization galaxies at $z\sim2$. Eleven out of these 14 galaxies have Ly$\alpha$ EW $\geqslant20\mbox{\AA}$, with a median Ly$\alpha$ EW of 36 \mbox{\AA}. A comparably high fraction of strong Ly$\alpha$ emission is also seen among EELGs in the nearby universe. \citet{Yang2017} investigated a sample of 43 compact, star-forming galaxies with strong [\textrm{O}~\textsc{iii}]$\lambda5007$ emission lines at $z\sim0.2$, and found that two thirds of the galaxies show intense (rest-frame EW $\geqslant20\mbox{\AA}$) Ly$\alpha$ emission. Finally and most relevant to the current work, \citet{Tang2018} studied a sample of 227 galaxies selected based on their intense [\textrm{O}~\textsc{iii}]$\lambda\lambda4959,5007$ emission at $1.3<z<2.4$. The targets were selected such that their [\textrm{O}~\textsc{iii}]$+$H$\beta$ EWs span the typical range observed at $z>6$ \citep{Labbe2013}. By modeling rest-optical line fluxes and broadband photometry simultaneously, these authors found that the hydrogen ionizing production efficiency increases with increasing [\textrm{O}~\textsc{iii}] EW. Their result suggests that the large ionizing efficiency at high [\textrm{O}~\textsc{iii}] EW is expected to boost the intrinsic Ly$\alpha$ EW, leading to a positive correlation between [\textrm{O}~\textsc{iii}] and Ly$\alpha$ EWs.

While \citet{Tang2018} presented a large rest-optical spectroscopic sample of $z>6.5$ analogs, rest-UV spectroscopy was not available for those targets. Accordingly, it was not possible to directly connect extreme ionization conditions in the rest optical and the properties of rest-UV metal Ly$\alpha$ emission properties. While \citet{Mainali2019} present both rest-optical and rest-UV spectroscopy for their four gravitationally-lensed $z>6.5$ analogs, the sample was too small to investigate systematically the connection between rest-optical and rest-UV spectra. In this paper, we assemble and present the first statistical sample of pre-selected $z>6.5$ analogs at $z\sim2$ with rest-UV spectroscopy. The targets were selected based on their strong [\textrm{O}~\textsc{iii}]$\lambda\lambda4959,5007$ EW or blue rest-frame $\ub$ color, both of which are shown to be linked with strong \textrm{C}~\textsc{iii}] emission \citep{Stark2014,Berg2016,Jaskot2016,Du2017} and therefore hint an extreme ionization conditions. By selecting galaxies with [\textrm{O}~\textsc{iii}] EW comparable to those at $z>6.5$ \citep{Labbe2013}, we aim to conduct a comprehensive examination of how rest-optical and rest-UV emission line properties relate. Our sample provides a unique window into modeling the physical conditions (e.g., Ly$\alpha$ escape fractions) of these $z>6.5$ analogs, and indicates a path forward for targeting such objects at lower redshifts in future observations.

This paper is organized as follows. In Section~\ref{sec:data}, we discuss the sample selection, observations, redshift estimates, and the final sample properties. We describe the measurements in Section~\ref{sec:measure}, including SED modeling, systemic redshift determination, construction of composite spectra, and measurements of spectral lines. We present the Ly$\alpha$ and \textrm{C}~\textsc{iii}] EW distributions in Section \ref{sec:results}, along with the correlations between [\textrm{O}~\textsc{iii}]$\lambda\lambda4959,5007$ and rest-UV emission lines (i.e., Ly$\alpha$, \textrm{He}~\textsc{ii}, \textrm{O}~\textsc{iii}], and \textrm{C}~\textsc{iii}]), and between metallicity and Ly$\alpha$ and \textrm{C}~\textsc{iii}] emission line strengths. We also report two objects with extreme rest-UV emission properties in Section~\ref{sec:results}. Finally, we
discuss the implications of our results in Section~\ref{sec:disc} and summarize the key findings in Section~\ref{sec:sum}.

Throughout this paper, we adopt a standard $\Lambda$CDM model
with $\Omega_{m}=0.3$, $\Omega_{\Lambda}=0.7$, and $H_{0}=70$ \kms. All wavelengths are measured in vacuum. Magnitudes and colors are on the AB system.

\section{Sample and Observations}
\label{sec:data}

To characterize the physical properties of the galaxies likely responsible for cosmic reionization, we aim to construct a sample of $z>6.5$ analogs during the peak epoch of cosmic star formation. In this section, we describe the criteria used for selecting galaxies at $z=1.2-2.3$ that potentially show strong emission lines in the rest-UV (Section \ref{sec:selection}), the collection and reduction of new spectroscopic data with the Low Resolution Imager and Spectrometer \citep[LRIS,][]{Oke1995,Steidel2004} on the Keck~I telescope (Section \ref{sec:LRIS_spec}), the estimate of galaxy redshifts based on rest-UV spectral features (Section \ref{sec:z_meas}), and the sample properties (Section \ref{sec:sample}).

\subsection{Selection of Potential $z>6.5$ Analogs}
\label{sec:selection}

Previous work has indicated multiple paths towards searching for galaxies with strong nebular emission properties. For example, the strengths of rest-UV and rest-optical emission lines are highly correlated. Galaxies with strong rest-UV emission lines (such as Ly$\alpha$ and \textrm{C}~\textsc{iii}]$\lambda\lambda1907,1909$) tend to show large
[\textrm{O}~\textsc{iii}]$\lambda\lambda4959, 5007$ or H$\alpha$ EW, as suggested by both photoionization models \citep[e.g.,][]{Jaskot2016} and observations \citep[e.g.,][]{Stark2014,Senchyna2017,Maseda2017,Harikane2018}. Additionally, Ly$\alpha$, \textrm{C}~\textsc{iii}]$\lambda\lambda1907,1909$, and other rest-UV emission lines are found to be stronger in galaxies with 
bluer rest-optical or rest-UV colors \citep{Du2017,Shapley2003,Stark2010,Stark2014}.

In light of these correlations, we selected targets primarily based on their strong rest-frame [\textrm{O}~\textsc{iii}]$\lambda\lambda4959, 5007$ or H$\alpha$ emission. These objects are considered ``Extreme Emission-Line Galaxies" (EELGs) that represent the high-EW tail of the distribution of star-forming galaxies. We drew our sample from the COSMOS and AEGIS fields covered by the 3D-HST Survey \citep{Momcheva2016}, which provides a multi-dimensional description of each galaxy. Multi-wavelength imaging (spanning from the X-ray to radio) is publicly available for these fields, as well as optical and IR photometry, grism spectra, emission-line measurements, rest-frame colors, grism and photometric redshifts, and stellar population parameters. The WFC3/G141 grism has a wavelength range of $1.1-1.7\mu$m, providing the spectral coverage of [\textrm{O}~\textsc{iii}]$\lambda\lambda$4959,5007 at $z=1.2-2.3$.

Following the criteria adopted in \citet{Tang2018}, we selected targets with $\mbox{EW}_{[\textrm{O}~\textsc{iii}]\lambda\lambda4959, 5007}\geqslant300\mbox{\AA}$ or $\mbox{EW}_{H\alpha}\geqslant300\mbox{\AA}$ in the 3D-HST grism catalog \citep{Momcheva2016} in the COSMOS field. For targets in the AEGIS field, we adopted slightly more stringent criteria, with $\mbox{EW}_{[\textrm{O}~\textsc{iii}]\lambda\lambda4959, 5007}\geqslant500\mbox{\AA}$ or $\mbox{EW}_{H\alpha}\geqslant400\mbox{\AA}$\footnote{In fact, we started with the more stringent target selection criteria for both AEGIS and COSMOS fields while designing LRIS slit masks. However, due to the smaller number of resulting EELG targets in the COSMOS field, we relaxed the criteria in COSMOS for [\textrm{O}~\textsc{iii}] and H$\alpha$ EW to $300\mbox{\AA}$ to achieve a comparable number of targets to that in the AEGIS field. We note that both the COSMOS and AEGIS EELG subsamples provide fair representations of the parent sample from \citet{Tang2018}.}. A total of 112 EELGs were selected in this manner based on their [\textrm{O}~\textsc{iii}] or H$\alpha$ EW, 33 of which we were able to assign slits to on LRIS masks (as described in Section~\ref{sec:LRIS_spec}). In practice, all but one of the objects in our sample had $\mbox{EW}_{[\textrm{O}~\textsc{iii}]\lambda\lambda4959, 5007}\geqslant300\mbox{\AA}$; therefore, our sample is effectively [\textrm{O}~\textsc{iii}]-selected. We note that our EELG targets have galaxy properties resembling those described in \citet{Tang2018}, with a slightly higher median [\textrm{O}~\textsc{iii}] EW, i.e.,   $\mbox{EW}_{[\textrm{O}~\textsc{iii}]\lambda\lambda4959, 5007,{\rm med}}=758\mbox{\AA}$ (Figure \ref{fig:galprop}).

The analysis of \citet{Du2017} suggested that the strength of rest-UV emission lines is correlated with rest-frame $\ub$ color. Given that we don't completely fill an LRIS mask with emission-line selected galaxies, we adopted blue rest-frame $\ub$ color as a secondary selection criterion. \citet{Du2017}
showed based on both individual and composite spectra
that \textrm{C}~\textsc{iii}]$\lambda\lambda1907,1909$ emission is the strongest in the bluest ($\ub\lesssim0.4$ mag) galaxies.
We therefore adopted $\ub<0.4$ mag as the secondary selection criterion to maximize the possibility of selecting strong \textrm{C}~\textsc{iii}] emitters and therefore potential EELGs. 
Twenty-two objects were selected based on their rest-frame $\ub$ color listed in the 3D-HST photometric catalog \citep{Skelton2014}, and the median $\ub$ color of this sample is 0.32 mag. We list the number of EELGs and $\ub$ objects in each field, along with the number of targets with redshift measurements in each category, in Table \ref{tab:selection}.

\subsection{Data and Observations}

\subsubsection{LRIS Spectroscopy}
\label{sec:LRIS_spec}

Multi-slit spectroscopy was obtained in the COSMOS and AEGIS fields using Keck/LRIS, a dichroic spectrograph. Fifty five galaxies were targeted (28 in the COSMOS field, and 27 in the AEGIS field), and details of these objects are listed in Table \ref{tab:ew}. LRIS data were collected on 29-30 March 2017 using multi-object slitmasks with $1.''2$ slits. 
The targets were observed with the 400 lines $\mbox{mm}^{-1}$ grism blazed at 3400$\mbox{\AA}$ (435 $\mbox{km s}^{-1}$ FWHM) on the blue side, the d500 dichroic, and the 600 lines $\mbox{mm}^{-1}$ red grating blazed at 5000$\mbox{\AA}$ (220 $\mbox{km s}^{-1}$ FWHM) on the red side. This setup enabled continuous spectral coverage between the blue and red spectra across the dichroic. As a result, the LRIS spectra cover a large set of rest-UV emission lines, including Ly$\alpha$, \textrm{C}~\textsc{iv}$\lambda\lambda$1548,1550, \textrm{He}~\textsc{ii}$\lambda$1640,  \textrm{O}~\textsc{iii}]$\lambda\lambda$1661,1665, and \textrm{C}~\textsc{iii}]$\lambda\lambda$1907,1909. Conditions during the observations were fair and stable with moderate ($0.''8$-$1.''0$) seeing. The integration time for the masks is 9 hours for the COSMOS field and 8.3 hours for the AEGIS field.

The data were primarily reduced using {\it IRAF} scripts, following similar procedures described in \citet{Steidel2003}. The multi-object slitmask images were cut up into individual slitlets, flat fielded using twilight flats for the blue side and dome flats for the red side, cleaned of cosmic rays, and background subtracted. Individual exposures were then combined to create the final, two-dimensional (2D) science spectrum for each object, which was extracted into one dimension (1D), wavelength and flux calibrated, and shifted into the vacuum frame. To avoid potential oversubtraction of the background, we followed the procedures outlined in \citet{Shapley2006}, excluding the region of the central trace (where the object continuum locates) when estimating the background. The initial wavelength solution was derived by fitting a 4th-order polymonial to the arc lamp spectra, with a typical residual of $0.2-0.3$~\AA\ on the blue side, and $\sim0.05$~\AA| on the red side. The initial wavelength solution was then shifted in zeropoint so that bright sky lines appeared at the correct wavelengths. Flux calibration was performed in two stages. The spectra were first calibrated using the observations of a spectrophotometric standard star, observed through a long slit and the 400/3400 grism (600/5000 grating) on the blue (red) side. We then used the CFHT/MegaCam $u$-band transmission curve for the blue side and {\it HST}/ACS $F606W$ transmission curve for the red side to perform bandpass spectrophotometry on the flux-calibrated spectra. The blue- and red-side spectra for each galaxy were scaled such that their flux levels matched the $u$-band and $F606W$ total fluxes, respectively, listed for the galaxy in the 3D-HST catalog. The blue- and red-side continuum in the resulting spectra in general agree well near $\sim5000\mbox{\AA}$ in the observed frame, where they connect at the dichroic cut-off. In one case (21918 in the AEGIS field) where the blue- and red-side continuum were offset from one other at observed-frame $\sim5000\mbox{\AA}$, we scaled both sides of the spectra to match the best-fit SED model inferred from BEAGLE (see Section~\ref{sec:sed}), and adopted the final scaled version for measurements. 

\begin{table}
\centering
\begin{threeparttable}
  \caption{LRIS Targets}
  \label{tab:selection}
{\renewcommand{\arraystretch}{1.5}
  \begin{tabular}{ccccc}
  \hline
  \hline
    Field & $\mbox{N}$\textsubscript{EELG} \tnote{1} & $\mbox{N}$\textsubscript{EELG,$z$}\tnote{2} & $\mbox{N}$\textsubscript{\ub}\tnote{1} & $\mbox{N}$\textsubscript{\ub,$z$}\tnote{2} \\
  \hline
  COSMOS & 15 & 15 & 13 & 11 \\ 
 \hline
  AEGIS & 18 & 17 & 9 & 6 \\
\hline
 \end{tabular}}
\begin{tablenotes}
\item[1] Number of EELG or $\ub$ targets observed.
\item[2] Number of EELG or $\ub$ targets with LRIS redshift measurements.
 \end{tablenotes}
\end{threeparttable}
\end{table}

\subsubsection{MOSFIRE Spectroscopy}
\label{sec:mosfire}

Out of 55 LRIS targets, 7 (5 EELGs and 2 $\ub$ objects) have available rest-optical spectra obtained from the Multi-object Spectrometer
for Infrared Exploration \citep[MOSFIRE;][]{McLean2012} on the Keck I telescope. These 7 targets (two at $z=1.5-1.6$ and five at $z=2.1-2.2$) were observed as part of the MOSFIRE Deep Evolution Field (MOSDEF) survey \citep{Kriek2015}. Six targets have spectral coverage in the $J$, $H$, and $K$ filters and exposure times of $\sim 2$ hours per filter, and one target has only $J$ and $H$ coverage and exposure times of $\sim 1$ hour per filter. For the targets at $z=2.1-2.2$, [\textrm{O}~\textsc{ii}] is covered in $J$,
H$\beta$ and [\textrm{O}~\textsc{iii}] are covered in $H$, and H$\alpha$, [\textrm{N}~\textsc{ii}], and [\textrm{S}~\textsc{ii}] are covered in $K$.
For the targets at $z=1.5-1.6$, H$\beta$ and [\textrm{O}~\textsc{iii}] are covered in $J$
and H$\alpha$, [\textrm{N}~\textsc{ii}], and [\textrm{S}~\textsc{ii}] are covered in $H$.
The MOSFIRE spectral resolution is $R=$3000, 3650, and 3600, respectively, for the $J$, $H$, and $K$ bands. The MOSFIRE spectra were reduced, optimally extracted, and placed on an absolute flux scale as described in \citet{Kriek2015}.

\subsubsection{MMIRS Spectroscopy}
\label{sec:MMIRS}

For a subset of 15 objects in our LRIS sample (14 EELGs and 1 $\ub$ object), there are rest-optical spectra from the Magellan Infrared Spectrograph \citep[MMIRS;][]{McLeod2012} on the MMT telescope. Among these 15 objects, two also have available MOSFIRE spectra (Section \ref{sec:mosfire}). Eleven out of 15 MMIRS spectra have coverage of [\textrm{O}~\textsc{iii}]$\lambda\lambda4959,5007$, and were used for the rest-optical line measurements presented in Section \ref{sec:oiii}. The spectra were taken using three configurations: $J$ grism with $zJ$ filter, $H3000$ grism with $H$ filter, and $K3000$ grism with $Kspec$ filter, with the resolving power of $R=$960, 1200, and 1200, respectively. 
These three grism and filter sets have a spectral coverage of $0.95-1.50\mu$m, $1.50-1.79\mu$m, and $1.95-2.45\mu$m, respectively, where the full suite of strong rest-optical emission features ([\textrm{O}~\textsc{ii}], H$\beta$, [\textrm{O}~\textsc{iii}], and H$\alpha$) can be probed for galaxies at $1.55\leqslant z \leqslant 1.70$ and $2.09\leqslant z \leqslant 2.48$.
The total integration time ranges from 1 hr to 8 hr per filter.
A full description of the MMT/MMIRS observations and data reduction can be found in \citet{Tang2018} and \citet{Chiling2015}.

\subsection{Redshift Measurements}
\label{sec:z_meas}

Galaxy redshifts can be inferred from multiple types of spectral features, including Ly$\alpha$ emission, low-ionization interstellar (LIS) absorption lines, and nebular emission lines (e.g., $\textrm{C}~\textsc{iii}]\lambda\lambda1907,1909$ and [$\textrm{O}~\textsc{iii}]\lambda\lambda4959,5007$). Accurately determining galaxy redshifts is crucial for establishing a systemic rest frame for each galaxy, and for constructing composite spectra. In this section, we describe how we estimated galaxy redshifts from Ly$\alpha$ emission, LIS absorption, and \textrm{C}~\textsc{iii}] emission features. In Section \ref{sec:v_sys}, we further explain how galaxy systemic redshifts were determined from these measurements (i.e., from $z_{Ly\alpha}$, $z_{LIS}$, and $z_{C III]}$, respectively). 

We adopted the methods described in Topping in prep. for estimating  redshifts based on rest-far-UV lines. First, we obtained the grism redshift from the 3D-HST catalog as an initial guess for redshift for each object. We then simultaneously modeled Ly$\alpha$ and six interstellar absorption features, namely, \textrm{Si}~\textsc{ii}$\lambda$1260, \textrm{O}~\textsc{i}$\lambda$1302+\textrm{Si}~\textsc{ii}$\lambda$1304, \textrm{C}~\textsc{ii}$\lambda$1334,  \textrm{Si}~\textsc{ii}$\lambda$1526, \textrm{Fe}~\textsc{ii}$\lambda$1608, and \textrm{Al}~\textsc{ii}$\lambda$1670. The science spectrum of each object was perturbed 100 times by its corresponding error spectrum. For each ``fake" spectrum, Ly$\alpha$ emission was fit with a Gaussian on a quadratic continuum (in order to better describe the curved continuum near the feature), while individual LIS absorption lines were fit with a Gaussian profile on a linear continuum. The resulting Ly$\alpha$ redshift was calculated based on the average Ly$\alpha$ centroids from those 100 fake spectra, and the redshifts measured from individual LIS absorption lines were based on the averaged centroids of corresponding lines. The uncertainties on $z_{Ly\alpha}$ and each individual LIS feature contributing to $z_{LIS}$ were derived from the standard deviation of the 100 centroid measurements of each respective feature. Before finalizing the redshift measurements, we visually inspected the fits of individual spectral features (Ly$\alpha$ and six LIS absorption lines) for each object, and excluded the lines with unreasonable fits or those not covered in the spectrum. In the process of calculating the overall $z_{LIS}$, we assigned different priorities to those six LIS absorption lines. The three lines that were considered ``kinematically accurate" are \textrm{Si}~\textsc{ii}$\lambda$1260, \textrm{C}~\textsc{ii}$\lambda$1334, and \textrm{Si}~\textsc{ii}$\lambda$1526 , which do not have potential contamination from adjacent features to possibly bias the line centroid measurement \citep{Shapley2003}. Whenever one or more of these lines were available, the weighted mean redshift was taken as the final $z_{LIS}$. If none of these preferred lines was available, the redshift of \textrm{Al}~\textsc{ii}$\lambda$1670 was subsequently adopted. The uncertainty on the resulting $z_{LIS}$ was calculated through error propagation of the redshift uncertainty of each contributing LIS feature.

The \textrm{C}~\textsc{iii}] redshift was derived using a different approach. Due to the moderate resolution of the LRIS data, the \textrm{C}~\textsc{iii}] doublet is typically not resolved in the blue-side spectra, although, in principle, the higher-resolution red-side spectra could resolve it. In fact, in only a few cases of red-side \textrm{C}~\textsc{iii}] spectra do we significantly detect a double-peaked profile (EW $S/N>3$). Consequently, we adopted the single-component Gaussian fit as a simple but sufficient model to characterize the \textrm{C}~\textsc{iii}] emission profile, and verified that separately modeling the doublet members in the ``resolved" cases (where double peaks are shown in detected \textrm{C}~\textsc{iii}] lines) on average gives us the same measurement in both EW and centroid within the 1$\sigma$ uncertainty. 

To measure $z_{C III]}$, we adopted $z_{Ly\alpha}$ (whenever available) or $z_{LIS}$ as the initial guess for redshift, and used the grism redshift when neither $z_{Ly\alpha}$ or $z_{LIS}$ was available. We estimated, in the flux-calibrated spectra, the continuum flux level, continuum slope, line centroid, EW, and Gaussian FWHM of the \textrm{C}~\textsc{iii}] feature using the program $splot$ in IRAF. These parameters, along with individual error spectra, were then used as inputs to the IDL least-square-fitting program MPFIT \citep{Mark2009}. The fitting was performed over a wavelength range of $(1.0+z_{est})\times1900\mbox{\AA}$ to $(1.0+z_{est})\times1920\mbox{\AA}$, where $z_{est}$ is the estimated redshift from $z_{Ly\alpha}$, $z_{LIS}$, or the grism redshift. The best-fit line centroid and associated 1$\sigma$ uncertainty returned by MPFIT were used to determine $z_{C III]}$. For this analysis, we assumed a doublet ratio of $I_{[\textrm{C}~\textsc{iii}]\lambda1907}/I_{\textrm{C}~\textsc{iii}]\lambda1909}=1.53$, i.e., in the low-electron-density limit \citep{Oster2006}, based on recent, resolved observations of \textrm{C}~\textsc{iii}] \citep{Maseda2017}.

The above fitting methods result in 31 objects with $z_{Ly\alpha}$ (17 EELGs and 14 $\ub$ objects), 31 objects with $z_{LIS}$ (22 EELGs and 9 $\ub$ objects), and 20 objects with $z_{C III]}$ (19 EELGs and 1 $\ub$ object). All objects with a $z_{C III]}$ estimate were required to have a significantly detected \textrm{C}~\textsc{iii}] feature ($\geqslant3\sigma$ EW) to make sure that $z_{C III]}$ is reliable. In comparison, the visual inspection of Ly$\alpha$ in determining the availability of $z_{Ly\alpha}$ corresponds to a $\geqslant3\sigma$ EW in Ly$\alpha$ emission\footnote{The Ly$\alpha$ EW estimated for the purpose of validating $z_{Ly\alpha}$ is different from that derived from the method described in Section \ref{sec:lya}, which accounts for the absorption bluewards of the Ly$\alpha$ emission peak. As $z_{Ly\alpha}$ was measured solely on Ly$\alpha$ emission, we calculated the EW significance by only including the flux density between the bases of the Ly$\alpha$ emission profile.} for 22 out of 31 objects with $z_{Ly\alpha}$ (27 objects have a Ly$\alpha$ EW $\geqslant2\sigma$). As for $z_{LIS}$, the visual inspection results in an equivalent $\geqslant3\sigma$ EW measurement of combined LIS features for 22 out of 31 objects with $z_{LIS}$ (29 objects have a combined LIS EW $\geqslant2\sigma$). For the remaining objects where $z_{Ly\alpha}$ (or $z_{LIS}$) was determined based on a $<2\sigma$ Ly$\alpha$ (or LIS) measurement, the validity of $z_{Ly\alpha}$ (or $z_{LIS}$) was verified by available $z_{C III]}$ or additional inspection of the spectra.

\subsection{Sample}
\label{sec:sample}

\begin{figure*}
\includegraphics[width=0.5\linewidth]{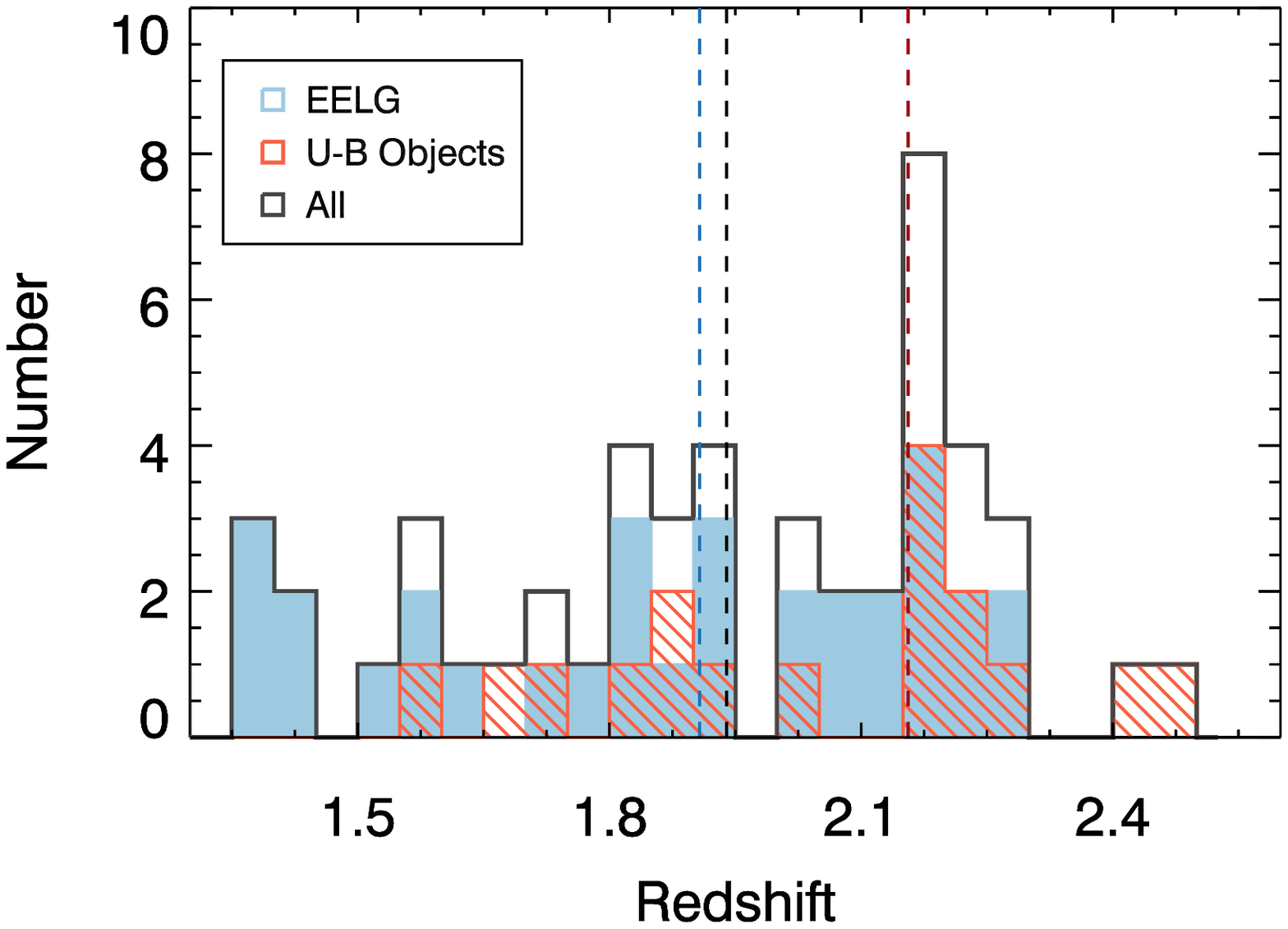}
\includegraphics[width=0.5\linewidth]{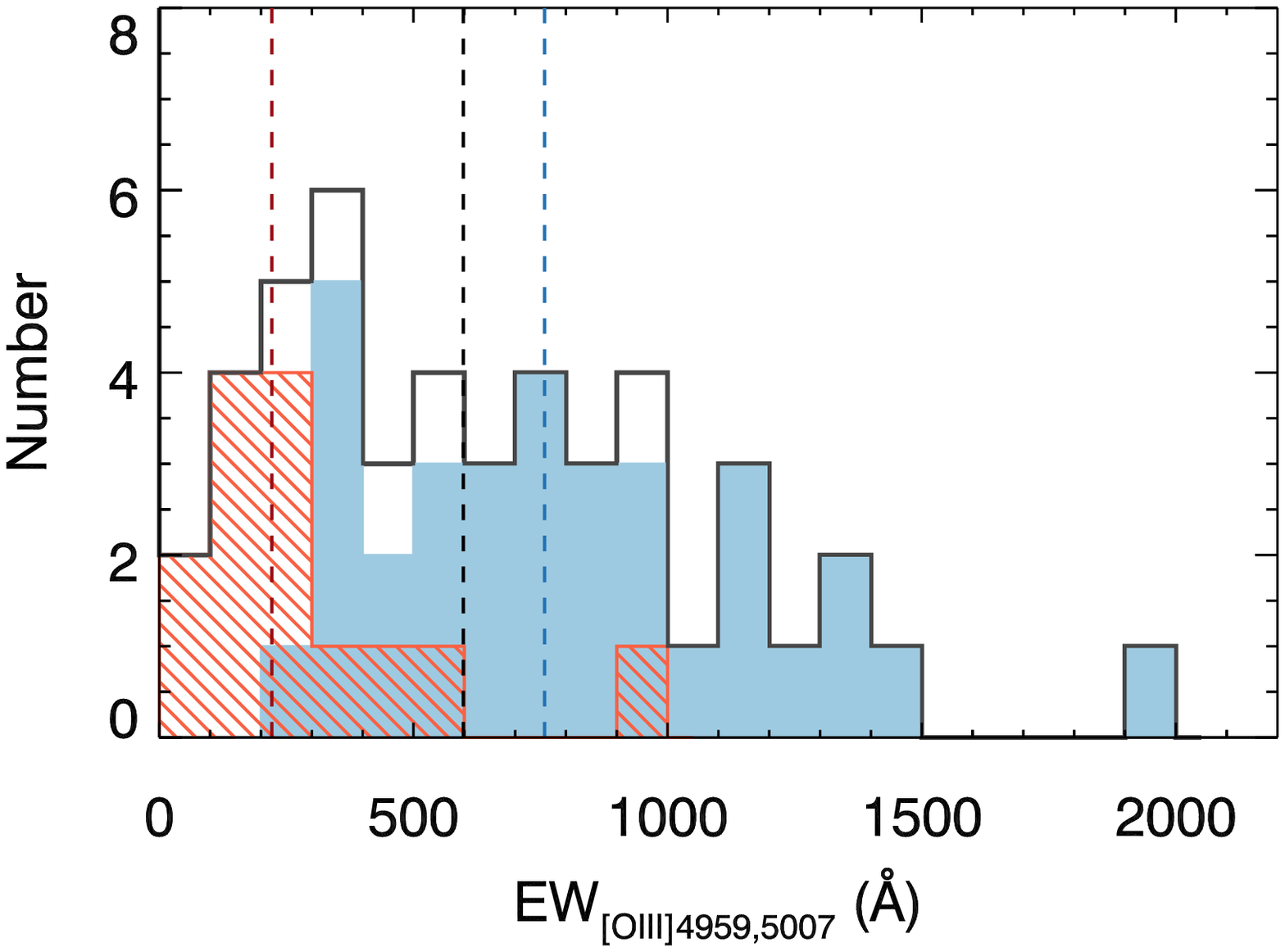}
\includegraphics[width=0.5\linewidth]{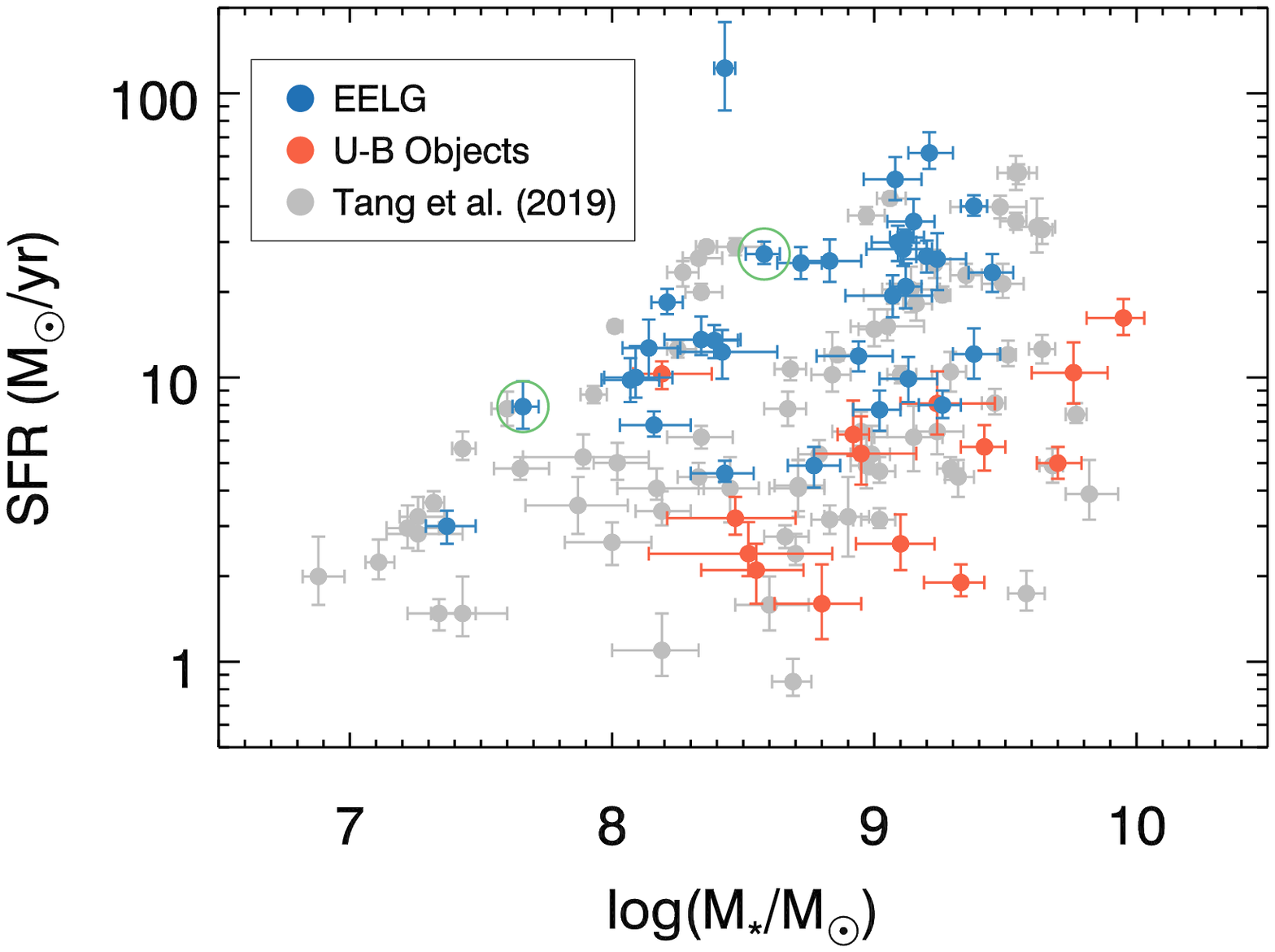}
\includegraphics[width=0.5\linewidth]{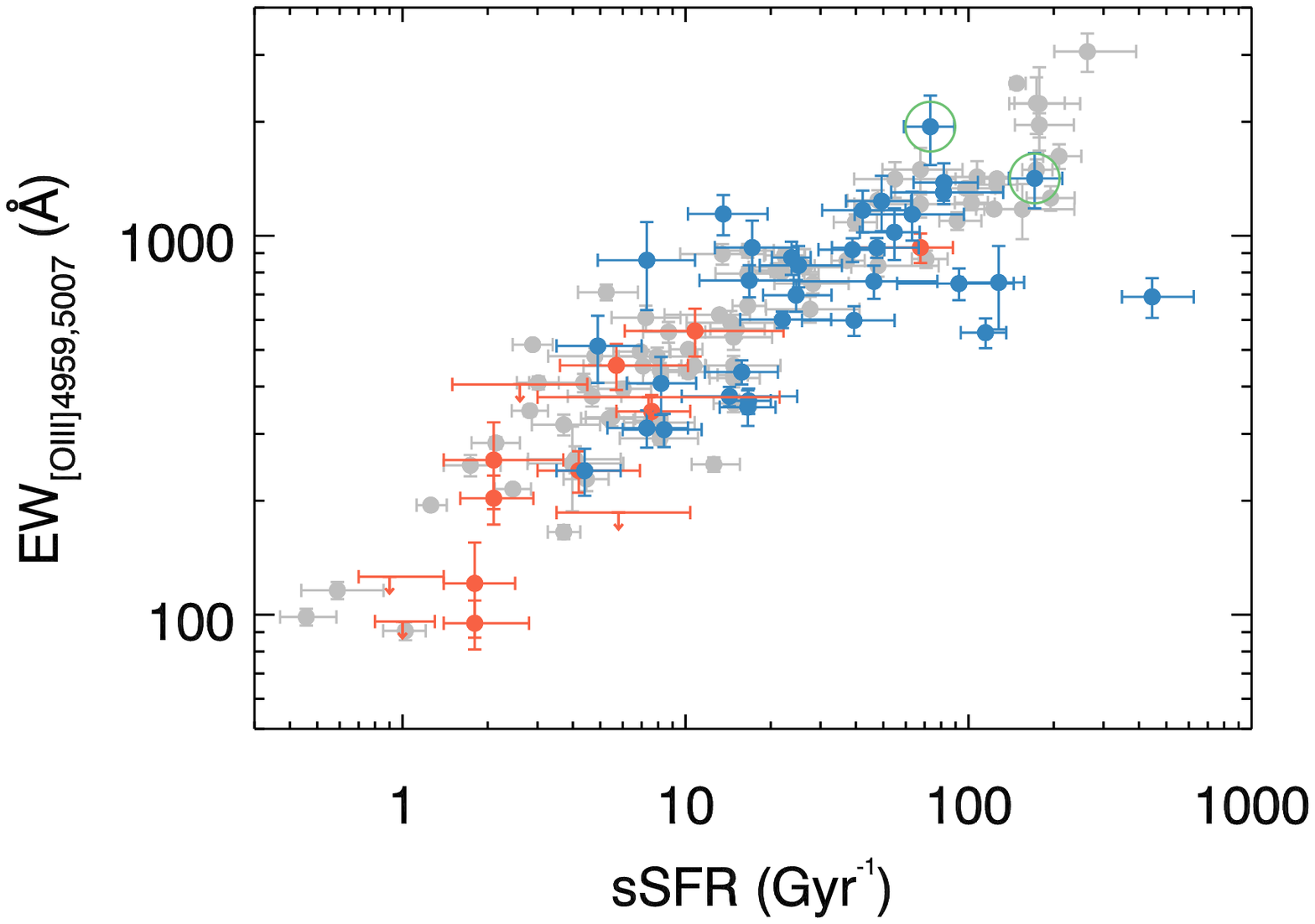}

\caption{Galaxy properties of the LRIS sample. \textbf{Upper Left:} Redshift distribution of LRIS targets. Solid blue, dashed red, and open black bars represent EELGs, $\ub$ objects, and the sum of the two, respectively. \textbf{Upper Right:} Rest-frame [\textrm{O}~\textsc{iii}]$\lambda\lambda4959, 5007$ EW distribution of the LRIS sample. Color coding of the histograms is the same as in the top left panel. The median value of each subgroup is shown with a dashed vertical line in the corresponding color. \textbf{Bottom Left:} SFR vs. stellar mass. Both SFR and stellar mass are derived from SED modeling assuming a Chabrier IMF as described in Section \ref{sec:sed}. Blue and red circles represent EELGs and $\ub$ objects in the LRIS sample, respectively, while light grey circles indicate the EELG sample from \citet{Tang2018}. Error bars in SFR and stellar mass indicate the associated 68$\%$ confidence intervals. The two objects marked with open green circles are COSMOS-7686 and  COSMOS-4064, which show exceptionally strong emission lines in both the rest-UV and rest-optical (see Section \ref{sec:case}). The parent EELG sample from \citet{Tang2018} lies on a similar SFR versus mass relation to that of the $z\sim6$ galaxies presented in \citet{Salmon2015}, although the $z\sim6$ galaxies in \citet{Salmon2015} have a higher typical stellar mass (log($M_{*}$/$M_{\sun}$)$\sim9-10$). \textbf{Bottom Right:} Rest-frame [\textrm{O}~\textsc{iii}]$\lambda\lambda4959, 5007$ EW vs. sSFR. The sSFR is derived from SED fitting, and the [\textrm{O}~\textsc{iii}] EW is calculated as described in Section \ref{sec:oiii}. The [\textrm{O}~\textsc{iii}]$\lambda5007$ EW from \citet{Tang2018} was converted in the [\textrm{O}~\textsc{iii}]$\lambda\lambda4959, 5007$ EW assuming a doublet ratio EW$_{[\textrm{O}~\textsc{iii}]\lambda4959}:$EW$_{[\textrm{O}~\textsc{iii}]\lambda5007}=1:2.98$. Color coding of the symbols is the same as in the bottom left panel. Downward-pointing arrows mark $3\sigma$ upper limits for [\textrm{O}~\textsc{iii}] non-detections.}
\label{fig:galprop}
\end{figure*}

\begin{table*}
\centering
  \caption{Galaxy Properties of LRIS Targets}
  \label{tab:galprop}
  {\renewcommand{\arraystretch}{1.5}
  \begin{tabular}{ccccccccccc}
  \hline
  \hline
    \multirow{3}{*}{Subgroup} &
      \multicolumn{2}{c}{Stellar Mass} &
      \multicolumn{2}{c}{SFR} &
      \multicolumn{2}{c}{sSFR} & 
      \multicolumn{2}{c}{Age} & \multicolumn{2}{c}{$\tau_{V}$} \\ 
      & \multicolumn{2}{c}{(log($M_{*}$/$M_{\sun}$))} &
      \multicolumn{2}{c}{(M$_{\sun}$yr$^{-1}$)} &
      \multicolumn{2}{c}{(Gyr$^{-1}$)} & 
      \multicolumn{2}{c}{(log($t_{age}$/yr))} &  \multicolumn{2}{c}{} \\
      \cline{2-11}
      & {Range} & {Median} & {Range} & {Median} & {Range} & {Median} & {Range} & {Median} & {Range} & {Median}  \\ 
  \hline
  EELG & 7.37$-$9.45 & 8.94 & 3.0$-$122.6 & 18.4 & 4.4$-$455.3 & 25.1 & 6.35$-$8.36 & 7.60 & 0.013$-$1.001 & 0.481 \\ 
 \hline
  $\ub$ & 6.94$-$9.95 & 8.94 & 1.6$-$18.6 & 4.9 & 0.9$-$552.2 & 5.5 & 6.26$-$9.05 & 8.28 & 0.053$-$0.551 & 0.235 \\
\hline
 \end{tabular}}
\end{table*}

As listed in Table \ref{tab:ew}, our target sample consists of 33 EELGs and 22 $\ub$-selected objects. We measured LRIS redshifts for 32 EELGs and 17 $\ub$-selected objects. The confirmed LRIS sample of 49 objects spans in redshift from $z=1.36-2.49$, with a median of $z_{\rm{med}}=1.94$. The EELGs have a slightly lower median redshift ($z_{\rm{med}}=1.91$) compared to the $\ub$ objects ($z_{\rm{med}}=2.16$).  

The rest-frame [\textrm{O}~\textsc{iii}]$\lambda\lambda4959, 5007$ EW was measured following the methods described in Section \ref{sec:oiii}. The EELGs range in [\textrm{O}~\textsc{iii}] EW from 240 to 1940 $\mbox{\AA}$ with a median of 758 $\mbox{\AA}$, while the $\ub$ objects range from 38 to 931 $\mbox{\AA}$ with a median of 222 $\mbox{\AA}$. In fact, 4 $\ub$ objects in our sample now also satisfy the ``EELGs" according to their updated $\geqslant300\mbox{\AA}$ [\textrm{O}~\textsc{iii}]$\lambda\lambda4959, 5007$ EW measurements (derived from the more robust method described in Section \ref{sec:oiii}). However, due to the lack of robust EW measurements from the grism catalog at the time of mask design, these objects were not initially designated as EELGs. 
In this work, we retain them in the $\ub$ category to reflect how they were initially selected. The redshift and [\textrm{O}~\textsc{iii}] EW distributions of the LRIS sample are shown in the top panels of Figure \ref{fig:galprop}.

\section{Measurements}
\label{sec:measure}

\subsection{SED Modeling}
\label{sec:sed}

To derive the integrated galaxy properties, we fit the broadband fluxes using BayEsian  Analysis  of GaLaxysEds \citep[BEAGLE;][]{Chevallard2016}, a tool that consistently models the production of stellar radiation and its transfer through the ISM and IGM. Here we provide an overview of the modeling procedure, and refer the readers to \citet{Tang2018} for a detailed description. The optical and near-IR broadband photometry of the targets was obtained from the publicly available 3D-HST catalog \citep{Skelton2014}. Employing \citet{BC03} stellar population templates and the photoionization modeling code CLOUDY \citep{Ferland2013}, BEAGLE simultaneously fits the broadband fluxes for each galaxy and the available strong emission lines in the rest-frame optical (e.g., [\textrm{O}~\textsc{ii}]$\lambda\lambda3727, 3729$, H$\beta$, [\textrm{O}~\textsc{ii}]$\lambda\lambda4959, 5007$, and H$\alpha$). We adopted a constant star-formation history, assuming a \citet{Chabrier2003} initial mass function and \citet{Calzetti2000} extinction curve. Additional parameterization of the BEAGLE model includes the stellar metallicity, log($Z_{*}$/$Z_{\sun}$), which is allowed to range within $-2.2\leqslant$log($Z_{*}$/$Z_{\sun}$)$\leqslant0.25$; interstellar metallicity, $Z_{ISM}$, which is assumed to be the same as $Z_{*}$; dimensionless \textrm{H}~\textsc{ii} region ionization parameter, $U=n_{LyC}/n_H$, where $n_{LyC}$ is the volume density of hydrogen-ionizing photons and $n_H$ is the hydrogen gas volume density, and the log of $U$ varies between $-4.0$ and $-1.0$; and dust-to-metal mass ratio, $\xi_{d}$, which spans the range $0.1-0.5$. We also fixed the C/O ratio at 0.44, equal to the standard value in nearby galaxies \citep{Gutkin2016}, and assumed that $40\%$ of dust optical
depth arises from grains in the diffuse ISM. 

The output galaxy properties from BEAGLE include stellar mass, age, SFR, sSFR, and $V$-band dust attenuation optical depth. For each parameter, the median value of the posterior probability distribution was adopted as the best-fit value, and the 1$\sigma$ error bar indicates the 68$\%$ confidence interval. We list the range and median value of best-fit galaxy properties for the EELG and $\ub$ subsets in Table \ref{tab:galprop}. On average, the EELGs have comparable stellar mass, much larger SFR and sSFR, slightly younger age, and larger $V$-band dust attenuation optical depth ($\tau_{V}$) than the $\ub$ objects. Figure \ref{fig:galprop} shows the stellar mass, SFR, and sSFR of the LRIS targets in comparison with the sample presented in \citet{Tang2018}.

\begin{deluxetable*}{rrccccc}
\tablewidth{0pt}
  \tablecaption{Spectral Line Measurements of LRIS Targets}
  \tablehead{
    \colhead{Field} &
    \colhead{ID} &
    \colhead{Redshift} &
    \colhead{Category} &
    \colhead{EW$_{[\textrm{O}~\textsc{iii}]\lambda\lambda4959, 5007}$} &
    \colhead{EW$_{\textrm{Ly}\alpha}$} &
    \colhead{EW$_{\textrm{C}~\textsc{iii}]\lambda\lambda1907, 1909}$} \\
    \colhead{} &
    \colhead{} &
    \colhead{} &
    \colhead{} &
    \colhead{($\mbox{\AA}$)} &
    \colhead{($\mbox{\AA}$)} &
    \colhead{($\mbox{\AA}$)} \\
    }
  \startdata
COSMOS & 521 & 2.1996 & $\ub$  & \nodata & 5.3  $\pm$  1.7  & 1.1  $\pm$ 0.3  \\ 
COSMOS & 1151 & 2.2475 & $\ub$  & \nodata & 2.7  $\pm$  2.2  & 1.5  $\pm$ 0.5  \\ 
COSMOS & 1619 & \nodata & $\ub$  & \nodata & \nodata & \nodata \\ 
COSMOS & 1762 & 1.4093 & EELG & 1022 $\pm$ 159 & \nodata & 5.1  $\pm$ 0.7  \\ 
COSMOS & 2129 & 2.4205 & $\ub$  & 111 $\pm$ 62 & 0.8  $\pm$  1.6  & $<$ 3.0 \\ 
COSMOS & 2814 & 1.7052 & $\ub$  & 38 $\pm$ 32 & -8.6 $\pm$  4.2  & $<$ 1.5  \\ 
COSMOS & 3164 & 2.0435 & $\ub$  & \nodata & -7.1 $\pm$  3.2  & $<$ 0.2  \\ 
COSMOS & 3403 & 2.0413 & EELG & 862 $\pm$ 225 & 19.4 $\pm$  10.6 & $<$ 4.8  \\ 
COSMOS & 3480 & 1.6474 & EELG & 311 $\pm$ 35 & 32.1 $\pm$  5.2  & $<$ 1.8  \\ 
COSMOS & 3839 & \nodata & $\ub$  & \nodata & \nodata & \nodata \\ 
COSMOS & 4064 & 1.5058 & EELG & 1940 $\pm$ 403 & \nodata & 6.9  $\pm$ 0.3  \\ 
COSMOS & 4156 & 2.1904 & EELG & 1234 $\pm$ 205 & 22.2 $\pm$  1.3  & 7.8  $\pm$ 0.5  \\ 
COSMOS & 4205 & 1.8400 & EELG & 555 $\pm$ 50 & 33.1 $\pm$  3.8  & \nodata \\ 
COSMOS & 4788 & 1.4083 & EELG & 240 $\pm$ 34 & \nodata & 2.1  $\pm$ 0.5  \\ 
COSMOS & 5281 & 1.8326 & EELG & 353 $\pm$ 38 & -24.3 $\pm$  1.7  & 7.9  $\pm$ 1.7  \\ 
COSMOS & 5283 & 2.1771 & EELG & 919 $\pm$ 66 & 12.5 $\pm$  0.4  & 2.9  $\pm$ 0.2  \\ 
COSMOS & 5593 & 2.1040 & EELG & 601 $\pm$ 30 & 10.4 $\pm$  1.3  & 3.0  $\pm$ 0.4  \\ 
COSMOS & 6283 & 2.2263 & EELG & 367 $\pm$ 28 & -11.5 $\pm$  2.8  & 1.0  $\pm$ 0.3  \\ 
COSMOS & 6332 & 2.1769 & $\ub$  & 180 $\pm$ 125 & 9.0  $\pm$  1.6  & $<$ 3.4 \\ 
COSMOS & 6348 & 2.0983 & EELG & 758 $\pm$ 76 & -0.8 $\pm$  3.4  & 5.1  $\pm$ 1.0  \\ 
COSMOS & 7672 & 2.1945 & $\ub$  & \nodata & 7.9  $\pm$  1.2  & 1.3  $\pm$ 0.4  \\ 
COSMOS & 7686 & 1.8026 & EELG & 1417 $\pm$ 236 & 131.6 $\pm$  10.9 & 13.2 $\pm$ 0.8  \\ 
COSMOS & 7883 & 2.1565 & EELG & 598 $\pm$ 54 & 17.7 $\pm$  1.6  & 5.0  $\pm$ 0.9  \\ 
COSMOS & 8067 & 2.2016 & $\ub$  & 121 $\pm$ 34 & 22.7 $\pm$  2.3  & $<$ 1.1  \\ 
COSMOS & 8383 & 2.0933 & EELG & 1302 $\pm$ 65 & 51.5 $\pm$  4.3  & $<$ 2.2  \\ 
COSMOS & 8711 & 1.8073 & $\ub$  & 100 $\pm$ 42 & 3.9  $\pm$  3.2  & $<$ 1.2  \\ 
COSMOS & 9983 & 1.8525 & $\ub$  & 256 $\pm$ 66 & 22.1 $\pm$  4.3  & $<$ 1.2  \\ 
COSMOS & 10155 & 2.4945 & $\ub$  & \nodata & 3.4  $\pm$  2.5  & $<$ 2.0  \\ 
AEGIS & 3057 & 2.2808 & EELG & 690 $\pm$ 83 & 3.3  $\pm$  1.0  & 3.5  $\pm$ 0.6  \\ 
AEGIS & 3455 & 1.9397 & $\ub$  & 240 $\pm$ 30 & 15.4 $\pm$  1.2  & 4.4  $\pm$ 0.7  \\ 
AEGIS & 4656 & 1.3588 & EELG & 931 $\pm$ 165 & \nodata & 3.7  $\pm$ 0.6  \\ 
AEGIS & 8907 & 1.5909 & $\ub$  & 95 $\pm$ 14 & 3.8  $\pm$  6.4  & 1.3  $\pm$ 0.3  \\ 
AEGIS & 9939 & 1.9328 & EELG & 835 $\pm$ 104 & 4.7  $\pm$  1.9  & $<$ 0.8  \\ 
AEGIS & 12032 & 1.8615 & EELG & 1168 $\pm$ 147 & 2.5  $\pm$  1.5  & \nodata \\ 
AEGIS & 13602 & \nodata & $\ub$  & 291 $\pm$ 135 & \nodata & \nodata \\ 
AEGIS & 14156 & 1.6757 & $\ub$  & 344 $\pm$ 36 & 37.6 $\pm$  14.9 & 8.2  $\pm$ 2.0  \\ 
AEGIS & 16874 & 1.8883 & $\ub$  & 455 $\pm$ 63 & 6.3  $\pm$  3.8  & $<$ 1.3  \\ 
AEGIS & 17160 & 2.1561 & $\ub$  & 561 $\pm$ 81 & 12.4 $\pm$  1.8  & 3.6  $\pm$ 0.7  \\ 
AEGIS & 17514 & 2.0157 & EELG & 748 $\pm$ 72 & -4.4 $\pm$  0.7  & 2.7  $\pm$ 0.3  \\ 
AEGIS & 17842 & 2.2949 & $\ub$  & 931 $\pm$ 83 & 10.7 $\pm$  1.9  & 10.5 $\pm$ 1.4  \\ 
AEGIS & 18543 & 2.1421 & EELG & 876 $\pm$ 88 & -0.8 $\pm$  0.3  & 2.5  $\pm$ 0.3  \\ 
AEGIS & 18729 & 2.2125 & EELG & 377 $\pm$ 22 & 8.0  $\pm$  0.7  & 3.6  $\pm$ 0.4  \\ 
AEGIS & 19696 & \nodata & $\ub$  & 203 $\pm$ 30 & \nodata & \nodata \\ 
AEGIS & 20987 & 2.1617 & EELG & 512 $\pm$ 103 & -5.4 $\pm$  2.0  & $<$ 1.0 \\ 
AEGIS & 21918 & 1.9066 & EELG & 697 $\pm$ 67 & -55.8 $\pm$  22.8 & 1.8  $\pm$ 0.3  \\ 
AEGIS & 22858 & 1.3983 & EELG & 753 $\pm$ 187 & \nodata & 6.9  $\pm$ 1.7  \\ 
AEGIS & 24181 & 1.3928 & EELG & 930 $\pm$ 56 & \nodata & 6.2  $\pm$ 0.8  \\ 
AEGIS & 24314 & \nodata & EELG & 1139 $\pm$ 168 & \nodata & \nodata \\ 
AEGIS & 24857 & 1.9083 & EELG & 1382 $\pm$ 170 & 1.4  $\pm$  1.0  & 6.6  $\pm$ 0.6  \\ 
AEGIS & 26531 & 1.5902 & EELG & 408 $\pm$ 71 & 13.8 $\pm$  7.1  & 4.1  $\pm$ 0.9  \\ 
AEGIS & 28358 & 1.5741 & EELG & 437 $\pm$ 32 & -7.9 $\pm$  4.1  & 5.7  $\pm$ 1.2  \\ 
AEGIS & 33462 & 2.2688 & EELG & 308 $\pm$ 31 & -5.2 $\pm$  2.3  & $<$ 1.2  \\ 
AEGIS & 33688 & 1.7176 & EELG & 1142 $\pm$ 139 & 35.8 $\pm$  2.3  & 3.8  $\pm$ 0.3  \\ 
AEGIS & 35021 & 1.7778 & EELG & 762 $\pm$ 74 & 3.2  $\pm$  2.8  & 2.2  $\pm$ 0.4  \\ 
AEGIS & 38677 & \nodata & $\ub$  & \nodata & \nodata & \nodata \\ 
 \enddata
\label{tab:ew}
\tablecomments{The EW values listed in this table are in the rest-frame. Individual \textrm{C}~\textsc{iii}] detections are listed with the best-fit EW value and associated $1\sigma$ uncertainty. For \textrm{C}~\textsc{iii}] features that are not significantly detected, a $3\sigma$ upper limit is reported.}
\end{deluxetable*}

\subsection{Systemic Redshift}
\label{sec:v_sys}

As described in Section \ref{sec:z_meas}, the redshift estimates for each object in our sample come from three types of features: Ly$\alpha$ emission, LIS absorption, and \textrm{C}~\textsc{iii}] emission. Ideally, we would be able to infer the systemic redshift of each object directly from the nebular emission lines (e.g., \textrm{C}~\textsc{iii}]). However, given that over half of the sample does not have a valid $z_{C III]}$, we used $z_{Ly\alpha}$ and $z_{LIS}$ as complementary redshift indicators. One caveat about Ly$\alpha$ and the LIS absorption lines is that these features do not appear at the galaxy systemic redshift because of the presence of neutral-gas outflows in galaxies. Ly$\alpha$ emission is typically redshifted with respect to the systemic redshift, while the LIS absorption lines are generally blueshifted. 

To compensate for the velocity shifts imprinted by outflows, we followed the procedures described in \citet{Rudie2012} and assumed that Ly$\alpha$ is redshifted by 300 \kms (for galaxies with rest-frame Ly$\alpha$ EW $<20\mbox{\AA}$) and the LIS absorption lines are blueshifted by 160 \kms. For galaxies with EW$_{rest,Ly\alpha}\geqslant20\mbox{\AA}$, Ly$\alpha$ emission is found to have a smaller offset ($\sim200$ \kms) redward of the systemic redshift \citep{Trainor2015}. Because of the prominence and associated higher S/N of the Ly$\alpha$ emission feature, we assigned higher priority to $z_{Ly\alpha}$ than to $z_{LIS}$. Accordingly, the Ly$\alpha$ velocity correction was applied when $z_{Ly\alpha}$ was available (31 objects), and the LIS correction was applied to those with a valid $z_{LIS}$ but either lacking coverage of Ly$\alpha$ or only showing Ly$\alpha$ in absorption (14 objects). 

For the 4 remaining objects with LRIS redshifts, neither $z_{Ly\alpha}$ nor $z_{LIS}$ was measured, but \textrm{C}~\textsc{iii}] was detected and used to estimate the systemic redshift. Somewhat unexpectedly, in cases where we had both LRIS \textrm{C}~\textsc{iii}] measurements and rest-optical [\textrm{O}~\textsc{iii}] measurements from either MMIRS or MOSFIRE, we found a typical blueshift of $\sim 200$\kms for  $z_{C III]}$ relative to the systemic redshift indicated by [\textrm{O}~\textsc{iii}]. For those galaxies with only  $z_{C III]}$ we therefore assumed the same blueshift of $200$\kms for \textrm{C}~\textsc{iii}]. While we could not fully pinpoint the origin of the velocity offset between \textrm{C}~\textsc{iii}] and [\textrm{O}~\textsc{iii}], we argue that the main focus of this work is on the strength instead of kinematic properties of rest-UV and rest-optical emission lines. Therefore, this potential kinematic discrepancy does not affect the results and conclusions we present here.

The scatter in the assumed velocity rules is the dominant factor in the uncertainty on the systemic redshift, which is estimated to be $\sim125$ \kms \citep{Steidel2010}. The systemic redshift, as derived above, was
used to transform each spectrum to the rest frame. While we did not verify the establishment of systemic redshift for individual galaxy spectra, the rest-frame composite spectra shown in Figure \ref{fig:spec} indicate a close alignment with the systemic velocity. Specifically, the measured rest-frame wavelength of the \textrm{C}~\textsc{iii}$\lambda1176$ stellar absorption feature is 1175.7\mbox{\AA}, which translates into a velocity offset of $\sim-80$ \kms. This offset is within the uncertainty on the systemic redshift.
We list the numbers for redshift measurements in each field and category in Table \ref{tab:selection}.

\subsection{Composite Spectra}
\label{sec:composite}

\begin{figure*}
\includegraphics[width=1.0\linewidth]{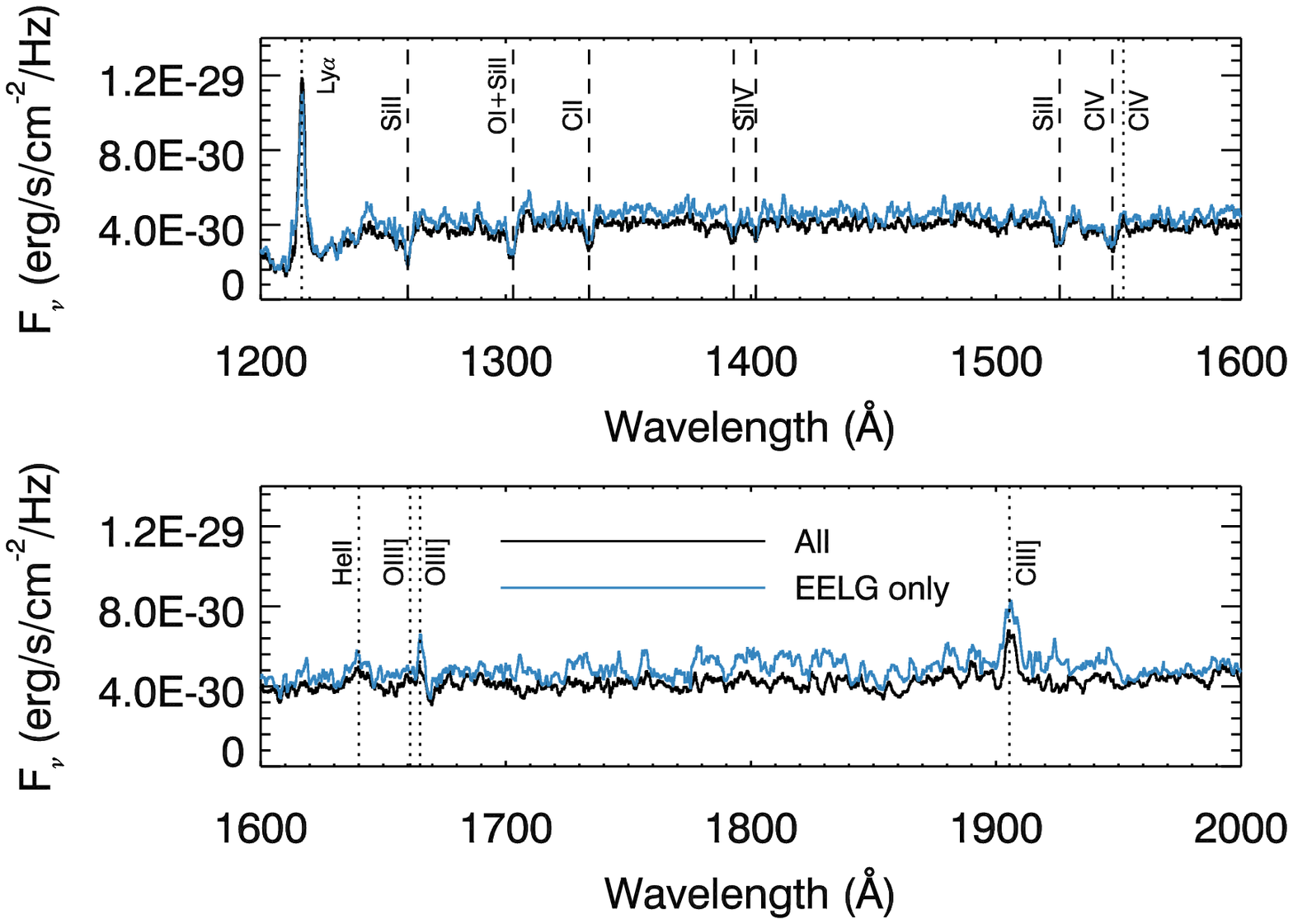}
\caption{Composite rest-frame UV spectra of 49 LRIS targets (black) and 32 EELGs (blue) with redshift measurements. Key emission (dotted lines) and absorption (dashed lines) features are labeled. At the resolution of the
LRIS spectra, the individual doublet members of \textrm{C}~\textsc{iv}$\lambda\lambda1548,1550$,  \textrm{C}~\textsc{iii}]$\lambda\lambda1907,1909$ are blended. The ``EELG-only'' composite is plotted in addition to the overall stack to show its potentially distinct features (e.g., more prominent \textrm{O}~\textsc{iii}]$\lambda1665$ emission.)}
\label{fig:spec}
\end{figure*}

To investigate how rest-UV emission properties correlate with rest-optical line strengths and oxygen abundance, we created composite spectra by dividing the sample (49 objects with redshift measurements) into 4 bins in most cases, with each bin containing nearly the same number of galaxies (but see Section \ref{sec:metal}). We chose [\textrm{O}~\textsc{iii}]$\lambda\lambda4959,5007$EW as the main sorting parameter to probe the emission properties of our LRIS galaxies, as the strength of this feature is known to correlate with that of multiple rest-UV emission lines \citep{Yang2017,Maseda2017}. In order to account for the potentially different nature of EELGs from the $\ub$ objects, we constructed two types of composite spectra: the ``EELG-only" stacks and the ``all" stacks, the latter of which include both EELGs and $\ub$ objects. One exception to the binning rules described above is in Section \ref{sec:metal}, where we divided the subset of objects with metallicity measurements into 2 bins to achieve higher S/N of the composite spectra.

As galaxies in the same bin have roughly the same spectral coverage in the observed frame, the difference in their redshifts results in different rest-frame coverage. In the composites we constructed, we required that the same set of objects contribute to all wavelengths of the measured spectral feature in each composite spectrum. For example, we only allowed objects with a minimum wavelength $\leqslant1200\mbox{\AA}$ to be included in the composites for the measurement of Ly$\alpha$. Similarly, we required a minimal coverage of $1500-1600\mbox{\AA}$ for the measurement of \textrm{C}~\textsc{iv}$\lambda\lambda$1448, 1550, $1600-1700\mbox{\AA}$ for the measurement of \textrm{He}~\textsc{ii}$\lambda1640$ and \textrm{O}~\textsc{iii}$\lambda\lambda$1661, 1665, and $1895-1920\mbox{\AA}$ for the measurement of \textrm{C}~\textsc{iii}$\lambda\lambda$1907, 1909. In general, this extra requirement regarding spectral coverage resulted in a minimal reduction of the sample size for line measurements within rest-frame $1500-2000\mbox{\AA}$, which is a natural outcome of the continuous spectral coverage between the blue and red spectra provided by our observational setup. In our sample, 6 out of 49 objects do not have Ly$\alpha$ coverage, and are therefore not included in the composites made to measure the rest-frame Ly$\alpha$ EW. 

We created the composite spectra by first smoothing the red-side spectra to the resolution of the blue-side spectra. Given that features beyond rest-frame $1430\mbox{\AA}$ (e.g., \textrm{C}~\textsc{iv}, \textrm{He}~\textsc{ii}, \textrm{O}~\textsc{iii}], and \textrm{C}~\textsc{iii}]) may either fall within the blue- or the red-side spectrum, depending on the redshift, we need to ensure that blue- and red-side spectra combined in spectral stacks at the same rest wavelength also have the same resolution. We then interpolated individual spectra onto a grid with $0.2\mbox{\AA}$ increments in wavelength, and performed median stacking to construct a composite spectrum. Figure \ref{fig:spec} shows the composite spectrum created by combining 32 individual EELG spectra with redshift measurements, as well as that constructed by including all 49 LRIS targets where valid redshifts were measured. Strong emission and absorption features are marked. 

Composite error spectra were created using the Monte Carlo method. In particular, we bootstrap-resampled the objects in each bin and perturbed each individual spectrum in the bootstrap sample according to its own error spectrum. The perturbed spectra in the bootstrap sample were then combined to create a new composite spectrum. The process was repeated 100 times and the standard deviation of these 100 fake composites at each wavelength was adopted as the composite error spectrum for each bin.

\subsection{Spectral Line Measurement}
\label{sec:line}

\subsubsection{[\textrm{O}~\textsc{iii}]$\lambda\lambda$4959, 5007}
\label{sec:oiii}

Given that not all $\ub$ objects in our sample have a valid [\textrm{O}~\textsc{iii}] EW listed in the 3D-HST catalog, we developed the following method to measure the [\textrm{O}~\textsc{iii}] EW for both EELGs and $\ub$ objects in a consistent manner. First, we calculated the total flux of [\textrm{O}~\textsc{iii}]$\lambda\lambda4959, 5007$ from the HST/G141 grism spectra (38 out of 49 objects), and from the MMIRS spectra whenever available (11 out of 49 objects) given its higher spectral resolution. The [\textrm{O}~\textsc{iii}] doublet members are resolved in the MMIRS spectra, but not in the grism spectra. Accordingly, we fit a single Gaussian profile to the unresolved [\textrm{O}~\textsc{iii}]$\lambda\lambda4959, 5007$ feature in the grism spectra, and to individual [\textrm{O}~\textsc{iii}] doublet members in the MMIRS spectra to determine the total flux of [\textrm{O}~\textsc{iii}]$\lambda\lambda4959, 5007$. The [\textrm{O}~\textsc{iii}]$\lambda\lambda4959, 5007$ EW was then derived by dividing the [\textrm{O}~\textsc{iii}] flux by the continuum flux density inferred from the best-fit SED model output by BEAGLE.

For EELGs showing a continuum $S/N\geqslant4$ in the vicinity of [\textrm{O}~\textsc{iii}], the [\textrm{O}~\textsc{iii}] EWs determined using the method described above (hereafter the SED-based EW) are in good agreement with those listed in the 3D-HST grism catalog. For two objects,  AEGIS-28358 and AEGIS-33462, the [\textrm{O}~\textsc{iii}] doublet resides near sky line residuals and the fluxes are therefore potentially contaminated. To that end, we adopted the SED-based [\textrm{O}~\textsc{iii}] EW measurements for these two objects, with the line fluxes estimated from the grism spectra. In summary, the final [\textrm{O}~\textsc{iii}]$\lambda\lambda4959, 5007$ EWs we used for this work were determined based on: (1) the SED-based [\textrm{O}~\textsc{iii}] EW for objects with MMIRS spectra (9 objects); (2) the grism [\textrm{O}~\textsc{iii}] EW from the 3D-HST catalog for EELGs without MMIRS spectra but with continuum near [\textrm{O}~\textsc{iii}] characterized by $S/N\geqslant4$ in the grism spectra (20 objects); and (3) the SED-based [\textrm{O}~\textsc{iii}] EW for the rest of the objects (EELGs with $S/N<4$ in the continuum in the grism spectra, all $\ub$ objects, and AEGIS-28358 and AEGIS-33462) where MMIRS spectra are not available (20 objects). The final sample of  [\textrm{O}~\textsc{iii}]$\lambda\lambda4959, 5007$ EWs is listed in Table \ref{tab:ew}.

\subsubsection{Ly$\alpha$}
\label{sec:lya}

We measured the rest-frame $\mbox{Ly}\alpha$ EW in both individual and composite spectra following the procedures described in \citet{Kornei2010} and \citet{Du2018}. Forty three (26 EELGs and 17 $\ub$ selected targets) have Ly$\alpha$ coverage and a secure redshift enabling the measurement of $\mbox{Ly}\alpha$ EW. The spectral morphology of $\mbox{Ly}\alpha$ in individual galaxy spectra is classified into 4 categories through visual inspection: ``emission," ``absorption," ``combination," and ``noise."  ``Emission" objects show dominant Ly$\alpha$ emission in the spectra, while for ``combination" objects the Ly$\alpha$ emission is superimposed on a large absorption trough. The Ly$\alpha$ morphology is classified as ``absorption" when a broad absorption trough resides around the rest-frame wavelength of Ly$\alpha$, and as ``noise" when the spectrum is featureless near Ly$\alpha$. 

Due to the nature of our sample selection, galaxies in our sample were mainly categorized as ``emission" and ``combination" objects (11 and 22 galaxies, respectively), with a small fraction being classified as ``absorption" (5 galaxies) and ``noise" objects (5 galaxies). For each object, regardless of their spectral morphology, the blue and red side continuum levels were estimated over the wavelength range of $1225-1255\mbox{\AA}$ and $1120-1180\mbox{\AA}$, respectively. For ``emission,"``combination," and ``absorption" objects, the $\mbox{Ly}\alpha$ flux was integrated between the blue and red wavelength ``boundaries," where the flux density level on either side of the $\mbox{Ly}\alpha$ feature (either emission or absorption) first meets the blue and red side continuum level, respectively. The blue boundary was fixed at $1208\mbox{\AA}$ for the ``emission" objects and forced to be no bluer than $1208\mbox{\AA}$ for the ``combination" objects \citep{Du2018}. For ``noise" objects, the Ly$\alpha$ flux was integrated over 1199.9 to 1228.8 $\mbox{\AA}$, the boundaries adopted in \citet{Kornei2010}. Finally, we computed $\mbox{Ly}\alpha$ EW by dividing the enclosed $\mbox{Ly}\alpha$ flux by the red side continuum flux-density  level. There are three objects that have spectral coverage less than $20\mbox{\AA}$ in the $1120-1180\mbox{\AA}$ window, making the estimation of the blue side continuum not reliable. Consequently, we measured, in each Ly$\alpha$ morphology category, the relative level of the blue and red side continua from objects with sufficient spectral coverage on both sides. The median blue-to-red continuum ratio in corresponding Ly$\alpha$ morphology category was then applied as a rough proxy of the blue continuum for those 3 objects lacking sufficient blue-side spectral coverage. We list the measured rest-frame Ly$\alpha$ EW in Table \ref{tab:ew}.

$\mbox{Ly}\alpha$ EW in the composite spectra was measured in the same manner as in the individual spectra, except that there are only ``emission" and ``combination" morphologies given the higher S/N of the composites and the predominant presence of Ly$\alpha$ emission in the individual spectra. We perturbed the composite science spectra 100 times with the corresponding composite error spectra, and measured the Ly$\alpha$ EW in the 100 fake composite spectra. The sigma-clipped average and standard deviation of those 100 measurements were adopted as the final Ly$\alpha$ EW and the $1\sigma$ uncertainty, respectively, for each composite spectrum.

\subsubsection{Other Rest-UV Emission Lines}
\label{sec:absorption lines}

Within the coverage of the LRIS spectra there are many rest-frame far-UV emission features in addition to Ly$\alpha$. These include \textrm{C}~\textsc{iv}$\lambda\lambda1548,1550$,  \textrm{He}~\textsc{ii}$\lambda1640$, \textrm{O}~\textsc{iii}]$\lambda\lambda$1661,1665, and  \textrm{C}~\textsc{iii}]$\lambda\lambda$1907,1909. The \textrm{C}~\textsc{iii}] emission doublet is the strongest feature in the far-UV after Ly$\alpha$, which enables us to model the line profile in individual spectra. The \textrm{C}~\textsc{iii}] EW measurement was performed on continuum-normalized spectra, using the method outlined in \citet{Du2018}. In short, the rest-frame spectra were continuum normalized using spectral regions identified in \citet{Rix2004} that are clean of spectral features, and with a $spline3$ function in the IRAF $continuum$ routine. We used order $=8$ in the $spline3$ function to model the continuum regions near the \textrm{C}~\textsc{iii}] feature. Additional spectral regions customized for each object were added to provide a reasonable fit to the continuum when the original regions failed to provide a proper description of the observed spectrum.

To fit the \textrm{C}~\textsc{iii}] emission profile, we used MPFIT with the initial values of continuum flux level, line centroid, EW and Gaussian FWHM estimated from the IRAF routine $splot$. The best-fit parameters were determined where the $\chi^{2}$ of the fit reached a minimum, and the 1$\sigma$ error bar associated with each parameter was derived from the covariance matrix. We then iterated the fitting over a narrower wavelength range: centroid$-4\sigma < \lambda < $centroid$+4\sigma$, where the centroid and $\sigma$ are, respectively, the returned central wavelength and standard deviation of the best-fit Gaussian profile from the initial MPFIT fit to the \textrm{C}~\textsc{iii}] profile over $\lambda_{rest}-10\mbox{\AA}$ to $\lambda_{rest}+10\mbox{\AA}$. We list the individual measurements of \textrm{C}~\textsc{iii}] EW in Table \ref{tab:ew}.

Given the weak nature of \textrm{C}~\textsc{iv}, \textrm{O}~\textsc{iii}], and \textrm{He}~\textsc{ii}, these features are not detected in most individual spectra. As a result, we only modeled these lines in composite spectra and in special cases where they are detected on an individual basis (e.g., COSMOS-4064 and COSMOS-7686, see Section \ref{sec:case} for details). We followed the same procedures described above for fitting the line profiles of \textrm{He}~\textsc{ii} and \textrm{O}~\textsc{iii}]$\lambda1665$ (\textrm{O}~\textsc{iii}]$\lambda1661$ is typically much weaker than \textrm{O}~\textsc{iii}]$\lambda1665$ in the spectra, and is therefore hardly detected even in composite spectra or in the spectra of COSMOS-4064 and COSMOS-7686). As for \textrm{C}~\textsc{iv}, the observed profile is a superposition of interstellar \textrm{C}~\textsc{iv} and a P-Cygni profile that originates from the stellar wind common
in O and B stars. The measurement of the interstellar \textrm{C}~\textsc{iv} emission may be biased without properly removing the underlying P-Cygni profile. Therefore, we determined and separated the 
stellar component following the method presented in \citet{Du2016}. In short, the blue wing of the observed \textrm{C}~\textsc{iv} absorption profile was bracketed by two spectra of synthetic stellar populations from the model grids of \citet{Leitherer2010}. The two bracketing models are at adjacent grid-points in metallicity. We produced a linear combination of these two model spectra such that the resulting combined spectrum produces the best fit to the blue wing of the observed \textrm{C}~\textsc{iv} profile. We then divided out the best-fit model from the observed spectrum. The remaining interstellar \textrm{C}~\textsc{iv} emission profile was modeled
with a single Gaussian, given that the \textrm{C}~\textsc{iv} doublet is not resolved in the LRIS spectra.

\section{Results}
\label{sec:results}

\subsection{Ly$\alpha$ and \textrm{C}~\textsc{iii}] EW Distributions}
\label{sec:ew_hist}

\begin{figure*}
\includegraphics[width=0.5\linewidth]{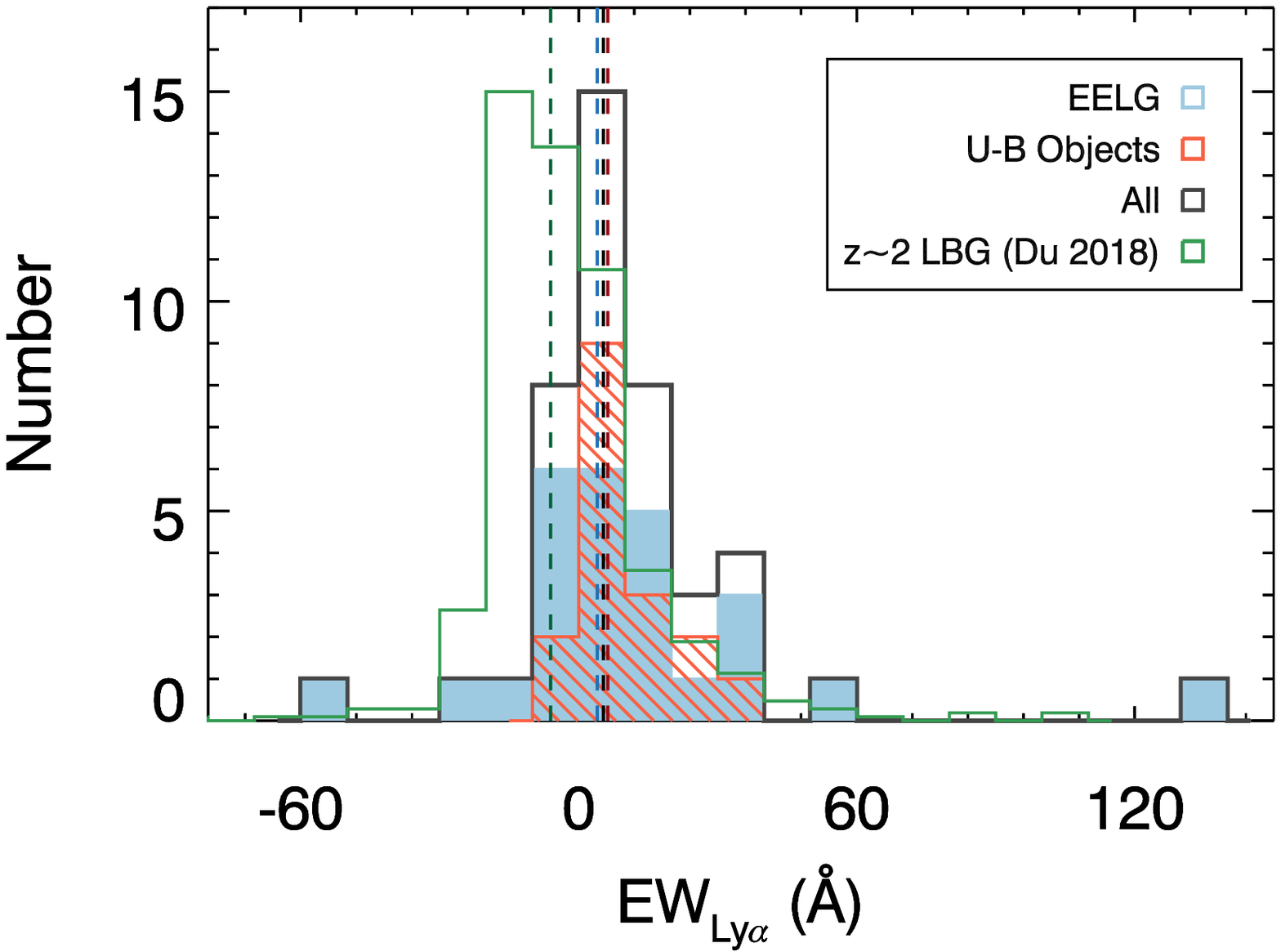}
\includegraphics[width=0.5\linewidth]{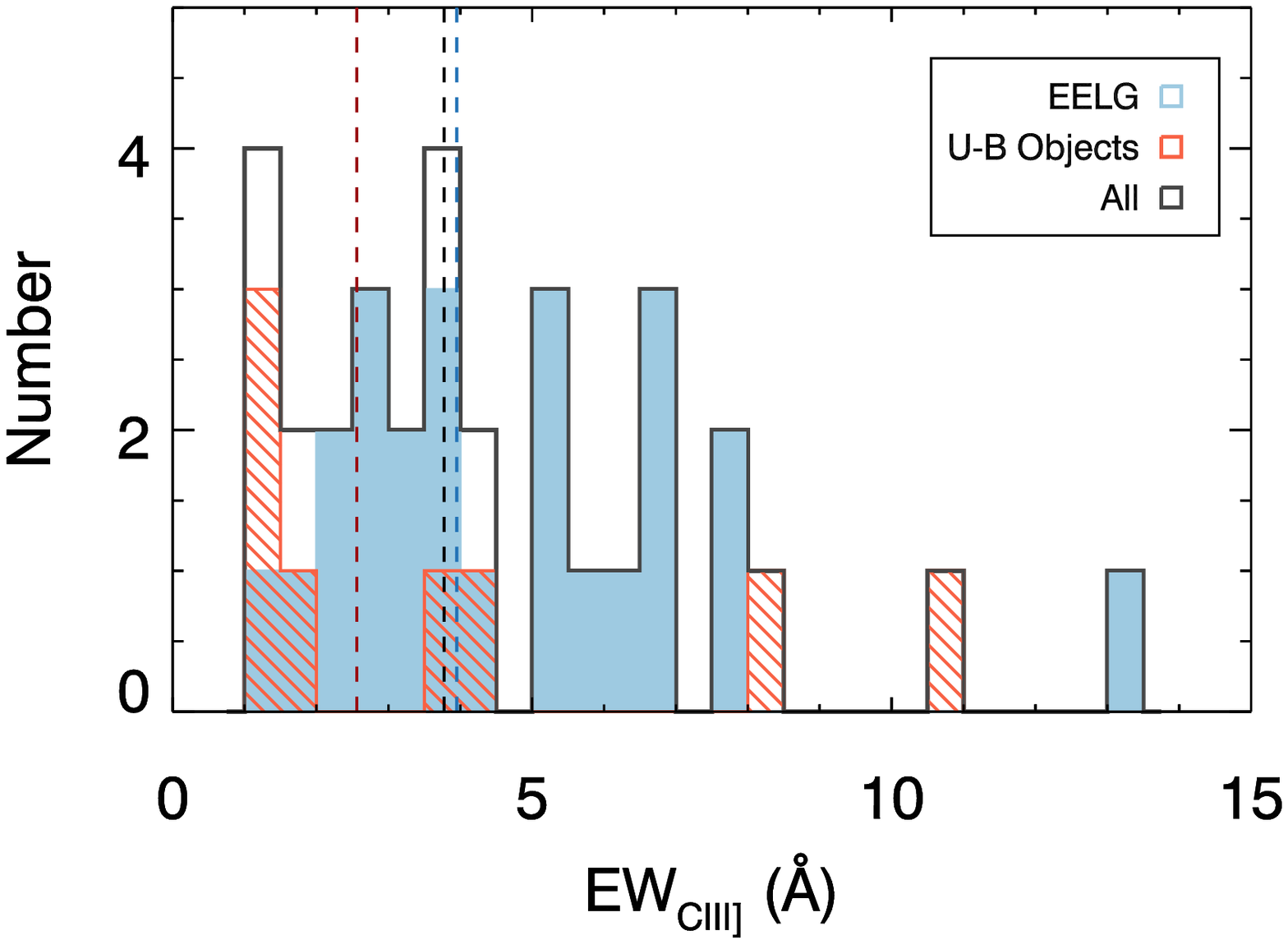}

\caption{\textbf{Left:} Rest-frame Ly$\alpha$ EW distribution of the LRIS sample. Color coding of the histograms is the same as in Figure \ref{fig:galprop}. The open green bar indicates the $z\sim2$ LBG sample presented in \citet{Du2018}, normalized to the same vertical scale as the overall LRIS sample (black open bar). The median Ly$\alpha$ EW of each subgroup is shown with a dashed vertical line in the corresponding color. The median EWs are 5.3 $\mbox{\AA}$, 4.0 $\mbox{\AA}$, and 6.3 $\mbox{\AA}$, respectively, for the overall LRIS sample (43 objects), EELGs (26 objects), and $\ub$ objects (17 objects) with Ly$\alpha$ EW measurements. Our new LRIS sample of EELGs and blue $\ub$ objects  has a systematically higher median Ly$\alpha$ EW than that of the continuum-selected $z\sim2$ LBGs (median EW -6.1 $\mbox{\AA}$). \textbf{Right:} Rest-frame \textrm{C}~\textsc{iii}]$\lambda\lambda1907,1909$ EW distribution of the LRIS sample. Objects shown in the histogram have $\geqslant3\sigma$ \textrm{C}~\textsc{iii}] EW detections. Color coding of the histograms and dashed lines is the same as in the left panel. The median \textrm{C}~\textsc{iii}] EWs of the overall LRIS sample (32 objects), EELGs (24 objects), and $\ub$ objects (8 objects) are 3.8 $\mbox{\AA}$, 4.0 $\mbox{\AA}$, and 2.6 $\mbox{\AA}$, respectively.}
\label{fig:ew_hist}
\end{figure*}

We measured Ly$\alpha$ EWs based on the method described in Section \ref{sec:lya} for 43 objects with valid redshift measurements and Ly$\alpha$ coverage in our sample. As shown in the left panel of Figure \ref{fig:ew_hist}, the median Ly$\alpha$ EW is very similar between EELGs and $\ub$ objects, although EELGs span a wider range in Ly$\alpha$ EW. The EELGs range in Ly$\alpha$ EW from $-55.8 \mbox{\AA}$ to 131.6 $\mbox{\AA}$ with a median of 4.0 $\mbox{\AA}$, while the $\ub$ objects range from $-8.6 \mbox{\AA}$ to 37.6 $\mbox{\AA}$ with a median of 6.3 $\mbox{\AA}$. Our LRIS sample has a systemically higher average Ly$\alpha$ EW compared to the $z\sim2$ Lyman Break Galaxy (LBG) sample presented in \citet{Du2018}, reflecting the different sample selection criteria (i.e., strong rest-optical nebular emission for EELGs vs. rest-UV continuum selection for LBGs). In terms of the fraction of Ly$\alpha$ emitters (LAEs; rest-frame EW$_{Ly\alpha}\geqslant20\mbox{\AA}$), the percentage is 23$\%$ for EELGs, and 18$\%$ for $\ub$ objects. This value is higher than the LAE percentage in the $z\sim2$ LBGs \citep[8$\%$; ][]{Du2018}, but very similar to that in the $z\sim3$ LBG sample \citep[25$\%$;][]{Shapley2003}.

Forty-eight objects in the LRIS sample have spectral coverage of \textrm{C}~\textsc{iii}]$\lambda\lambda1907,1909$. All of these objects have a \textrm{C}~\textsc{iii}] measurement except for AEGIS-12032, the red-side spectrum of which is contaminated by a nearby source. In the right panel of Figure \ref{fig:ew_hist}, we plot 32 objects with significant \textrm{C}~\textsc{iii}] detection (EW$_{\textrm{C}~\textsc{iii}]}\geqslant3\sigma$). The EELGs have stronger typical \textrm{C}~\textsc{iii}] emission than the $\ub$ objects, spanning in \textrm{C}~\textsc{iii}] EW from 1.0 $\mbox{\AA}$ to 13.2 $\mbox{\AA}$ with a median of 4.0 $\mbox{\AA}$. The $\ub$ objects, on the other hand, range in \textrm{C}~\textsc{iii}] EW from 1.1 $\mbox{\AA}$ to 10.5 $\mbox{\AA}$, with a median of 2.6 $\mbox{\AA}$. 37$\%$ EELGs and 12$\%$ $\ub$ objects with \textrm{C}~\textsc{iii}] measurements have a rest-frame \textrm{C}~\textsc{iii}] EW $\geqslant5\mbox{\AA}$, and therefore qualify as \textrm{C}~\textsc{iii}] emitters \citep[as defined in][]{Rigby2015}. In comparison, $z\sim1$ star-forming galaxies have a median \textrm{C}~\textsc{iii}] EW of $\sim1\mbox{\AA}$ with no single \textrm{C}~\textsc{iii}] emitter identified out of 183 \textrm{C}~\textsc{iii}] detections \citep{Du2017}. As for the $z\sim2$ LBG sample presented in \citet{Du2018}, while individual \textrm{C}~\textsc{iii}] EW measurements were not available, \textrm{C}~\textsc{iii}] in the composite made out of the highest-Ly$\alpha$ quartile only has an EW of $3.4\mbox{\AA}$. Both the median \textrm{C}~\textsc{iii}] EW and the detection rate of \textrm{C}~\textsc{iii}] emitters indicate that the EELG/$\ub$ objects we selected here are distinct from the typical star-forming galaxy population at similar redshifts. However, we note that our LRIS sample has a much lower characteristic \textrm{C}~\textsc{iii}] EW compared to those observed at $z>6.5$, which are typically $\gtrsim10\mbox{\AA}$ \citep{Zitrin2015,Stark2015,Stark2017,Mainali2018,Hutchison2019}. It is worth keeping in mind that these individual, large \textrm{C}~\textsc{iii}] EW measurements at $z>6.5$ are likely not representative of the average \textrm{C}~\textsc{iii}] strength among $z>6.5$ star-forming galaxies, due to observational bias. The fact that only 2 out of 32 LRIS objects with \textrm{C}~\textsc{iii}] detections show a $\geqslant10\mbox{\AA}$ \textrm{C}~\textsc{iii}] EW indeed signals the rarity of these intense \textrm{C}~\textsc{iii}] emitters. In Section~\ref{sec:search}, we present a more detailed discussion of how representative our LRIS sample is, in the context of extreme and typical star-forming galaxies at $z>6.5$.

\subsection{Rest-UV Spectra vs. [\textrm{O}~\textsc{iii}]}
\label{sec:line_corr}

\begin{figure*}

\includegraphics[width=0.5\linewidth]{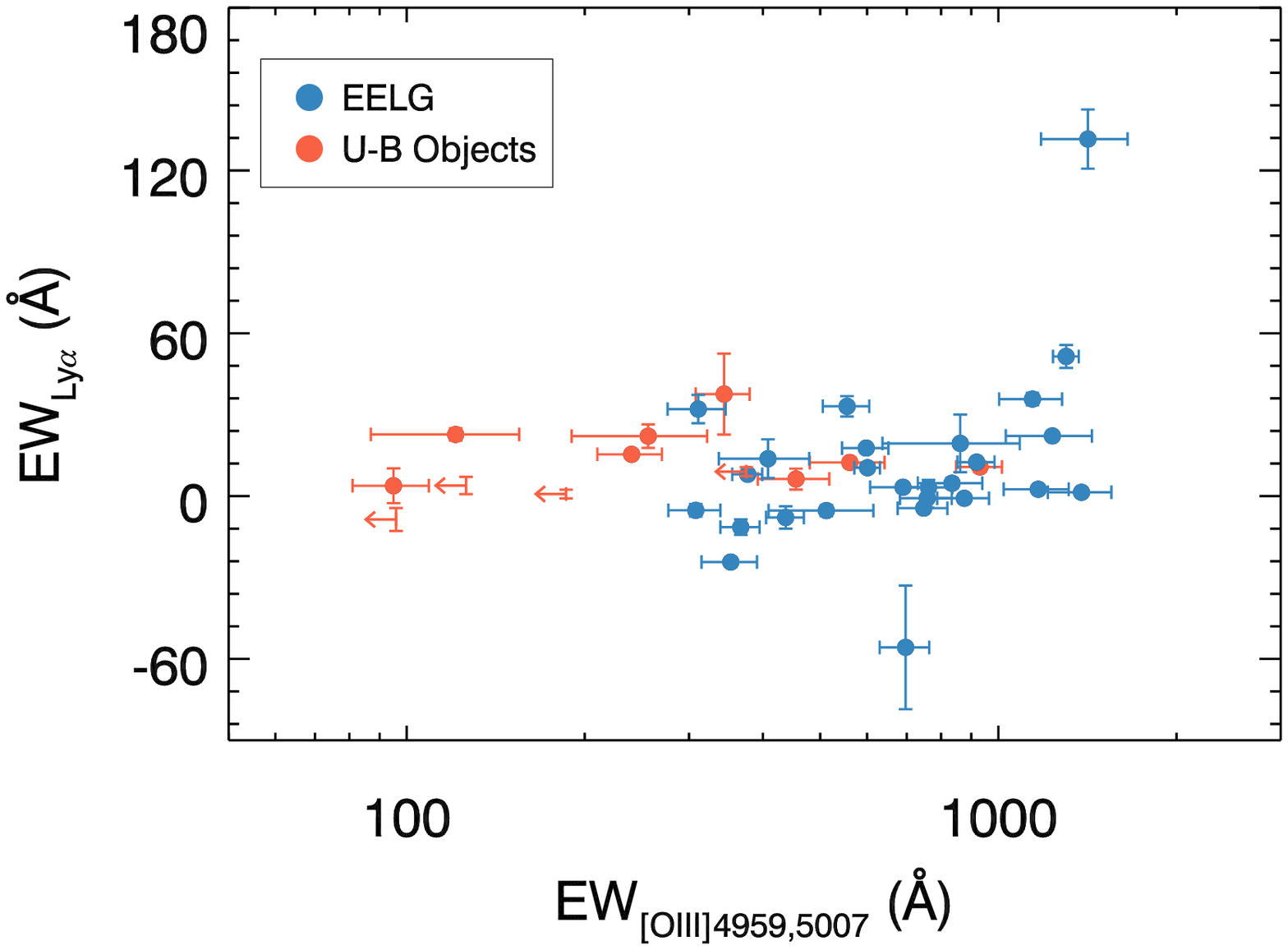}
\includegraphics[width=0.5\linewidth]{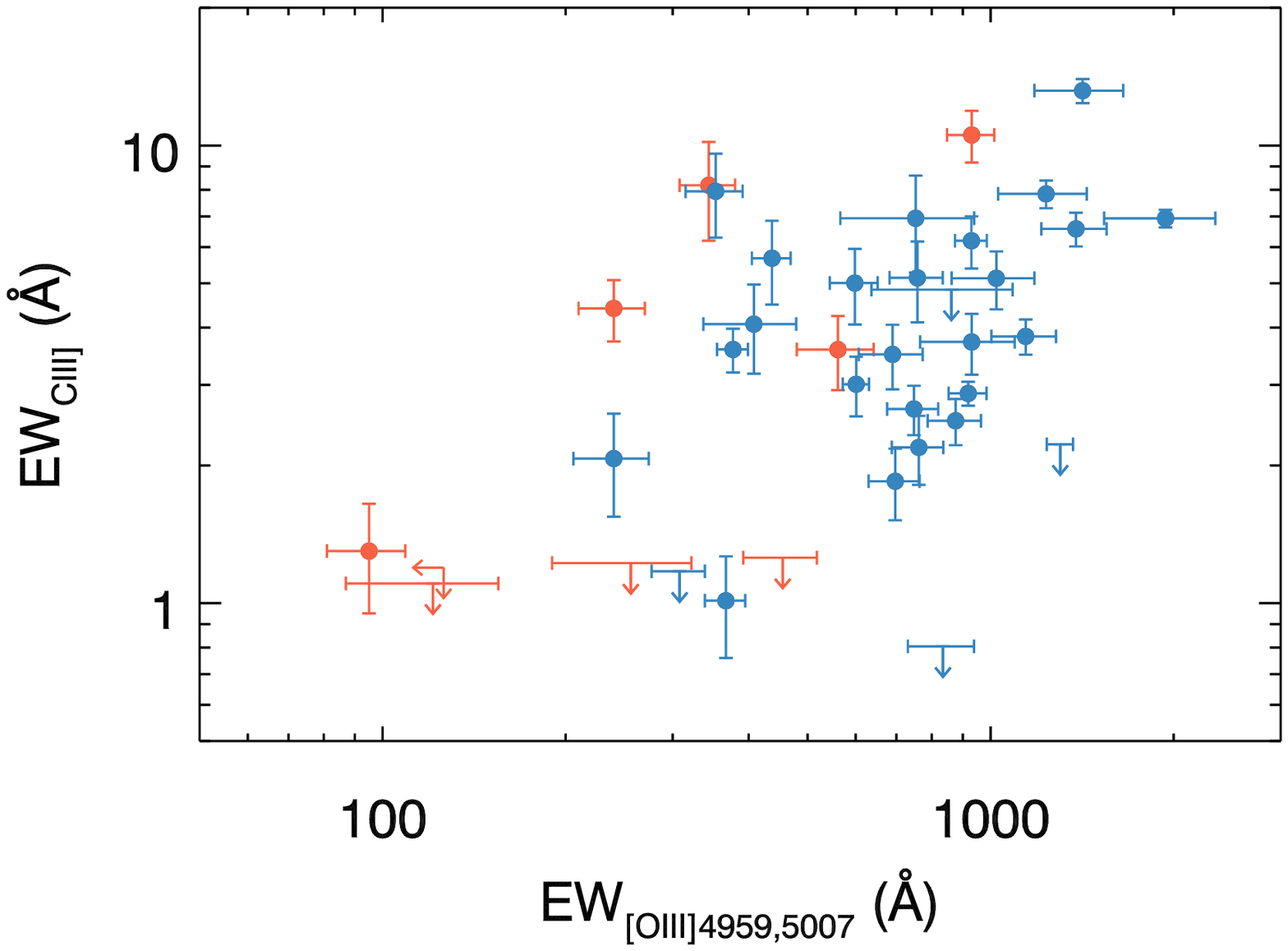}
\includegraphics[width=0.5\linewidth]{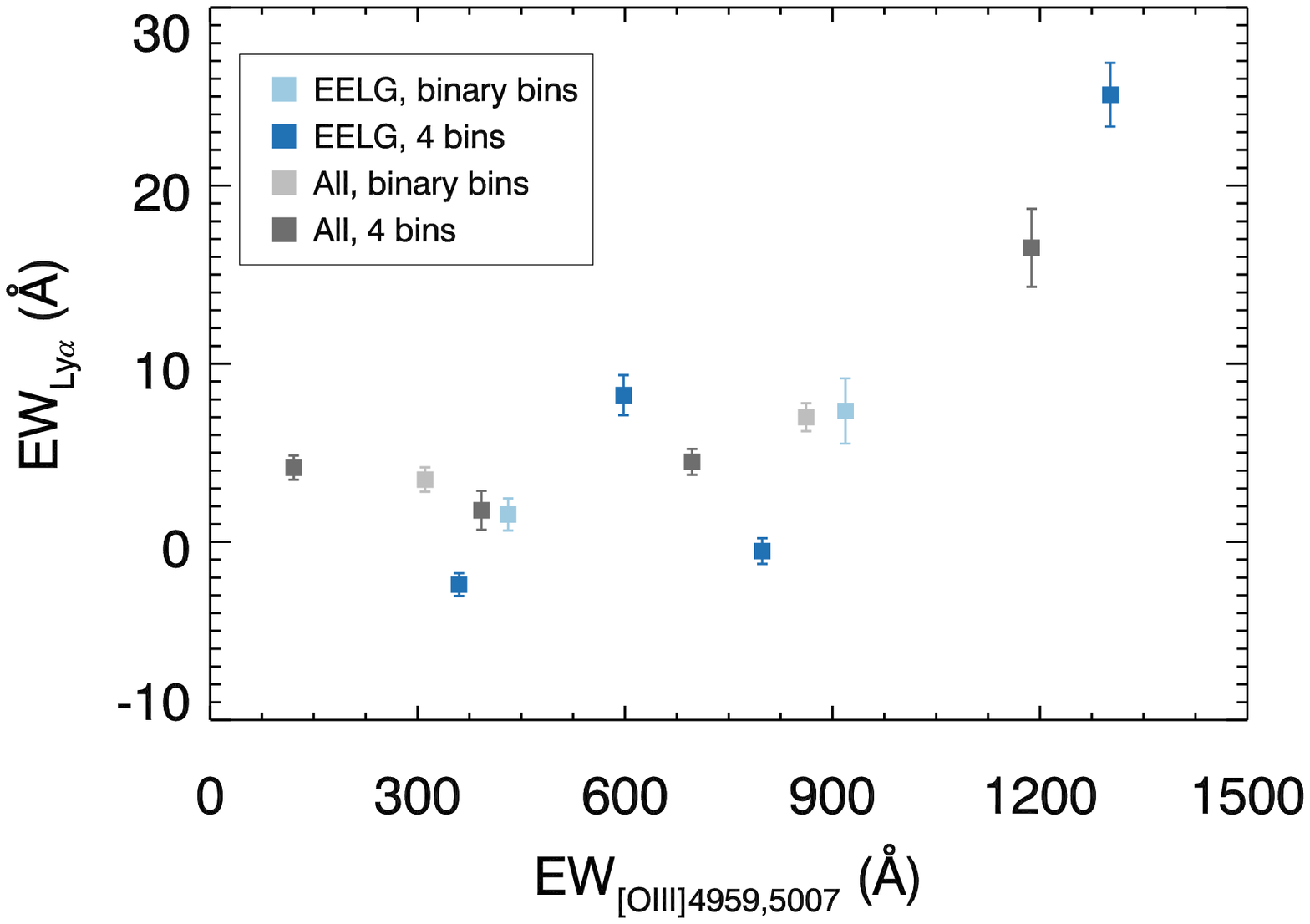}
\includegraphics[width=0.5\linewidth]{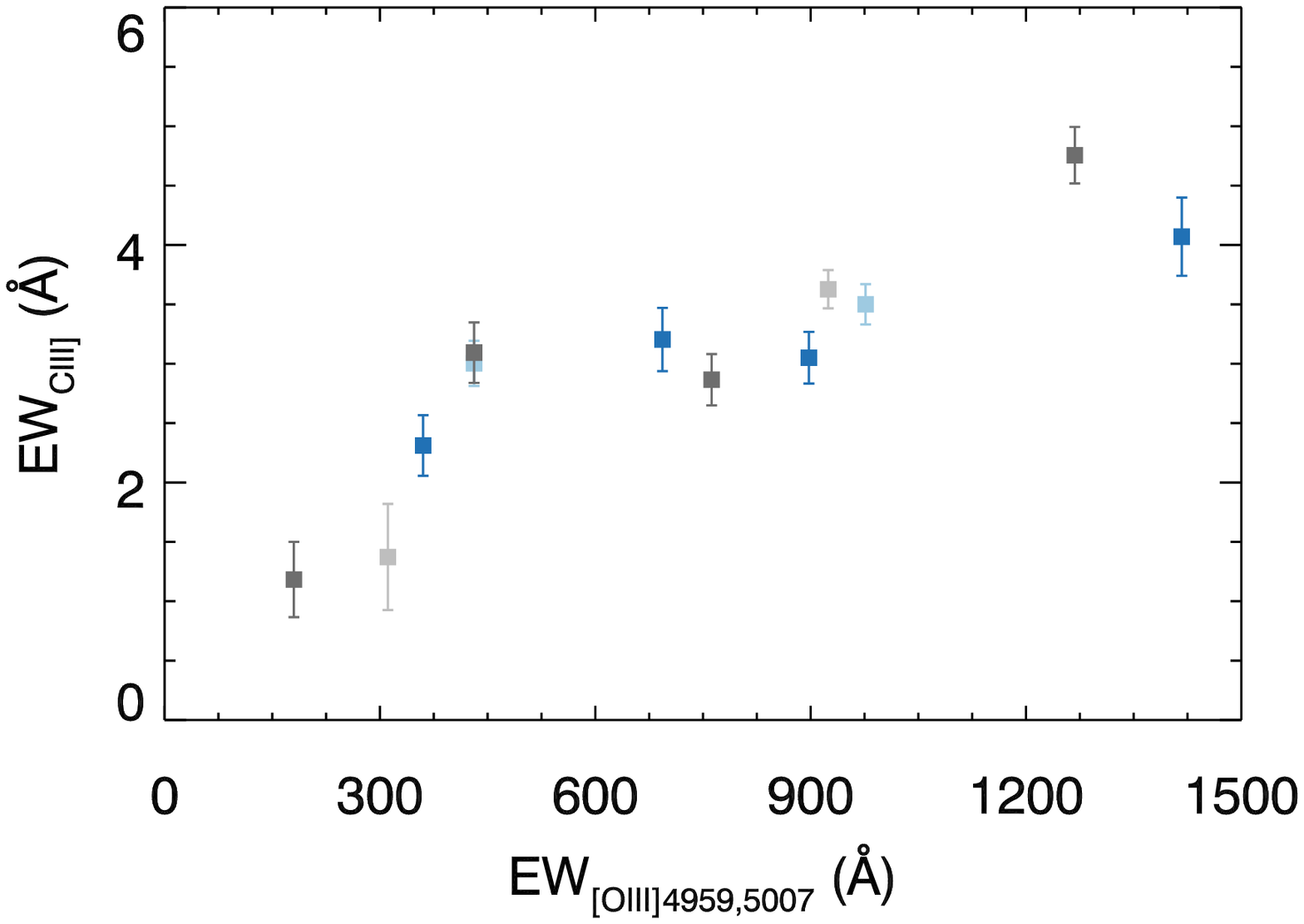}

\caption{Ly$\alpha$ EW (left) and \textrm{C}~\textsc{iii}] EW (right) vs. [\textrm{O}~\textsc{iii}]$\lambda\lambda4959,5007$ EW from individual measurements (top) and composite spectra binned according to [\textrm{O}~\textsc{iii}] EW (bottom). Color coding of the upper panels is the same as in Figure \ref{fig:galprop}. Error bars shown in the figures indicate $1\sigma$ uncertainties for detected lines (EW $\geqslant3\sigma$), while left- and downward-pointing arrows mark $3\sigma$ upper limits for non-detections. The bottom panels show the Ly$\alpha$ (left) and \textrm{C}~\textsc{iii}] EWs measured from composites including all LRIS targets (black) and only EELGs (blue) with Ly$\alpha$ or \textrm{C}~\textsc{iii}] coverage. Measurements from binary bins (quartiles) are shown in lighter (darker) corresponding colors. The median [\textrm{O}~\textsc{iii}] EW in individual bins is plotted.}
\label{fig:strem_oiiistack}
\end{figure*}

One of the key goals of this work is to 
test the correspondence between the strength of rest-optical emission lines (e.g., [\textrm{O}~\textsc{iii}]) and those in the rest-UV (e.g., Ly$\alpha$, \textrm{C}~\textsc{iv}, \textrm{He}~\textsc{ii}, \textrm{O}~\textsc{iii}], and \textrm{C}~\textsc{iii}]),
In particular, we examined how the EWs of Ly$\alpha$ and \textrm{C}~\textsc{iii}], the two most prominent emission features in the rest-UV, correlate with the strength of [\textrm{O}~\textsc{iii}]$\lambda\lambda4959,5007$. In Figure \ref{fig:strem_oiiistack}, we plot Ly$\alpha$ (left) and \textrm{C}~\textsc{iii}] EWs (right) versus [\textrm{O}~\textsc{iii}] EW, in both individual measurements and from composite spectra. We performed two types of stacking, binary and quartile, binned according to [\textrm{O}~\textsc{iii}] EW. Bins were constructed to contain roughly equal numbers of galaxies.  Binary stacks have higher S/N because a larger number of galaxies are included in each bin, and are a good tracer of the overall trend between the line strengths within the dynamic range probed. The quartile composites, on the other hand, are useful in revealing underlying trends, if any, that are obscured by a simple binary division. We also created the composite spectra using the entire LRIS sample (both EELGs and $\ub$ objects with valid redshifts and available [\textrm{O}~\textsc{iii}]$\lambda\lambda4959,5007$ measurements; 44 objects) and EELGs only (32 objects with reliable redshifts, and [\textrm{O}~\textsc{iii}] EW measurements satisfying the criteria described in Section~\ref{sec:selection}), in case EELGs and $\ub$ objects follow distinct relations between rest-optical and rest-UV lines.

The strengths of Ly$\alpha$ and \textrm{C}~\textsc{iii}] show strikingly different trends with [\textrm{O}~\textsc{iii}] EW. Both individual measurements and composite spectra suggest that the Ly$\alpha$ EW stays fairly flat with [\textrm{O}~\textsc{iii}] EW when EW$_{[\textrm{O}~\textsc{iii}]}\lesssim1000 \mbox{\AA}$. Among 9 objects with EW$_{\textrm{Ly}\alpha}>20 \mbox{\AA}$, roughly half of them (4 out of 9) have an [\textrm{O}~\textsc{iii}] EW $>1000 \mbox{\AA}$. 
The composite spectra also suggest that the Ly$\alpha$ emission does not become prominent ($\gtrsim 20 \mbox{\AA}$) in stacks except for one representing the strongest-[\textrm{O}~\textsc{iii}] quartile, where the median [\textrm{O}~\textsc{iii}] is well above 1000 $\mbox{\AA}$. This relatively flat trend between Ly$\alpha$ and [\textrm{O}~\textsc{iii}] at lower [\textrm{O}~\textsc{iii}] EW (EW$_{[\textrm{O}~\textsc{iii}]}\lesssim1000 \mbox{\AA}$) indicates that the observed Ly$\alpha$ and [\textrm{O}~\textsc{iii}] EWs are not correlated on the lower [\textrm{O}~\textsc{iii}]-EW end. Although the intrinsic production rate of Ly$\alpha$ photons in general increases with increasing [\textrm{O}~\textsc{iii}] EW (as a 
result of the stronger ionizing radiation fields produced by the associated young, metal-poor stars), the dominant factor modulating the strength of Ly$\alpha$ emission in this regime appears to be the resonant scattering of Ly$\alpha$ photons by the neutral hydrogen gas in the galaxies. This effect is reflected in the well-explored anti-correlation between Ly$\alpha$ and LIS EWs \citep{Shapley2003,Du2018}, along with the fact that the LIS EW is a proxy for the covering fraction of neutral gas \citep{Reddy2016}. 
Our results suggest that, at EW$_{[\textrm{O}~\textsc{iii}]}\lesssim1000 \mbox{\AA}$, differences in the neutral gas covering fraction along the line-of-sight play a more important role than differences in the intrinsic Ly$\alpha$ production rate in determining the variations in the observed Ly$\alpha$ EW.

Galaxies in the quartile with the highest [\textrm{O}~\textsc{iii}] EW are among the lowest metallicity systems probed within the EELG sample, as suggested by the negative correlation between oxygen abundance and [\textrm{O}~\textsc{iii}] EW found in previous studies \citep[e.g.,][]{Jones2015} and Figure \ref{fig:metal_scatter}. At the same time, environments favorable for [\textrm{O}~\textsc{iii}] production are also conducive to higher Ly$\alpha$ production and large Ly$\alpha$ escape fractions ($f_{esc,Ly\alpha}$). The hard ionizing spectrum associated with metal-poor star formation in these galaxies not only boosts the intrinsic Ly$\alpha$ production, but also further ionizes the neutral HI gas in the ISM. This effect in turn reduces the covering fraction of neutral gas, enhancing the escape of Ly$\alpha$ photons and resulting in a larger Ly$\alpha$ EW \citep{Trainor2015,Erb2016}. Accordingly, at the high-[\textrm{O}~\textsc{iii}] extreme of the EELG sample, the increase in average Ly$\alpha$ EW traces an increase in the intrinsic  Ly$\alpha$ production rate -- not only a difference in the neutral gas covering fraction.

When examining the relation between \textrm{C}~\textsc{iii}] and [\textrm{O}~\textsc{iii}] EWs, we find that stronger \textrm{C}~\textsc{iii}] appears in galaxies with stronger [\textrm{O}~\textsc{iii}], as shown in both individual measurements and the stacks. This monotonic trend between \textrm{C}~\textsc{iii}] and [\textrm{O}~\textsc{iii}] is not surprising, as both features are nebular, collisionally excited transitions whose strengths peak in \textrm{H}~\textsc{ii} regions with strong radiation fields and relatively low metallicities \citep{Jaskot2016,Senchyna2017,Maseda2017,Nakajima2018aa, Mainali2019}. The similar dependence on \textrm{H}~\textsc{ii} region properties makes [\textrm{O}~\textsc{iii}] a good tracer of \textrm{C}~\textsc{iii}], providing an important pathway in searching for potential \textrm{C}~\textsc{iii}] emitters based on rest-optical spectra.

\begin{figure*}
\includegraphics[width=0.5\linewidth]{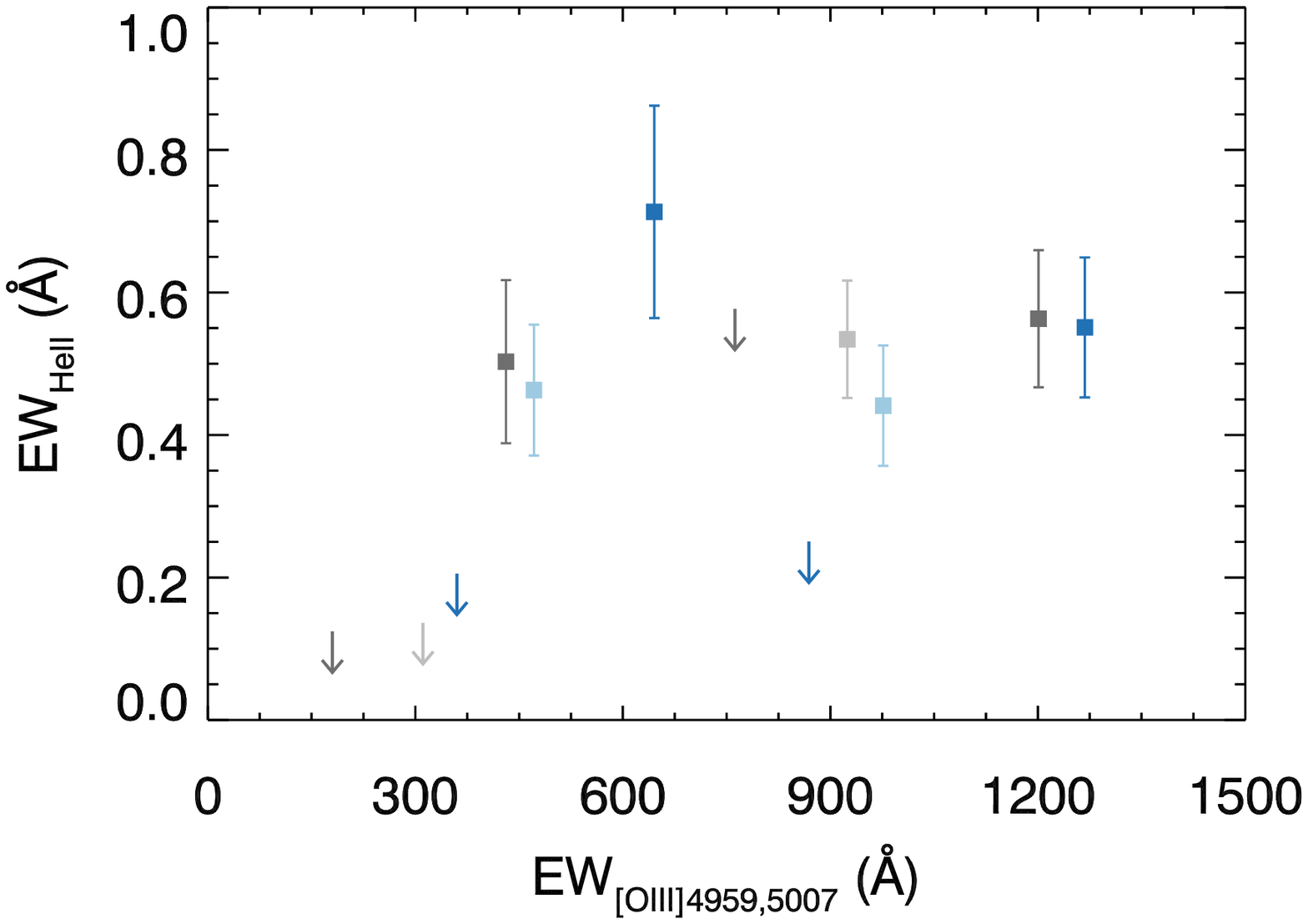}
\includegraphics[width=0.5\linewidth]{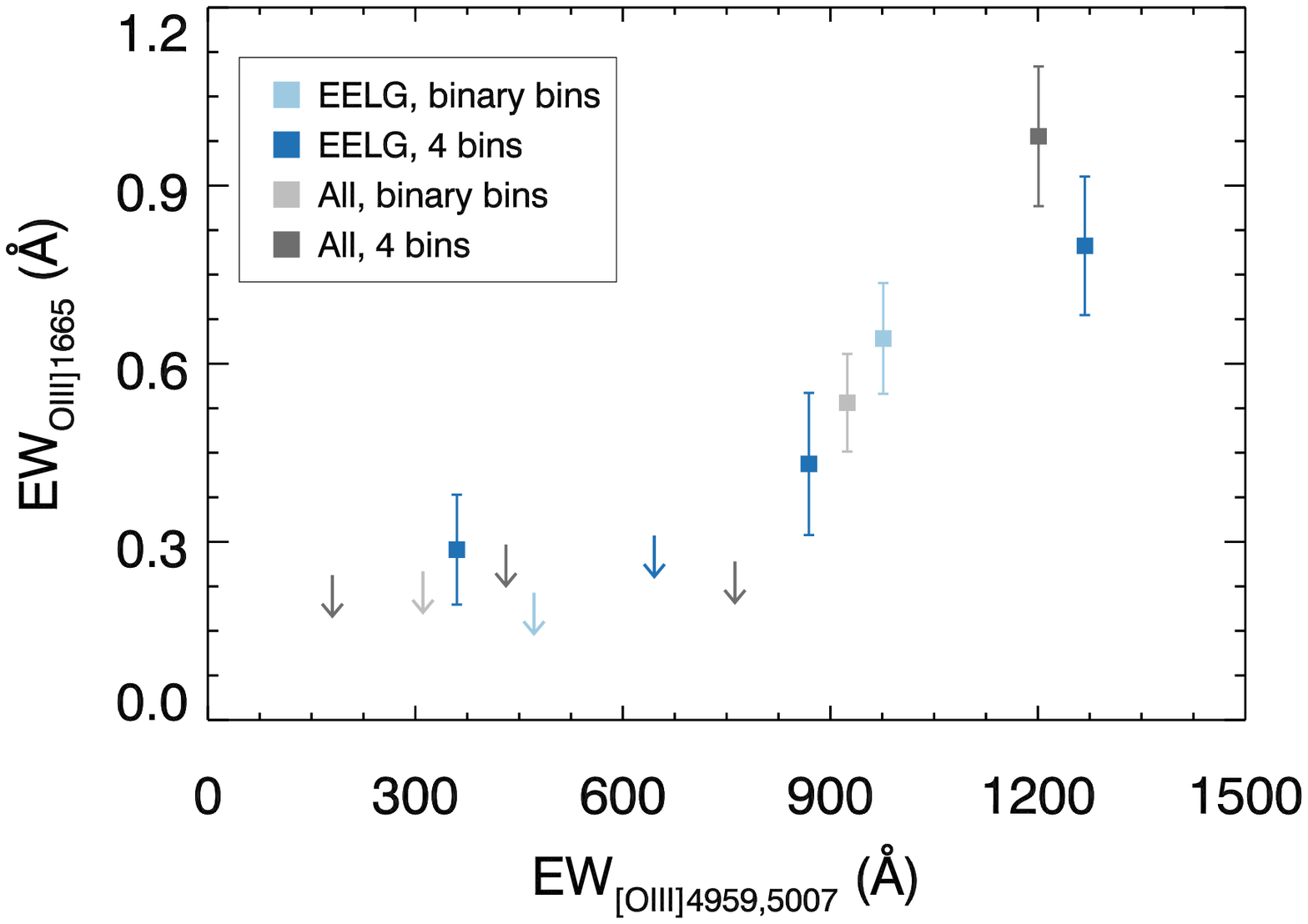}

\caption{\textrm{He}~\textsc{ii} and \textrm{O}~\textsc{iii}]$\lambda1665$ EWs vs. [\textrm{O}~\textsc{iii}]$\lambda\lambda4959,5007$ EW in composite spectra binned according to [\textrm{O}~\textsc{iii}]$\lambda\lambda4959,5007$ EW. Color coding of the symbols is the same as in the bottom panels of Figure \ref{fig:strem_oiiistack}. Downward-pointing arrows mark $3\sigma$ upper limits for \textrm{He}~\textsc{ii} or \textrm{O}~\textsc{iii}]$\lambda1665$ non-detections in the composites.}
\label{fig:weakem_oiiistack}
\end{figure*}

Lastly, we explored how the strengths of weaker rest-UV transitions, i.e., \textrm{He}~\textsc{ii} and \textrm{O}~\textsc{iii}]$\lambda\lambda1661,1665$, depend on [\textrm{O}~\textsc{iii}]$\lambda\lambda4959,5007$ EW. Considering that these weak lines are not typically detected on an individual basis, we measured the EW of \textrm{He}~\textsc{ii} and \textrm{O}~\textsc{iii}] only in composite spectra. As shown in Figure \ref{fig:weakem_oiiistack}, \textrm{He}~\textsc{ii} is not significantly detected in stacks with a median EW$_{[\textrm{O}~\textsc{iii}]}\lesssim450 \mbox{\AA}$. The EW of detected \textrm{He}~\textsc{ii} stays largely the same ($\sim0.5 \mbox{\AA}$) with increasing [\textrm{O}~\textsc{iii}] EW above the $\sim450 \mbox{\AA}$ threshold. As the \textrm{He}~\textsc{ii} profile is a superposition of a narrow, nebular component and a broad component originating from stellar winds, EW measurements alone cannot capture the detailed change in the line profile. \citet{Senchyna2017} demonstrate that the profile of \textrm{He}~\textsc{ii} varies with gas-phase metallicity, such that at lower metallicity (higher [\textrm{O}~\textsc{iii}]$\lambda\lambda4959,5007$ EW), the nebular component becomes stronger while the stellar component becomes weaker. We do not observe an apparent, monotonic change in the width of the \textrm{He}~\textsc{ii} emission line across the [\textrm{O}~\textsc{iii}] stacks, which possibly results from the fact that the metallicity regime 12+log(O/H)$\lesssim8.0$ is not being probed in our LRIS sample (also see Section \ref{sec:metal}). 

As for the measurement of \textrm{O}~\textsc{iii}]$\lambda\lambda1661,1665$, the weaker doublet member, \textrm{O}~\textsc{iii}]$\lambda1661$, is not detected in all [\textrm{O}~\textsc{iii}] stacks. Hence, we report the results based on \textrm{O}~\textsc{iii}]$\lambda1665$ in Figure \ref{fig:weakem_oiiistack}\footnote{Given the fixed ratio of 1:2.5 for \textrm{O}~\textsc{iii}]$\lambda 1661$ and \textrm{O}~\textsc{iii}]$\lambda1665$, this correlation also holds for the sum of the rest-UV \textrm{O}~\textsc{iii}] doublet, up to a scaling factor.}. Similar to \textrm{He}~\textsc{ii}, \textrm{O}~\textsc{iii}]$\lambda1665$ is not detected in most cases below a certain threshold, EW$_{[\textrm{O}~\textsc{iii}]}\lesssim900 \mbox{\AA}$. However, the EW of \textrm{O}~\textsc{iii}]$\lambda1665$ increases with that of [\textrm{O}~\textsc{iii}]$\lambda\lambda4959,5007$ above the threshold, and reaches $\sim 1.0\mbox{\AA}$ in the highest-[\textrm{O}~\textsc{iii}] quartile. The positive trend between \textrm{O}~\textsc{iii}]$\lambda1665$ and [\textrm{O}~\textsc{iii}]$\lambda\lambda4959,5007$ stems from the fact that these features are both enhanced in lower-metallicity environments with both harder stellar ionizing spectra and hotter electron temperatures.

\subsection{Rest-UV Spectra vs. Metallicity}
\label{sec:metal}

\begin{figure}
\includegraphics[width=1.0\linewidth]{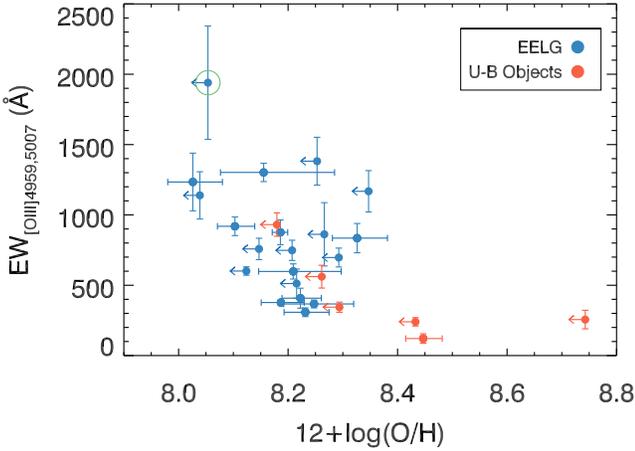}
\caption{[\textrm{O}~\textsc{iii}] EW vs. oxygen abundance from individual measurements in the LRIS sample. Color coding of the symbols is the same as in Figure \ref{fig:strem_oiiistack}. The green open circle marks COSMOS-4064.}
\label{fig:metal_scatter}
\end{figure}

\begin{figure}
\includegraphics[width=1.0\linewidth]{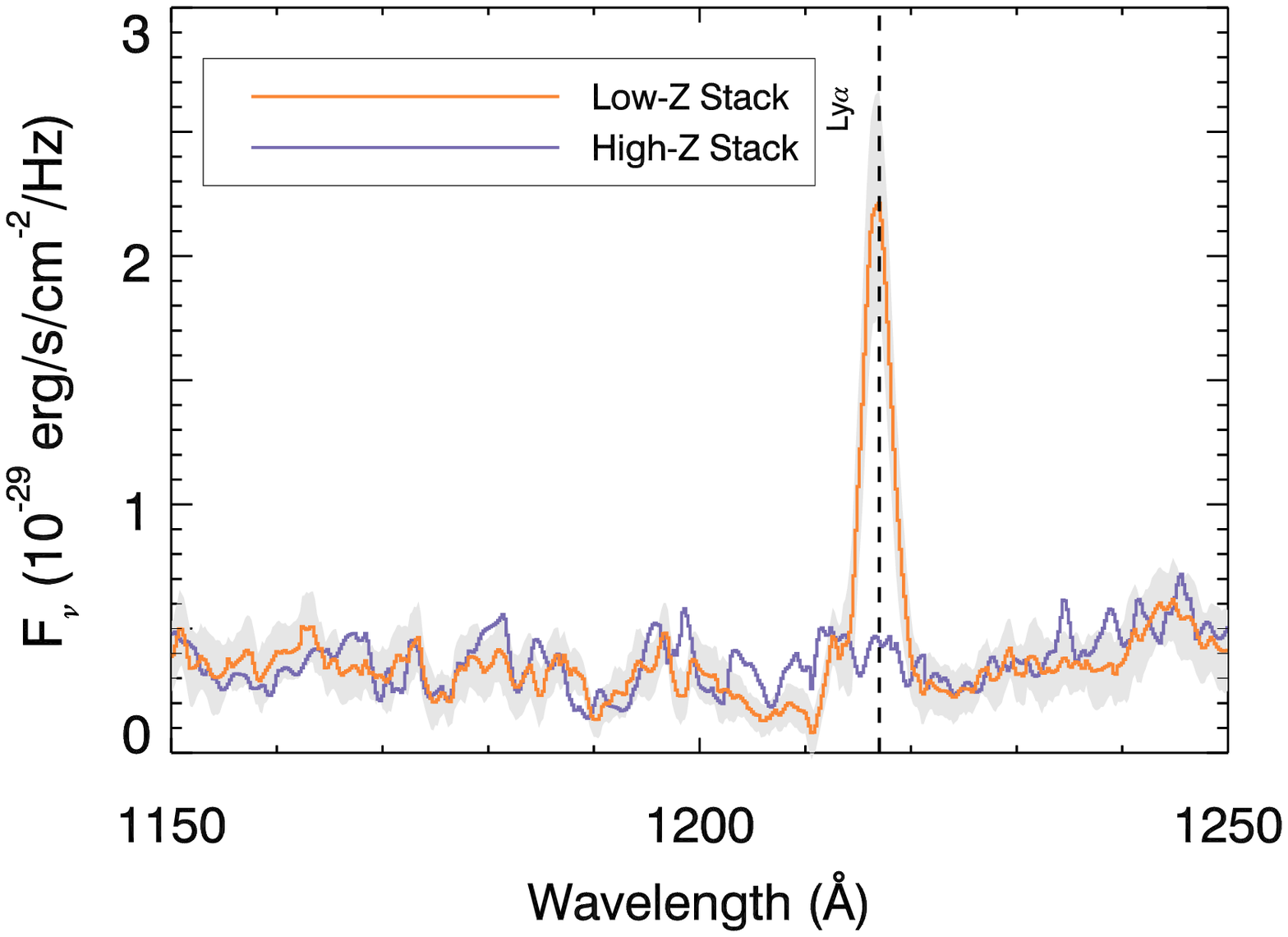}
\includegraphics[width=1.0\linewidth]{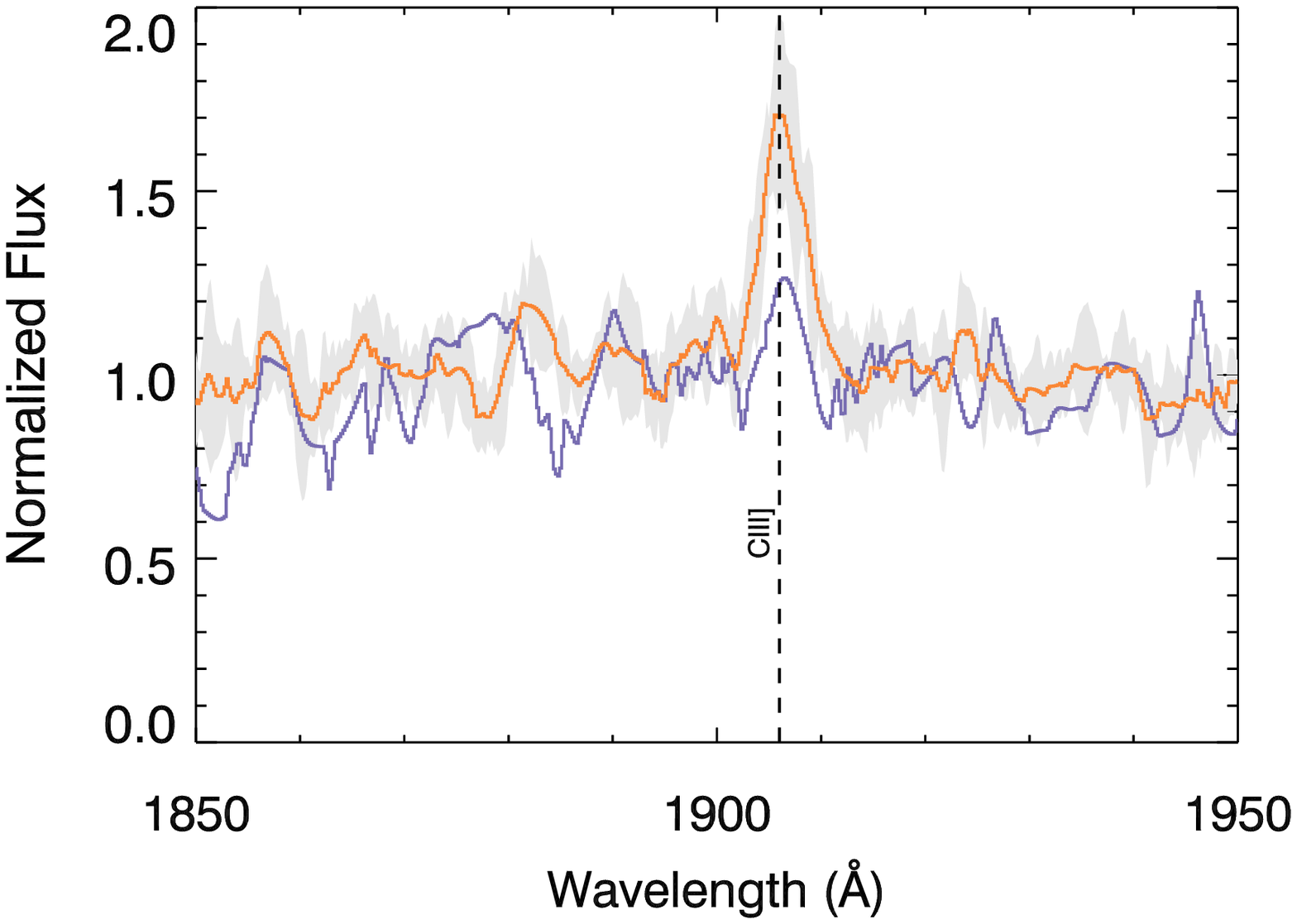}
\caption{Comparison of the rest-UV spectra for the low- (orange) and high-metallicity (purple) stacks, zoomed in on regions near Ly$\alpha$ (top) and \textrm{C}~\textsc{iii}] (bottom). The region near Ly$\alpha$ is shown in flux density space, while that near \textrm{C}~\textsc{iii}] is shown in normalized flux. Dashed black lines mark the rest-frame wavelengths of key spectral features. The grey shaded area indicates the $1\sigma$ uncertainty range of the low-metallicity stack.}
\label{fig:metal_compar}
\end{figure}

Previous work has suggested that gas-phase oxygen abundance moderates the strength of the most prominent rest-UV emission lines, including Ly$\alpha$ \citep[e.g.,][]{Yang2017,Erb2016,Trainor2016,Nakajima2018mnras} and \textrm{C}~\textsc{iii}] \citep[e.g.,][]{Senchyna2017,Maseda2017,Senchyna2019,Nakajima2018aa,Jaskot2016}. These emission features are observed to be stronger in galaxies with lower metallicities. To investigate whether the same trend is observed in our LRIS sample, we measured metallicity from individual targets in our sample, stacked the spectra according to metallicity, and compared the rest-UV spectra of the stacks.

For a subset of the sample where the objects fall in the redshift range of $z=1.95-2.39$, [\textrm{O}~\textsc{ii}]$\lambda\lambda3727,3729$, $\mbox{H}\beta$, and [\textrm{O}~\textsc{iii}]$\lambda\lambda4959,5007$ are simultaneously covered by the WFC3/G141 grism, enabling measurements of gas-phase metallicity. To maximize the number of objects from which metallicity can be measured, we used the indicator O32$\equiv I_{[\textrm{O}~\textsc{iii}]\lambda5007}/I_{[\textrm{O}~\textsc{ii}]\lambda\lambda3727,3729}$ to estimate nebular oxygen abundance. O32 is most directly tied to the ionization parameter, but, due to the well-known anti-correlation between ionization parameter and metallicity \citep{PM2014}, O32 displays a negative, monotonic trend with oxygen abundance across a wide range in metallicity \citep[$7.8<12+\log(O/H)<8.6$;][]{Jones2015}, making it a reliable metallicity indicator. MMIRS or MOSFIRE line flux measurements were used whenever available, given the high quality and resolution of the spectra. When both MMIRS and MOSFIRE spectra are available, the flux measurements from MOSFIRE spectra were adopted. For objects where only the grism spectrum is available, an intrinsic doublet ratio of $I_{[\textrm{O}~\textsc{iii}]\lambda4959}$:$I_{[\textrm{O}~\textsc{iii}]\lambda5007}$=1:2.98 was assumed in order to estimate the dust-corrected [\textrm{O}~\textsc{iii}]$\lambda5007$ line flux. 

Twenty five objects have spectral coverage of both [\textrm{O}~\textsc{iii}]$\lambda5007$ and [\textrm{O}~\textsc{ii}]$\lambda3727,3729$ features. The [\textrm{O}~\textsc{iii}] flux is detected at $\geqslant3\sigma$ across the entire subsample. We obtained an O32 value for 11 objects with $\geqslant3\sigma$ [\textrm{O}~\textsc{ii}] flux measurements, and an O32 lower limit for 14 objects where [\textrm{O}~\textsc{ii}] is not significantly detected (flux $\leqslant3\sigma$) by adopting the $3\sigma$ upper limit of the [\textrm{O}~\textsc{ii}] flux. Due to differential dust extinction, the observed emission line fluxes need to be corrected for reddening. We assumed that $E(B-V)_{gas}$ inferred from a \citet{Cardelli1989} law is equal to $E(B-V)_{star}$ derived from a \citet{Calzetti2000} law. Our analysis of the EELG sample from \citet{Tang2018} suggests that $E(B-V)_{gas}$ is overall consistent with $E(B-V)_{star}$, with a scatter of $\sim0.1$ dex at lower ($\lesssim450 \mbox{\AA}$) [\textrm{O}~\textsc{iii}] EW \citep[but see][]{Reddy2015} . Furthermore, given that the LRIS objects in general contain little dust (median $E(B-V)_{gas}=E(B-V)_{star}=0.09$), a different relation between $E(B-V)_{gas}$ and $E(B-V)_{star}$ does not significantly change the results presented here.

Accordingly, we corrected [\textrm{O}~\textsc{ii}] and [\textrm{O}~\textsc{iii}] line fluxes with  $E(B-V)_{gas}=E(B-V)_{star}$, where $E(B-V)_{star}$ is the best-fit stellar extinction parameter from BEAGLE. This way, we obtained dust-corrected O32 values and lower limits for the 25 objects in our sample with [\textrm{O}~\textsc{iii}] and [\textrm{O}~\textsc{ii}] coverage. The $1\sigma$ uncertainty on the O32 value (where both [\textrm{O}~\textsc{ii}] and [\textrm{O}~\textsc{iii}] are significantly detected) was estimated using the Monte Carlo method. More specifically, we perturbed both [\textrm{O}~\textsc{ii}] and [\textrm{O}~\textsc{iii}] line fluxes according to their corresponding uncertainties 100 times, dust corrected the ``fake'' line fluxes in each realization, and adopted the standard deviation of the distribution of simulated, dust-corrected O32 as the 68th-percentile confidence
interval. Finally, we placed the O32 values, confidence intervals, and lower limits on the oxygen abundance scale following the calibration of \citet{Jones2015}.

Figure \ref{fig:metal_scatter} shows the relation between [\textrm{O}~\textsc{iii}] EW and oxygen abundance for 25 objects with O32 values and upper limits, as well as COSMOS-4064 with oxygen abundance estimated based on O3N2. We note that neither of COSMOS-4064 nor COSMOS-7686, the two objects that show extremely strong rest-UV emission lines, has a metallicity measurement based on O32 due to the lack of [\textrm{O}~\textsc{ii}] coverage. Therefore, we determined their oxygen abundances using other metallicity indicators. COSMOS-4064 has available [\textrm{O}~\textsc{iii}]$\lambda5007$, H$\beta$, H$\alpha$, and [\textrm{N}~\textsc{ii}]$\lambda6584$ line fluxes measured from MOSFIRE spectroscopy, hence the metallicity was estimated based on O3N2$\equiv$ O3/N2=$([\textrm{O}~\textsc{iii}]/\mbox{H}\beta)/([\textrm{N}~\textsc{ii}]/\mbox{H}\alpha)$ \citep{Pettini2004}. For COSMOS-4064, all lines but [\textrm{N}~\textsc{ii}] are detected at the $\geqslant3\sigma$ level. Accordingly, we obtained a lower limit of 130.0 on O3N2, translating into an upper limit in oxygen abundance 12+log(O/H)$=8.05$ using the \citet{Jones2015} calibration. As for COSMOS-7686, given the lack of high-resolution MOSFIRE or MMIRS spectra, the line fluxes were measured from the grism spectra, where [\textrm{N}~\textsc{ii}] is blended with H$\alpha$. Instead, we performed a coarse ``metallicity" estimate based on O3$\equiv$[\textrm{O}~\textsc{iii}]/H$\beta$. The O3 value for COSMOS-7686 is 6.32, which is close to the peak in the log(O3) versus 12+log(O/H) relation in \citet{Maiolino2008}, corresponding to 12+log(O/H)$\sim8.0$. We note that although there may be potential systematic offsets among the O32, O3N2, and O3 indicators, we can still infer that COSMOS-4064 and COSMOS-7686 are among the lowest metallicity objects in the LRIS sample, as also suggested by the SED-fitting results that take into account the observed rest-UV line EWs (see Section \ref{sec:case}).

Individual galaxies in Figure \ref{fig:metal_scatter} display a negative correlation between [\textrm{O}~\textsc{iii}] EW and gas-phase metallicity. This relationship is consistent with the anti-correlation observed between [\textrm{O}~\textsc{iii}] EW and nebular oxygen abundance in the $z\sim 2$ MOSDEF sample \citep{Reddy2018}, and can be explained in terms of the harder stellar ionizing spectrum, hotter \textrm{H}~\textsc{ii} region temperature, and higher nebular excitation in lower-metallicity star-forming regions.

Our main motivation here is to examine how metallicity affects the rest-far-UV spectrum of distant star-forming galaxies, both in terms of strong emission features (e.g., Ly$\alpha$, \textrm{C}~\textsc{iii}]) and weaker ones (e.g., \textrm{He}~\textsc{ii}, \textrm{O}~\textsc{iii}]). As we do not typically detect weak lines at $>3\sigma$ level in the vast majority of the sample, except for COSMOS-4064 and COSMOS-7686, we created binary stacks in metallicity to compare the strength of different emission features. The high-metallicity stack includes 5 objects with 12+log(O/H) values, and has a median metallicity of 12+log(O/H)$=8.25$. The low-metallicity composite contains 6 objects with 12+log(O/H) values, and 4 objects with 12+log(O/H) upper limits that are smaller than the lowest 12+log(O/H) value in the high-metallicity bin. The stacks were constructed in this manner to (1) equally divide the 12+log(O/H) values into the low- and high-metallicity bins, (2) achieve a higher S/N by including the 12+log(O/H) upper limits, and (3) ensure that the 12+log(O/H) upper limits included in the composite indeed belong to the lower half of the subsample in metallicity). The resulting median metallicity for the low-metallicity bin, after accounting for the 12+log(O/H) upper limits, is 12+log(O/H)$<8.17$.

\begin{table*}
\centering
\begin{threeparttable}
  \caption{Rest-UV Emission Line EW Measurements in Metallicity Stacks}
  \label{tab:metal}
{\renewcommand{\arraystretch}{1.5}
  \begin{tabular}{cccccc}
  \hline
  \hline
    Stack & Median 12+log(O/H) & Ly$\alpha$ & \textrm{He}~\textsc{ii} & \textrm{O}~\textsc{iii}]$\lambda1665$ & \textrm{C}~\textsc{iii}]$\lambda\lambda1907,1909$ \\
     &  & ($\mbox{\AA}$) & ($\mbox{\AA}$) & ($\mbox{\AA}$) & ($\mbox{\AA}$) \\
  \hline
  Low-metallicity & $<8.17$ & 12$\pm$1.05 & 0.44$\pm$0.12 & $<0.41$ & 3.84$\pm$0.24	 \\ 
 \hline
  High-metallicity & 8.25 & $-0.03\pm$1.45 & $<0.52$ & $<0.84$ & 1.71$\pm$0.24 \\
\hline
 \end{tabular}}
\end{threeparttable}
\end{table*}

Figure \ref{fig:metal_compar} shows portions of the rest-UV spectra of low- and high-metallicity stacks, zoomed in on Ly$\alpha$ and \textrm{C}~\textsc{iii}]. The low-metallicity composite shows a significantly stronger Ly$\alpha$ and \textrm{C}~\textsc{iii}] than its high-metallicity counterpart. We do not detect a $S/N\geqslant3$ \textrm{He}~\textsc{ii} line in the high-metallicity stack or a $S/N\geqslant3$ \textrm{O}~\textsc{iii}]$\lambda1665$ in either stack. The line profiles as well as EWs of \textrm{He}~\textsc{ii} and (rest-UV) \textrm{O}~\textsc{iii}] show no apparent differences with metallicity\footnote{Although there is a small but significant \textrm{He}~\textsc{ii} detection in the low-metallicity stack, the EW of the \textrm{He}~\textsc{ii} detection is consistent with the \textrm{He}~\textsc{ii} EW upper limit in the high-metallicity stack.}. We summarize the EW measurements of the rest-UV lines from the metallicity stacks in Table \ref{tab:metal}.

\subsection{Two Extreme Objects}
\label{sec:case}

\begin{figure*}
\includegraphics[width=1.0\linewidth]{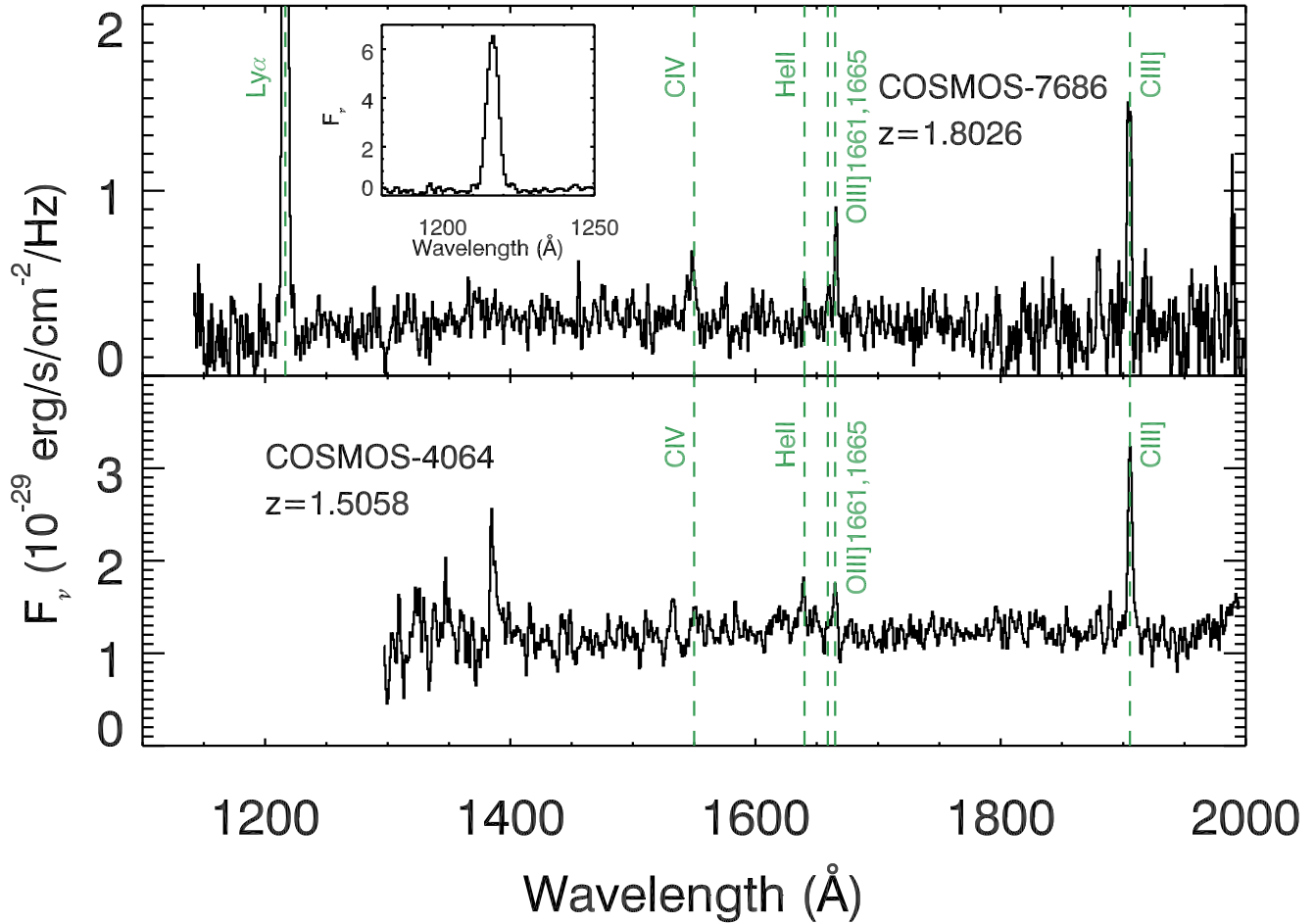}
\caption{Flux-calibrated rest-UV spectra of COSMOS-7686 ($z=1.8026$; top) and COSMOS-4064 ($z=1.5058$; bottom). Strong emission features are identified with green dashed lines. The inset panel shows the full Ly$\alpha$ profile of COSMOS-7686, using the same y-axis units as in the main panels. The emission feature at a rest wavelength of   1385\AA\ in the spectrum of COSMOS-4064 corresponds to unidentified serendipitous emission at an observed wavelength of 3470\AA. If this feature is Ly$\alpha$, it corresponds to a redshift of $z=1.85$, but we have not identified the foreground emitter in the {\it HST} image.} 
\label{fig:indiv_case}
\end{figure*}

Strong rest-UV line emission from metals and \textrm{He}~\textsc{ii} appears to be fairly common at $z>6.5$ \citep{Stark2015,Hutchison2019}. Therefore, the $z>6.5$ analogs at lower redshifts are also expected to show not only prominent Ly$\alpha$ and \textrm{C}~\textsc{iii}] emission, but also weaker lines such as \textrm{C}~\textsc{iv}$\lambda\lambda1548,1550$, \textrm{He}~\textsc{ii}$\lambda1640$, and \textrm{O}~\textsc{iii}]$\lambda\lambda1661,1665$. Within our LRIS sample, 9 objects (6 EELGs and 3 $\ub$ targets) are LAEs, and 13 objects (11 EELGs and 2 $\ub$ targets) are \textrm{C}~\textsc{iii}] emitters. Among these, only two objects (COSMOS-4064 and COSMOS-7686, both EELGs) show detectable \textrm{C}~\textsc{iv}$\lambda\lambda1548,1550$, \textrm{He}~\textsc{ii}$\lambda1640$, and \textrm{O}~\textsc{iii}]$\lambda\lambda1661,1665$, as observed in a handful of spectroscopically confirmed $z>6.5$ galaxies \citep[e.g.,][]{Stark2015}.

COSMOS-4064 and COSMOS-7686 first stand out because of their extremely strong emission in [\textrm{O}~\textsc{iii}], Ly$\alpha$, and \textrm{C}~\textsc{iii}]. COSMOS-7686, with an EW$_{[\textrm{O}~\textsc{iii}]\lambda\lambda4959,5007}=1417\pm236\mbox{\AA}$, has the strongest emission in Ly$\alpha$ (EW$_{Ly\alpha}=131.6\pm10.9\mbox{\AA}$) and \textrm{C}~\textsc{iii}] (EW$_{\textrm{C}~\textsc{iii}]\lambda\lambda1907,1909}=13.2\pm0.8\mbox{\AA}$) among the LRIS targets. As for COSMOS-4064, although the LRIS spectrum does not cover Ly$\alpha$, its [\textrm{O}~\textsc{iii}]$\lambda\lambda4959,5007$ EW ($1940\pm403\mbox{\AA}$) is the largest in our LRIS sample. We note that neither COSMOS-4064 nor COSMOS-7686 is identified as an X-ray source in the Chandra COSMOS Legacy Survey \citep{Civano2016}. In the discussion that follows, we interpret the strong emission lines in these two sources as due primarily to active star formation. As shown in the lower panels of Figure \ref{fig:galprop}, both objects fall on the high-[\textrm{O}~\textsc{iii}]-EW and high-sSFR tail of galaxy distributions, revealing their unusual nature even among the EELGs.

For both objects, we additionally identified \textrm{C}~\textsc{iv}, \textrm{He}~\textsc{ii}, and \textrm{O}~\textsc{iii}] emission in the spectra, and measured the EWs of these features following the procedures described in Section \ref{sec:line}. We refined the SED fitting results for these two objects by including the observed rest-UV metal line EWs in the BEAGLE model, in addition to the available rest-optical line fluxes and broadband photometry.\footnote{We did not attempt to fit the \textrm{He}~\textsc{ii} emission, as the BEAGLE models are known to be incapable of reproducing strong nebular \textrm{He}~\textsc{ii} emission \citep{Chevallard2018}.} The rest-UV EW measurements and the resulting integrated galaxy properties are summarized in Table \ref{tab:case}. 

Considering the low-metallicity nature of COSMOS-7686 and COSMOS-4064, their C/O abundance ratios are expected to be lower than (C/O)$_{\sun}$ given the dependence of C/O on metallicity \citep{Erb2010}. Therefore, instead of assuming C/O=(C/O)$_{\sun}$ as described in Section \ref{sec:sed}, we experimented with a range of lower C/O abundance ratios, finding that the model with $\mbox{C/O}=0.52\mbox{(C/O)}_{\sun}=0.23$ yields the best fitting results for both objects. The best-fit model can well produce the rest-optical line fluxes and the galaxy SED, with the predicted and observed line fluxes agreeing typically within $1\sigma$. Furthermore, the \textrm{O}~\textsc{iii}]$\lambda1665$ and \textrm{C}~\textsc{iii}] EWs predicted by the best-fit model match with the observed values within the $1\sigma$ uncertainties for COSMOS-7686, although the \textrm{C}~\textsc{iv} EW is not reproduced (significantly underestimated)\footnote{While using models with lower C/O ratios (C/O=0.27 (C/O)$_{\sun}$ or 0.1(C/O)$_{\sun}$) can increase the predicted \textrm{C}~\textsc{iv} EW, the predictions of \textrm{C}~\textsc{iii}]/\textrm{O}~\textsc{iii}]$\lambda1665$ and \textrm{C}~\textsc{iv}/\textrm{O}~\textsc{iii}]$\lambda1665$ ratios are systemically lower than the observed values in these models. As a result, we still consider the C/O=0.52 (C/O)$_{\sun}$ model as the best-fit model, despite the discrepancy between the observed and predicted \textrm{C}~\textsc{iv} EWs.}. As for COSMOS-4064, all of \textrm{C}~\textsc{iv}, \textrm{O}~\textsc{iii}]$\lambda1665$, and \textrm{C}~\textsc{iii}] can be reproduced within the $1- 2\sigma$ uncertainties.

As listed in Table \ref{tab:case}, COSMOS-4064 and COSMOS-7686 share some similarities in integrated galaxy properties. They are both of low stellar mass with exceptionally high sSFRs and extremely young stellar populations. Moreover, they are both characterized by high ionization parameters and are especially metal poor. We note that the inferred absolute value of the gas-phase oxygen abundance for these two galaxies depends on whether using rest-optical strong-line metallicity indicators or BEAGLE fitting to broadband photometry, rest-UV and rest-optical spectroscopy. However, in a relative sense, based on rest-optical strong-line ratios, COSMOS-7686 and COSMOS-4064 are at the high-excitation, low-metallicity extreme of the LRIS sample.

The galaxy properties inferred for COSMOS-4064 and COSMOS-7686 are consistent with the emerging physical picture for EELGs \citep{Tang2018,Stark2014}. Accordingly, in EELGs with extreme emission from  high-ionization nebular features the hard radiation field generated by young, metal-poor stars ionizes the ISM and elevates the electron temperature in the \textrm{H}~\textsc{ii} regions \citep{Stark2014}. With large sSFRs, the free electron density increases as a result of more ionizing photons being produced per unit mass, giving rise to large EWs in \textrm{O}~\textsc{iii}]$\lambda\lambda1661,1665$, \textrm{C}~\textsc{iii}]$\lambda\lambda1907,1909$, and [\textrm{O}~\textsc{iii}]$\lambda\lambda4959,5007$ \citep{Stark2014,Jaskot2016}. In the meantime, the hard ionizing spectrum in these galaxies leads to a larger fraction of high-energy, ionizing photons, resulting in a high ionization state of carbon and strong emission in \textrm{C}~\textsc{iii}] and \textrm{C}~\textsc{iv}. This physical picture can further explain the more prominent \textrm{C}~\textsc{iii}] and \textrm{C}~\textsc{iv} emission observed in COSMOS-7686 than in COSMOS-4064, as a result of its significantly lower gas-phase metallicity and younger stellar populations \citep{Senchyna2019}.
Finally, although COSMOS-4064 and COSMOS-7686 show the most intense \textrm{C}~\textsc{iii}] and \textrm{C}~\textsc{iv} emission within our $z\sim2$ LRIS sample, the EWs of these features are still not quite as high (i.e., rest-frame EW $\gtrsim20\mbox{\AA}$) as those of the strongest metal emission lines detected at $z>6.5$ \citep{Stark2015b,Stark2015,Stark2017}, perhaps indicating even more extreme conditions in such $z>6.5$ galaxies.

\begin{table*}
\tablewidth{0pt}
\centering
\begin{threeparttable}
  \caption{Rest-UV EW and Integrated Galaxy Properties of COSMOS-7686 and COSMOS-4064}
  \label{tab:case}
{\renewcommand{\arraystretch}{1.5}
  \begin{tabular}{cccccc}
  \hline
  \hline
    Object & 
    Ly$\alpha$ & 
    \textrm{C}~\textsc{iv} \tnote{1} & 
    \textrm{He}~\textsc{ii} & 
    \textrm{O}~\textsc{iii}]$\lambda1665$ & 
    \textrm{C}~\textsc{iii}] \\
    & 
    ($\mbox{\AA}$) & 
    ($\mbox{\AA}$) & 
    ($\mbox{\AA}$) & 
    ($\mbox{\AA}$) & 
    ($\mbox{\AA}$) \\
    \hline
    COSMOS-7686 & $131.6\pm10.9$ & 8.8$\pm$1.0 & $<1.41$ & 5.3$\pm$0.9 & 13.2$\pm$0.8 \\
    \hline
    COSMOS-4064 & \nodata & 1.5$\pm$0.2 & 1.7$\pm$0.3 & 1.5$\pm$0.3 & 6.9$\pm$0.3 \\
    \hline
    \hline
    Object & 
    Stellar Mass &
    Stellar age &
    sSFR &
    Ionization Parameter &
    Gas-phase Oxygen Abundance \\
    & 
    (log(M$_{*}$/M$_{\sun}$)) & 
    (Myr) & 
    (Gyr$^{-1}$) & 
    (log(U)) &
    (12+log(O/H)) \\
  \hline
 COSMOS-7686 & $7.86^{0.06}_{0.06}$ & $8.0^{1.0}_{1.0}$ & $120^{11}_{10}$ & $-1.13^{0.09}_{0.11}$ & $7.62^{0.07}_{0.06}$ \\ 
 \hline
 COSMOS-4064 & $8.60^{0.05}_{0.07}$ & $16^{2}_{3}$ & $61^{11}_{6}$ & $-2.15^{0.05}_{0.06}$ & $8.13^{0.03}_{0.03}$  \\
 
\hline
 \end{tabular}}
  \begin{tablenotes}
\item[1] EW of interstellar \textrm{C}~\textsc{iv} emission, obtained by removing the P-Cygni stellar profile.
\end{tablenotes}
\end{threeparttable}
\end{table*}

\section{Discussion}
\label{sec:disc}

\subsection{A Closer Look at the Ly$\alpha$ vs. [\textrm{O}~\textsc{iii}]$\lambda$5007 relation}
\label{sec:corrcompar}

\begin{figure*}
\includegraphics[width=0.5\linewidth]{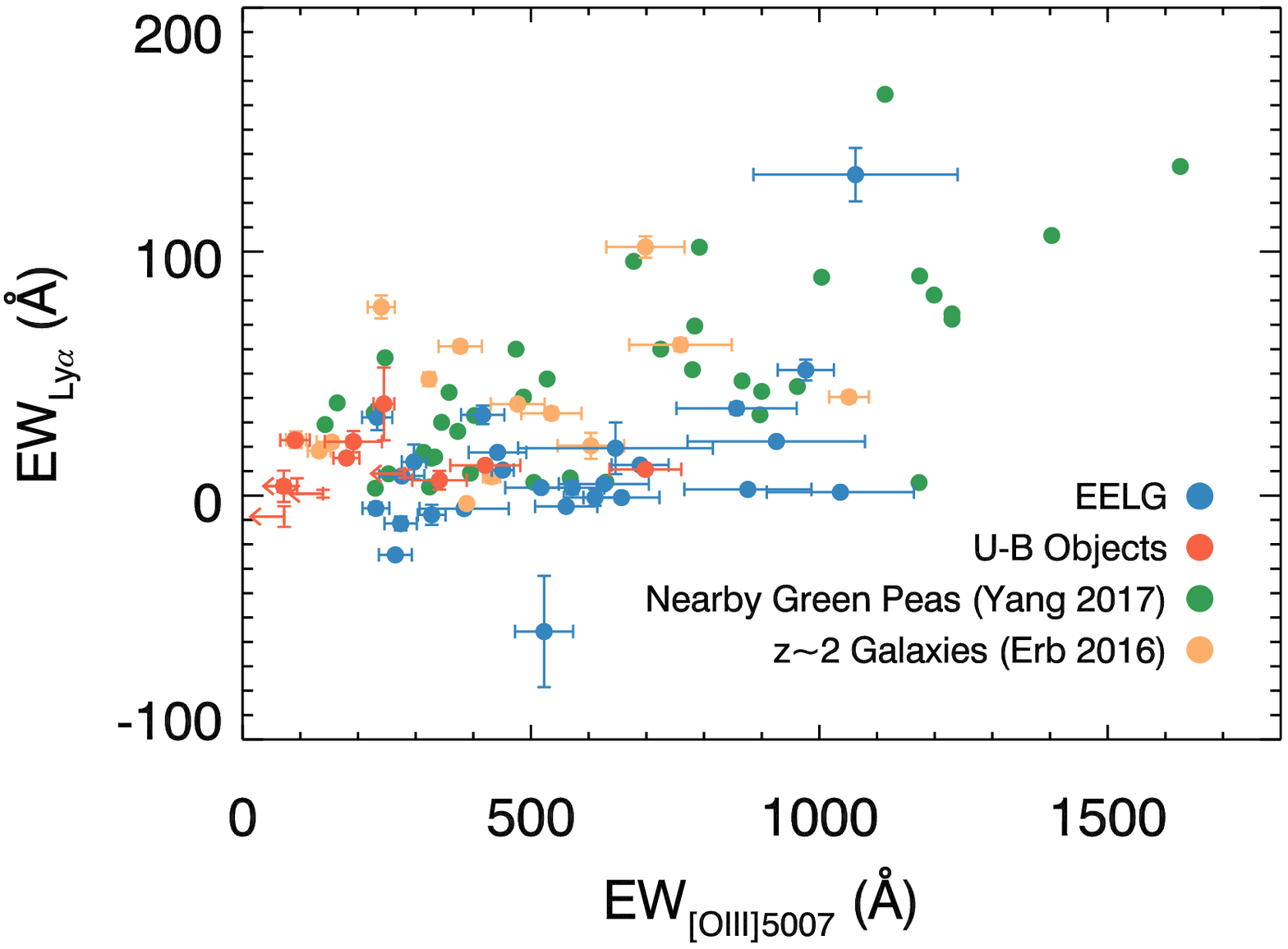}
\includegraphics[width=0.5\linewidth]{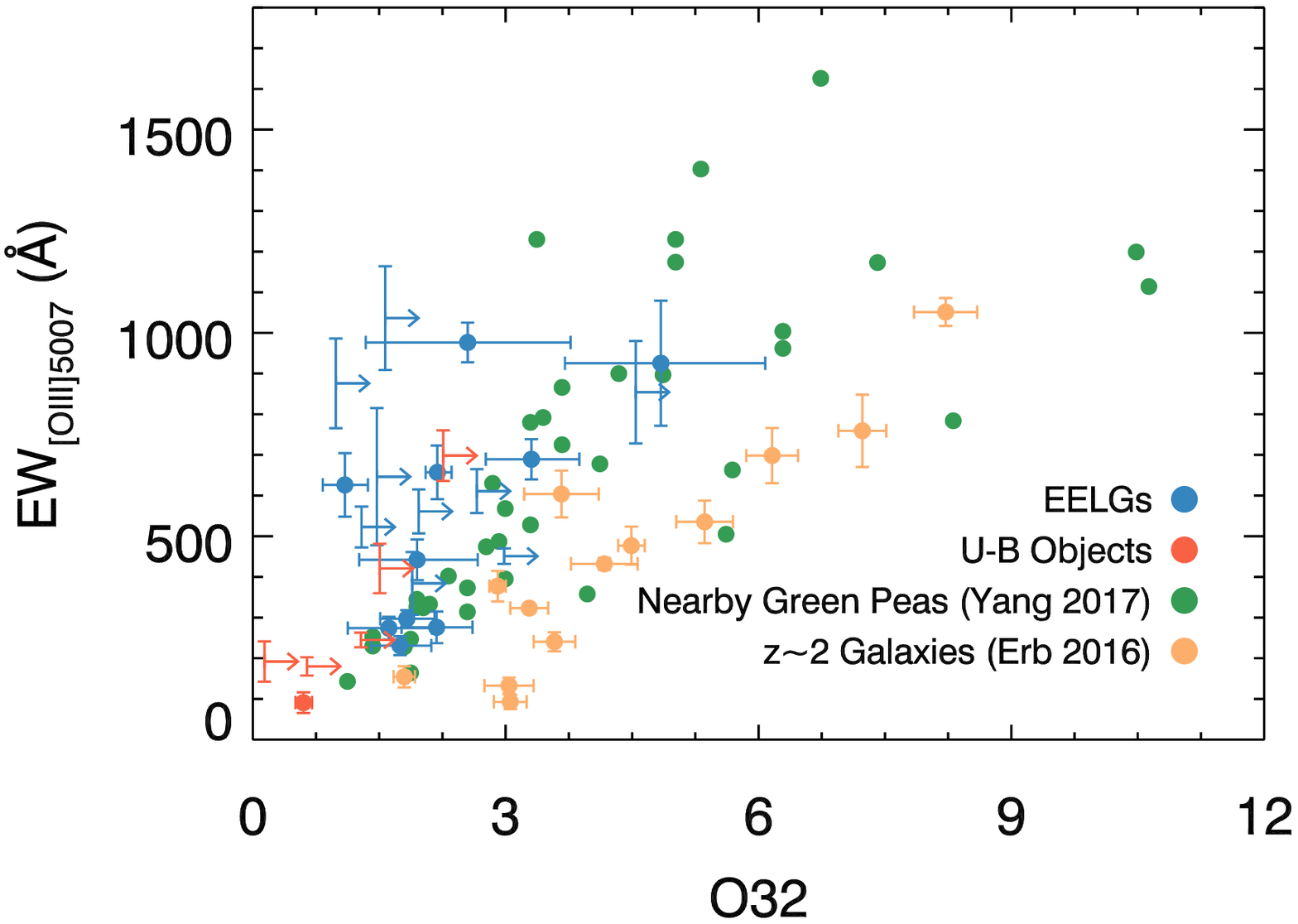}
\caption{\textbf{Left:} Ly$\alpha$ EW vs. [\textrm{O}~\textsc{iii}]$\lambda$5007 EW for different samples, with measurements drawn from the literature. Measurements from the EELGs and $\ub$ objects in our LRIS sample are shown with blue and red circles, respectively. Yellow circles represent $z\sim2$ BPT-selected galaxies presented in \citet{Erb2016}, while green circles indicate nearby green-pea galaxies presented in \citet{Yang2017}, respectively. Error bars shown in the figure represent 68$\%$ confidence intervals, while left-pointing arrows show $3\sigma$ upper limits in [\textrm{O}~\textsc{iii}]$\lambda$5007 EW. The [\textrm{O}~\textsc{iii}]$\lambda$5007 EW is directly measured from the high-resolution MMIRS spectra whenever available. A doublet ratio of 1:2.98 is assumed for estimating the [\textrm{O}~\textsc{iii}]$\lambda$5007 EW for objects that only have grism spectra. \textbf{Right:} [\textrm{O}~\textsc{iii}]$\lambda$5007 EW vs. O32 for the same set of samples. The O32 values are calculated based on dust-corrected flux ratio of [\textrm{O}~\textsc{iii}]$\lambda$5007/[\textrm{O}~\textsc{ii}]$\lambda\lambda$3727,3729. Legends are the same as in the left panel. Right-pointing arrows denote $3\sigma$ lower limits in O32.}
\label{fig:disc_plot}
\end{figure*}

Extensive effort has been made to characterize Ly$\alpha$ visibility during and prior to the reionization epoch, which potentially provides valuable constraints on the IGM neutral fraction \citep[e.g.,][]{Mason2018b,Mason2019,Hoag2019}.
Observationally, it has been suggested that the Ly$\alpha$ detection rate and EW at $z>6.5$ are enhanced in galaxies with large [\textrm{O}~\textsc{iii}] or [\textrm{O}~\textsc{iii}]+H$\beta$ EW, as inferred from the $Spitzer/IRAC$ $[3.6]$ and $[4.5]$-band photometry for galaxies at $z\sim 7$. \citet{Ono2012} examine 11 ``dropout'' galaxies at $z\sim7$, 3 of which have significantly detected Ly$\alpha$ emission, with a rest-frame EW of $\sim30-40\mbox{\AA}$. Among these three LAEs, a large [\textrm{O}~\textsc{iii}] EW that accounts for $\sim60\%$ of the $Spitzer/IRAC$ $[4.5]$ flux is inferred for one galaxy at $z\sim7.2$ (no information is available on [\textrm{O}~\textsc{iii}] EW for the other two LAES given their relatively lower redshifts). Similarly, \citet{Finkelstein2013} report a $S/N=7.8$ Ly$\alpha$ detection at $z=7.51$, accompanying an inferred [\textrm{O}~\textsc{iii}] EW of $560 - 640\mbox{\AA}$ from $Spitzer/IRAC$ photometry. More recently, \citet{RB2016} present a sample of 4 $z=7-9$ galaxies with inferred [\textrm{O}~\textsc{iii}]+H$\beta$ EWs $\gtrsim1500\mbox{\AA}$. All 4 galaxies show $S/N\gtrsim6$ Ly$\alpha$ detections in spectroscopic follow-up observations \citep{Oesch2015,Zitrin2015,Stark2017}.

The connection between Ly$\alpha$ and rest-optical emission lines has been considered at lower redshifts as well. For example, \citet{Erb2016} examine 14 $z\sim2$ low-metallicity (12+log(O/H)$\lesssim8.0$) galaxies selected based on their evidence for high nebular excitation in the [\textrm{O}~\textsc{iii}]$\lambda$5007/H$\beta$
versus [\textrm{N}~\textsc{ii}]$\lambda$6584/H$\alpha$ diagnostic diagram. This sample also contains a high fraction ($79\%$; 11 out of 14 objects) of LAEs. Conversely, \citet{Trainor2016} study the rest-optical spectroscopic properties of a sample of 60 LAEs at $\langle z \rangle \approx 2.56$, finding that their average [\textrm{O}~\textsc{iii}]$\lambda$5007/H$\beta$
and [\textrm{N}~\textsc{ii}]$\lambda$6584/H$\alpha$ ratios are consistent with the low-metallicity, high-excitation extreme of the continuum-selected star-forming galaxy population \citep{Steidel2014}. In other work, \citet{Hagen2016} compare the properties of LAEs and [\textrm{O}~\textsc{iii}]-selected galaxies at $z\sim2$ and claim no significant difference in galaxy properties between the two samples.
In the nearby universe, \citet{Yang2017} study the Ly$\alpha$ emission feature in the HST/COS spectra of 43 Green Pea galaxies at $z\sim0.2$, which are compact, star-forming galaxies that show strong [\textrm{O}~\textsc{iii}]$\lambda$5007 emission. These authors find that 28 out of 43 objects ($65\%$) have EW$_{Ly\alpha}\geqslant20\mbox{\AA}$, with a Ly$\alpha$ EW distribution similar to that of a high-redshift LAE sample at $z\sim2.8$ \citep{Zheng2016}. 
Our own results in Section~\ref{sec:line_corr} suggest a more nuanced relationship between [\textrm{O}~\textsc{iii}]$\lambda$5007 and Ly$\alpha$ emission, in which there is no significant positive correlation between Ly$\alpha$ and [\textrm{O}~\textsc{iii}]$\lambda$5007 emission strengths until the [\textrm{O}~\textsc{iii}]$\lambda$5007 exceeds
$\sim 1000$\AA.

To further examine how Ly$\alpha$ EW varies with the [\textrm{O}~\textsc{iii}] strength, we compiled data from the literature for galaxies with measurements of both of those features, to compare with our own Ly$\alpha$ and [\textrm{O}~\textsc{iii}]$\lambda$5007 measurements shown in the upper left panel in Figure \ref{fig:strem_oiiistack}. This comparison is shown in Figure \ref{fig:disc_plot}. Compared to the EELGs and $\ub$ objects in our LRIS sample, the targets in the \citet{Erb2016} and \citet{Yang2017} samples have systematically higher Ly$\alpha$ at fixed [\textrm{O}~\textsc{iii}]$\lambda$5007 EW (Figure \ref{fig:disc_plot}), left panel). This discrepancy is likely a result of the adoption of different selection criteria. By design, objects in the \citet{Erb2016} and \citet{Yang2017} samples were selected in the high-ionization tail of the star-forming sequence, as opposed to a simple [\textrm{O}~\textsc{iii}] criterion, and therefore they display a noticeably higher typical O32 value than the LRIS galaxies at fixed [\textrm{O}~\textsc{iii}]$\lambda$5007 EW (Figure \ref{fig:disc_plot}, right panel). In fact, no LRIS galaxies in the subset with O32 measurements have  O32 $>5$, while a decent fraction of nearby Green Peas and $z\sim2$ low-metallicity galaxies ($40\%$ and $50\%$, respectively\footnote{Restricting the [\textrm{O}~\textsc{iii}]$\lambda$5007 EW range of the \citet{Yang2017} and \citet{Erb2016} samples to that of the LRIS sample brings the corresponding high-O32 fractions down to $26\%$ and $46\%$, respectively. Yet the percentages of high ($>5$) O32 are still larger than that of the LRIS sample.}) have an O32 above such value, extending to O32 $\simeq10$.

Studies have shown that O32, as a proxy for the ionization parameter, is correlated with both intrinsic Ly$\alpha$ production \citep{Trainor2016,Shivaei2018,Nakajima2018mnras} and the escape fraction of Ly$\alpha$ photons through the neutral ISM \citep{Yang2017,Trainor2019}. A high O32 value is typically associated with a large ionizing photon production efficiency \citep{Shivaei2018,Tang2018}. Given that Ly$\alpha$ photons are produced by the reprocessing of hydrogen ionizing photons, environments with high ionizing efficiencies (and larger O32 values) should boost the intrinsic Ly$\alpha$ EW. On the other hand, several models have been proposed to explain how the ionization state of gas in \textrm{H}~\textsc{ii} regions is connected to the Ly$\alpha$ optical depth. These include effects from intense radiation and/or enhanced stellar feedback \citep{Clarke2002,Jaskot2013,Stark2017} and density-bounded nebula and gas geometry \citep{Nakajima2014,Jaskot2019}. In both scenarios, Ly$\alpha$ photons escape through passageways with low \textrm{H}~\textsc{i} column density and/or covering fraction. Collectively, these results suggest a positive correlation between O32 and the observed Ly$\alpha$ EW. Hence, we conclude that the larger fraction of LAEs in the samples of \citet{Erb2016} and \citet{Yang2017} relative to that in our LRIS EELG sample is likely a consequence of the different rest-optical selection criteria adopted, which correspond to different O32 distributions. In fact, galaxies with higher O32 values, i.e., in the samples of \citet{Erb2016} and \citet{Yang2017}, follow a different Ly$\alpha$ EW vs. [\textrm{O}~\textsc{iii}] EW relationship from that of the LRIS EELGs, such that Ly$\alpha$ is stronger in higher-O32 galaxies at fixed [\textrm{O}~\textsc{iii}] EW because of the combination of higher Ly$\alpha$ production efficiency and lower Ly$\alpha$ optical depth.

In summary, our results indicate that [\textrm{O}~\textsc{iii}] EW alone is not an effective predictor of Ly$\alpha$ EW. The Ly$\alpha$ versus [\textrm{O}~\textsc{iii}] relation observed among galaxies with strong rest-optical emission lines at $z\leq 2$ further poses a question on the nature of the intrinsic Ly$\alpha$ EW distribution in typical $z>6.5$ star-forming galaxies \citep[e.g.,][]{Labbe2013}. While all 3 $z\leq 2$ samples discussed above were selected based on rest-optical emission properties, we must determine if the intrinsic Ly$\alpha$ EW distribution at $z>6.5$ is closer to that of our $z\sim 2$ LRIS EELG sample, signaling a relatively lower LAE fraction and mean Ly$\alpha$ EW, or more similar to that of the $z\sim0$ Green Peas or $z\sim2$ BPT-selected galaxies, with a systematically larger typical Ly$\alpha$ EW. Rest-optical spectroscopic data obtained at $z>6.5$ will help differentiate the scenarios here, through detailed characterization of the O32 and oxygen abundance distributions of ``average'' $z>6.5$ star-forming galaxies.

\subsection{Extreme and Typical Galaxies at $z\sim6.5$}
\label{sec:search}

Strong rest-UV metal line (e.g., \textrm{N}~\textsc{v}, \textrm{C}~\textsc{iv}, \textrm{O}~\textsc{iii}], and \textrm{C}~\textsc{iii}]) and \textrm{He}~\textsc{ii} emission is commonly detected at $z>6.5$ in bright, lensed systems \citep{Stark2015b,Stark2015,Stark2017} and LAEs \citep{Sobral2015,Hutchison2019,Laporte2017,Shibuya2018}. Through photoionization modeling and SED fitting, these galaxies are found to have lower stellar masses, higher sSFRs, younger stellar ages, lower dust attenuations, harder radiation fields, and lower gas-phase metallicities compared to the typical star-forming galaxy populations at lower redshifts. These properties have been used to extend observational efforts in searching for and identifying intense rest-UV line emitters at $z
\sim 0-2$ \citep{Stark2014,Du2017,Maseda2017,Berg2019,Senchyna2019,Mainali2019}.

As noted earlier, star-forming galaxies at $z>6.5$ show prominent $[\textrm{O}~\textsc{iii}]+\mbox{H}\beta$ emission (rest-frame EW $\gtrsim670\mbox{\AA}$), as suggested by the $Spitzer/IRAC$ $[3.6]-[4.5]$ color \citep{Labbe2013,Smit2015}. The strong emission from [\textrm{O}~\textsc{iii}], as a result, indicates one possible path towards selecting $z>6.5$ analogs at lower redshifts. It is therefore of great interest to obtain rest-UV spectra of these analogs -- selected based on their [\textrm{O}~\textsc{iii}] EWs -- and characterize their physical properties using rest-UV spectral features.

In this work, we designed our observations to select galaxies at $z=1.2-2.3$, with properties similar to those at $z>6.5$, based on their rest-optical [\textrm{O}~\textsc{iii}]$\lambda\lambda$4959,5007 strength. We find that 13 out of 47 objects ($28\%$) with \textrm{C}~\textsc{iii}] EW measurements have \textrm{C}~\textsc{iii}] EW $\geqslant5\mbox{\AA}$. If matching in [\textrm{O}~\textsc{iii}] EW alone was a sufficient condition in selecting lower-redshift analogs, our results would suggest that most typical $z>6.5$ galaxies \citep[probed by their stellar continuum, as presented in][]{Labbe2013} will not exhibit strong ($\geqslant5\mbox{\AA}$) \textrm{C}~\textsc{iii}] emission -- given that the LRIS/EELG subsample has a very similar characteristic [\textrm{O}~\textsc{iii}] EW to that of the $z>6.5$ galaxies in \citet{Labbe2013}. Just as in the case of inferring the intrinsic Ly$\alpha$ distribution at $z>6.5$, we must allow for the possibility that, at fixed [\textrm{O}~\textsc{iii}] EW, galaxies at $z>6.5$ have larger O32 and lower oxygen abundance
than our $z\sim 2$ LRIS EELG sample. Such differences may translate into differences in the observed frequency of galaxies with strong ($\geqslant5\mbox{\AA}$) \textrm{C}~\textsc{iii}] emission. Additional information collected from the ``average'' galaxies at $z>6.5$, such as the excitation state of the gas (as probed by O32), will greatly help determine if [\textrm{O}~\textsc{iii}] EW alone is sufficient for selecting analogs at lower redshifts.

We note that those two intense rest-UV metal line emitters (COSMOS-7686 and COSMOS-4064) reported in Section \ref{sec:case} should not be taken as analogs of typical $z>6.5$ galaxies, as their inferred galaxy properties are instead comparable to -- though not quite as extreme as -- those of $z>6.5$ galaxies that show intense (rest-frame EW $\gtrsim20\mbox{\AA}$) emission in \textrm{C}~\textsc{iii}] and \textrm{C}~\textsc{iv} \citep{Stark2015b,Stark2015,Stark2017}. In fact, \citet{Mainali2018} finds that the fraction of $z>5.4$ galaxies with a $>20\mbox{\AA}$ \textrm{C}~\textsc{iii}] EW is only 14$\%$, and this percentage goes down to $\sim1\%$ at lower redshifts \citep[i.e., $z\sim2-3$;][]{LeFe2017}.

Although extreme, objects similar to COSMOS-7686 and COSMOS-4064 serve as unique laboratories for characterizing not only rest-UV metal emission features, but also Ly$\alpha$ and Lyman Continuum (LyC) escape fractions ($f_{esc,LyC}$) in systems with intense radiation fields. Such systems potentially shed light on the reionization of the early universe. It is therefore important to consider how to increase the success rate in searching for similar galaxies at $z\sim2$, during the peak epoch of cosmic star formation, where more extensive and higher S/N multi-wavelength data are available. Adopting a selection threshold of a few hundred angstroms in [\textrm{O}~\textsc{iii}]$\lambda\lambda$4959,5007 and H$\alpha$ EWs will result in a sample of analogs of the typical star-forming galaxies at $z>6.5$, where the fraction of objects similar to COSMOS-7686 and COSMOS-4064 is low. On the other hand, the results presented in Section \ref{sec:case} suggest that a selection criterion based on the high-EW tail (EW$_{[\textrm{O}~\textsc{iii}]\lambda\lambda4959,5007}\gtrsim1000\mbox{\AA}$) of the EELG distribution will maximize the chances of targeting galaxies with moderately low gas-phase metallicity (12+log(O/H)$\lesssim8.0$, or $\sim0.2Z_{\sun}$) and hard ionizing spectra. We note that this [\textrm{O}~\textsc{iii}] EW threshold is a necessary but insufficient condition in observing galaxies similar to $z\sim6.5$ extreme emitters, as out of 8 galaxies with [\textrm{O}~\textsc{iii}] EW $>1000\mbox{\AA}$ within the LRIS sample, not all are classified as LAEs or \textrm{C}~\textsc{iii}] emitters. However, within the subsample where objects have an [\textrm{O}~\textsc{iii}] EW $>1000\mbox{\AA}$, the fractions of detected LAEs and \textrm{C}~\textsc{iii}] emitters are significantly higher ($67\%$ and $71\%$, respectively) compared to the LRIS/EELG subset ($23\%$ and $37\%$, respectively) or the overall LRIS sample ($21\%$ and $28\%$, respectively). These differences indicate the effectiveness of the proposed higher threshold in [\textrm{O}~\textsc{iii}] EW.

Adoption of the [\textrm{O}~\textsc{iii}] EW $\sim1000\mbox{\AA}$ threshold in selecting galaxies comparable to those at $z>6.5$ with extreme rest-UV emission lines is further supported by the results of \citet{Tang2018}. These authors find that large ($\gtrsim6.5$) O32 value and moderate ($\gtrsim10\%$) $f_{esc,LyC}$ are commonly associated with galaxies showing [\textrm{O}~\textsc{iii}]$\lambda\lambda4959,5007\gtrsim1000\mbox{\AA}$. As $f_{esc,LyC}$ is typically lower than $f_{esc,Ly\alpha}$ \citep{Kimm2019}, $f_{esc,LyC}\gtrsim10\%$ corresponds to at least $f_{esc,Ly\alpha}\gtrsim10\%$, which translates into a $\gtrsim20\mbox{\AA}$ Ly$\alpha$ EW in the rest-frame (the definition of LAEs), based on an empirical relation between $f_{esc,Ly\alpha}$ and Ly$\alpha$ EW calibrated for LAEs \citep{Sobral2019}. This piece of evidence further justifies the criterion of [\textrm{O}~\textsc{iii}]$\lambda\lambda4959,5007\gtrsim1000\mbox{\AA}$ for targeting LAEs with much higher success rates. A larger sample of EELGs selected based on extreme ($\gtrsim1000\mbox{\AA}$) [\textrm{O}~\textsc{iii}] EW is needed to further investigate the relationships among [\textrm{O}~\textsc{iii}] EW, O32, Ly$\alpha$ EW, and $f_{esc,Ly\alpha}$ at $z>6.5$, ideally through lower-redshift analogs.

\section{Summary and Conclusions}
\label{sec:sum}

Effectively searching for and characterizing the lower-redshift analogs of $z\sim6.5$ star-forming galaxies greatly improves, by proxy, our knowledge of the properties of the  galaxies that contributed to the reionization of the universe. In this paper, we examine multiple correlations between rest-UV and rest-optical emission properties for a statistical sample of 49 galaxies at $z\sim1.2-2.3$, selected based on [\textrm{O}~\textsc{iii}]$\lambda\lambda4959,5007$ EW or rest-frame $\ub$ color. Our key results are listed below.

1. EELGs and the $\ub$ objects have similar overall Ly$\alpha$ EW distributions, with a median EW$_{Ly\alpha}$ of $4.0\mbox{\AA}$ and $6.3\mbox{\AA}$, respectively. The LRIS sample overall shows a much higher median Ly$\alpha$ EW compared to continuum-selected star-forming galaxies at similar redshifts. We find that 23$\%$ of $z\sim 2$ EELGs and 18$\%$ of $\ub$ objects are LAEs (rest-frame Ly$\alpha$ EW $\geqslant20\mbox{\AA}$). These percentages are very similar to the fraction of LAEs among $z\sim 3$ LBGs  \citep{Shapley2003}, but are higher than the LAE fraction within $z\sim 2$ UV-continuum-selected star-forming galaxies \citep{Du2018}.

2. The EELG subsample on average has stronger \textrm{C}~\textsc{iii}] emission than the $\ub$-selected subsample (with median EWs of 4.0$\mbox{\AA}$ and 2.6$\mbox{\AA}$, respectively), likely as a result of having more intense ionization conditions as suggested by the systematically larger [\textrm{O}~\textsc{iii}] EWs in the EELG subsample. Both EELG and $\ub$ subsamples are found to have greater median \textrm{C}~\textsc{iii}] EWs compared to typical continuum-selected star-forming galaxies at $z\sim1-2$. The fraction of \textrm{C}~\textsc{iii}] emitters (rest-frame \textrm{C}~\textsc{iii}] EW $\geqslant5\mbox{\AA}$) is 37$\%$ for EELGs and 12$\%$ for $\ub$ objects.

3. From both individual measurements and composite spectra, we found that Ly$\alpha$ does not display any apparent correlation with [\textrm{O}~\textsc{iii}] EW at EW$_{[\textrm{O}~\textsc{iii}]} \lesssim 1000\mbox{\AA}$, and only becomes prominent (EW$\gtrsim20\mbox{\AA}$) in the tail of high ($\gtrsim1000\mbox{\AA}$) [\textrm{O}~\textsc{iii}] EW. The observed trend supports a physical picture in which the emergent Ly$\alpha$ EW is modulated by both the neutral hydrogen covering fraction along the line of sight and the intrinsic Ly$\alpha$ production rate, which, respectively, dominate the low- and high-EW$_{[\textrm{O}~\textsc{iii}]}$ regimes of the Ly$\alpha$ vs. [\textrm{O}~\textsc{iii}] EW relation. 

4. The correlation between \textrm{C}~\textsc{iii}] and [\textrm{O}~\textsc{iii}] appears to be monotonic, as suggested by individual measurements and composite spectra. This correspondence suggests that environments that favor [\textrm{O}~\textsc{iii}] production (low gas-phase metallicity, hard radiation field, large ionization parameter, higher sSFR) also boost \textrm{C}~\textsc{iii}] emission, primarily by maintaining an elevated electron temperature (from reduced metal cooling) for the production of collisionally excited emission lines \citep{Jaskot2016}.

5. We used composite spectra to study the correlation between weak rest-UV emission lines, \textrm{He}~\textsc{ii}$\lambda1640$ and \textrm{O}~\textsc{iii}]$\lambda1665$, and (rest-optical) [\textrm{O}~\textsc{iii}]. Neither rest-UV metal emission-line feature is detected in stacks at the low [\textrm{O}~\textsc{iii}]-EW end (EW$_{[\textrm{O}~\textsc{iii}]} \lesssim 450\mbox{\AA}$ for \textrm{He}~\textsc{ii} and EW$_{[\textrm{O}~\textsc{iii}]} \lesssim 900\mbox{\AA}$ for \textrm{O}~\textsc{iii}]). Among the composites where the features are significantly detected, \textrm{He}~\textsc{ii} appears decoupled from [\textrm{O}~\textsc{iii}], while the strength of \textrm{O}~\textsc{iii}]$\lambda1665$ increases with increasing [\textrm{O}~\textsc{iii}] EW.

6. We estimated the oxygen abundance based on O32 in 25 galaxies with spectral coverage of [\textrm{O}~\textsc{ii}], [\textrm{O}~\textsc{iii}], and H$\beta$. A negative correspondence is observed between [\textrm{O}~\textsc{iii}] EW and oxygen abundance. By creating binary stacks in metallicity, we find that the low-metallicity stack shows significantly stronger Ly$\alpha$ and \textrm{C}~\textsc{iii}] emission than its high-metallicity counterpart. Ly$\alpha$ emission is enhanced in low-metallicity environments, as such systems tend to have higher ionization parameter and lower \textrm{H}~\textsc{i} optical depth. Our results are also in agreement with predictions from photoionization models, where the harder ionizing spectrum and higher nebular gas temperature at lower oxygen abundance lead to enhanced \textrm{C}~\textsc{iii}].

7. We reported two extreme emission galaxies in our LRIS sample, COSMOS-7686 and COSMOS-4064, the properties of which resemble those of the small sample of $z>6.5$ galaxies with detections of \textrm{C}~\textsc{iii}]- and \textrm{C}~\textsc{iv} emission lines \citep[e.g.,][]{Stark2015b,Stark2015,Stark2017}.  These two galaxies show noticeably strong emission from \textrm{C}~\textsc{iii}], \textrm{C}~\textsc{iv}, \textrm{He}~\textsc{ii}, and \textrm{O}~\textsc{iii}]$\lambda1665$. After incorporating the observed rest-UV line EWs into SED modeling, we discovered that both objects have low stellar masses
(log(M/M$_{\sun})\lesssim8.60$),
exceptionally young stellar populations ($<20$ Myr), 
large sSFRs ($>60$ Gyr$^{-1}$), high ionization parameters (log(U)$\gtrsim-2.0$), 
and low gas-phase metallicities ($\lesssim0.25Z_{\sun}$).

8. Objects in our LRIS sample have a systemically lower Ly$\alpha$ EW compared to other rest-optical-selected galaxies in \citet{Erb2016} and \citet{Yang2017}. We attribute this discrepancy to different selection criteria adopted for these samples. Compared to our LRIS sample, the nearby Green Peas in \citet{Yang2017} and $z\sim2$ low-metallicity galaxies in \citet{Erb2016} have significantly higher O32 values. An enhanced O32 is correlated not only with increased intrinsic Ly$\alpha$ production, but also with lower Ly$\alpha$ optical depth, increasing $f_{esc,Ly\alpha}$ and thus the emergent Ly$\alpha$ EW \citep{Stark2017,Shivaei2018}. The high-O32 galaxies, therefore, have higher Ly$\alpha$ EWs at fixed [\textrm{O}~\textsc{iii}] EW than the EELGs in our LRIS sample.

9. If matching in [\textrm{O}~\textsc{iii}] EW alone is a sufficient condition in selecting $z>6.5$ analogs at lower redshifts, our results then suggest that LAEs and \textrm{C}~\textsc{iii}] emitters are not common at $z>6.5$. However, if [\textrm{O}~\textsc{iii}] EW is not the only criterion needed for selecting $z>6.5$ analogs, we must take into account other physical parameters (e.g., O32) as suggested by the comparison to the \citet{Erb2016} and \citet{Yang2017} samples. In terms of effectively targeting galaxies comparable to those extreme ones at $z>6.5$ with intense emission in \textrm{C}~\textsc{iii}], \textrm{C}~\textsc{iv}, and Ly$\alpha$, we propose a selection criterion of [\textrm{O}~\textsc{iii}]$\lambda\lambda4959,5007$ EW $\gtrsim1000\mbox{\AA}$. Galaxies with [\textrm{O}~\textsc{iii}] EW above the proposed threshold are likely to have a large ($\gtrsim6.5$) O32 that is typically correlated with 
strong ($\gtrsim20\mbox{\AA}$ rest-frame EW)
Ly$\alpha$ emission.

Characterizing detailed physical conditions of $z>6.5$ galaxies and quantifying the fraction of extreme systems similar to COSMOS-7686 and COSMOS-4064 are key steps to a comprehensive understanding of the sources present during and possibly responsible for cosmic reionization. Due to observational biases, galaxies detected at $z>6.5$ so far with extremely strong rest-UV and rest-optical emission may not be representative of the underlying star-forming galaxy populations during this epoch. Furthermore, the information of typical star-forming galaxies at $z>6.5$ has largely been obtained through photometric measurements, which do not enable direct measurements of ionization and emission properties through spectral-line diagnostics. The near-IR capabilities of the upcoming {\it James Webb Space Telescope} will enable a large sample of typical star-forming galaxies at $z>6.5$ to be selected via the ``drop-out'' method, as well as followup spectroscopic observations in rest-UV and rest-optical \citep{Williams2018}. Such data will place unique constraints on the physical conditions of average galaxy populations at $z>6.5$ as well as those on the high-EW tail of the distribution in the early universe that are now just barely within reach of the largest current ground-based telescopes.

\acknowledgements 
We thank Allison Strom and Chuck Steidel for providing additional line measurements. 
We thank St\'{e}phane Charlot for assistance in
the use of BEAGLE.
AES, XD, and MWT acknowledge support from Program number HST-GO-15287, provided by NASA through a grant from the Space Telescope Science Institute, which is operated by the Association of Universities for Research in Astronomy, Incorporated, under NASA contract NAS5-26555. DPS acknowledges support from the National Science Foundation through the grant AST-1410155. XD and BM acknowledge support from the NASA MUREP Institutional Opportunity (MIRO) through the grant NNX15AP99A. We are grateful to the 3D-HST team for providing ancillary data on galaxy properties. We wish to extend special thanks to those of Hawaiian ancestry on whose sacred mountain we are privileged to be guests. Without their generous hospitality, most of the observations presented herein would not have been possible.

\bibliographystyle{apj}
\bibliography{DU19_new}

\begin{thebibliography}{}
\expandafter\ifx\csname natexlab\endcsname\relax\def\natexlab#1{#1}\fi

\bibitem[{{Amor{\'\i}n} {et~al.}(2017){Amor{\'\i}n}, {Fontana},
  {P{\'e}rez-Montero}, {Castellano}, {Guaita}, {Grazian}, {Le F{\`e}vre},
  {Ribeiro}, {Schaerer}, {Tasca}, {Thomas}, {Bardelli}, {Cassar{\`a}},
  {Cassata}, {Cimatti}, {Contini}, {de Barros}, {Garilli}, {Giavalisco},
  {Hathi}, {Koekemoer}, {Le Brun}, {Lemaux}, {Maccagni}, {Pentericci}, {Pforr},
  {Talia}, {Tresse}, {Vanzella}, {Vergani}, {Zamorani}, {Zucca}, \&
  {Merlin}}]{Amorin2017}
{Amor{\'\i}n}, R., {Fontana}, A., {P{\'e}rez-Montero}, E., {et~al.} 2017,
  Nature Astronomy, 1, 0052

\bibitem[{{Berg} {et~al.}(2019){Berg}, {Chisholm}, {Erb}, {Pogge}, {Henry}, \&
  {Olivier}}]{Berg2019}
{Berg}, D.~A., {Chisholm}, J., {Erb}, D.~K., {et~al.} 2019, The Astrophysical
  Journal, 878, L3

\bibitem[{{Berg} {et~al.}(2016){Berg}, {Skillman}, {Henry}, {Erb}, \&
  {Carigi}}]{Berg2016}
{Berg}, D.~A., {Skillman}, E.~D., {Henry}, R.~B.~C., {Erb}, D.~K., \& {Carigi},
  L. 2016, \apj, 827, 126

\bibitem[{{Bouwens} {et~al.}(2014){Bouwens}, {Bradley}, {Zitrin}, {Coe},
  {Franx}, {Zheng}, {Smit}, {Host}, {Postman}, {Moustakas}, {Labb{\'e}},
  {Carrasco}, {Molino}, {Donahue}, {Kelson}, {Meneghetti}, {Ben{\'\i}tez},
  {Lemze}, {Umetsu}, {Broadhurst}, {Moustakas}, {Rosati}, {Jouvel},
  {Bartelmann}, {Ford}, {Graves}, {Grillo}, {Infante}, {Jimenez-Teja}, {Lahav},
  {Maoz}, {Medezinski}, {Melchior}, {Merten}, {Nonino}, {Ogaz}, \&
  {Seitz}}]{Bouwens2014}
{Bouwens}, R.~J., {Bradley}, L., {Zitrin}, A., {et~al.} 2014, \apj, 795, 126

\bibitem[{{Bouwens} {et~al.}(2015){Bouwens}, {Illingworth}, {Oesch}, {Trenti},
  {Labb{\'e}}, {Bradley}, {Carollo}, {van Dokkum}, {Gonzalez}, {Holwerda},
  {Franx}, {Spitler}, {Smit}, \& {Magee}}]{Bouwens2015}
{Bouwens}, R.~J., {Illingworth}, G.~D., {Oesch}, P.~A., {et~al.} 2015, \apj,
  803, 34

\bibitem[{{Bruzual} \& {Charlot}(2003)}]{BC03}
{Bruzual}, G., \& {Charlot}, S. 2003, \mnras, 344, 1000

\bibitem[{{Calzetti} {et~al.}(2000){Calzetti}, {Armus}, {Bohlin}, {Kinney},
  {Koornneef}, \& {Storchi-Bergmann}}]{Calzetti2000}
{Calzetti}, D., {Armus}, L., {Bohlin}, R.~C., {et~al.} 2000, \apj, 533, 682

\bibitem[{{Cardelli} {et~al.}(1989){Cardelli}, {Clayton}, \&
  {Mathis}}]{Cardelli1989}
{Cardelli}, J.~A., {Clayton}, G.~C., \& {Mathis}, J.~S. 1989, \apj, 345, 245

\bibitem[{{Chabrier}(2003)}]{Chabrier2003}
{Chabrier}, G. 2003, \pasp, 115, 763

\bibitem[{{Chevallard} \& {Charlot}(2016)}]{Chevallard2016}
{Chevallard}, J., \& {Charlot}, S. 2016, \mnras, 462, 1415

\bibitem[{{Chevallard} {et~al.}(2018){Chevallard}, {Charlot}, {Senchyna},
  {Stark}, {Vidal-Garc{\'\i}a}, {Feltre}, {Gutkin}, {Jones}, {Mainali}, \&
  {Wofford}}]{Chevallard2018}
{Chevallard}, J., {Charlot}, S., {Senchyna}, P., {et~al.} 2018, \mnras, 479,
  3264

\bibitem[{{Chilingarian} {et~al.}(2015){Chilingarian}, {Beletsky}, {Moran},
  {Brown}, {McLeod}, \& {Fabricant}}]{Chiling2015}
{Chilingarian}, I., {Beletsky}, Y., {Moran}, S., {et~al.} 2015, \pasp, 127, 406

\bibitem[{{Civano} {et~al.}(2016){Civano}, {Marchesi}, {Comastri}, {Urry},
  {Elvis}, {Cappelluti}, {Puccetti}, {Brusa}, {Zamorani}, {Hasinger},
  {Aldcroft}, {Alexand er}, {Allevato}, {Brunner}, {Capak}, {Finoguenov},
  {Fiore}, {Fruscione}, {Gilli}, {Glotfelty}, {Griffiths}, {Hao}, {Harrison},
  {Jahnke}, {Kartaltepe}, {Karim}, {LaMassa}, {Lanzuisi}, {Miyaji}, {Ranalli},
  {Salvato}, {Sargent}, {Scoville}, {Schawinski}, {Schinnerer}, {Silverman},
  {Smolcic}, {Stern}, {Toft}, {Trakhtenbrot}, {Treister}, \&
  {Vignali}}]{Civano2016}
{Civano}, F., {Marchesi}, S., {Comastri}, A., {et~al.} 2016, \apj, 819, 62

\bibitem[{{Clarke} \& {Oey}(2002)}]{Clarke2002}
{Clarke}, C., \& {Oey}, M.~S. 2002, \mnras, 337, 1299

\bibitem[{{Du} {et~al.}(2016){Du}, {Shapley}, {Martin}, \& {Coil}}]{Du2016}
{Du}, X., {Shapley}, A.~E., {Martin}, C.~L., \& {Coil}, A.~L. 2016, \apj, 829,
  64

\bibitem[{{Du} {et~al.}(2017){Du}, {Shapley}, {Martin}, \& {Coil}}]{Du2017}
---. 2017, \apj, 838, 63

\bibitem[{{Du} {et~al.}(2018){Du}, {Shapley}, {Reddy}, {Jones}, {Stark},
  {Steidel}, {Strom}, {Rudie}, {Erb}, {Ellis}, \& {Pettini}}]{Du2018}
{Du}, X., {Shapley}, A.~E., {Reddy}, N.~A., {et~al.} 2018, \apj, 860, 75

\bibitem[{{Erb} {et~al.}(2010){Erb}, {Pettini}, {Shapley}, {Steidel}, {Law}, \&
  {Reddy}}]{Erb2010}
{Erb}, D.~K., {Pettini}, M., {Shapley}, A.~E., {et~al.} 2010, \apj, 719, 1168

\bibitem[{{Erb} {et~al.}(2016){Erb}, {Pettini}, {Steidel}, {Strom}, {Rudie},
  {Trainor}, {Shapley}, \& {Reddy}}]{Erb2016}
{Erb}, D.~K., {Pettini}, M., {Steidel}, C.~C., {et~al.} 2016, \apj, 830, 52

\bibitem[{{Ferland} {et~al.}(2013){Ferland}, {Porter}, {van Hoof}, {Williams},
  {Abel}, {Lykins}, {Shaw}, {Henney}, \& {Stancil}}]{Ferland2013}
{Ferland}, G.~J., {Porter}, R.~L., {van Hoof}, P.~A.~M., {et~al.} 2013, \rmxaa,
  49, 137

\bibitem[{{Finkelstein} {et~al.}(2013){Finkelstein}, {Papovich}, {Dickinson},
  {Song}, {Tilvi}, {Koekemoer}, {Finkelstein}, {Mobasher}, {Ferguson},
  {Giavalisco}, {Reddy}, {Ashby}, {Dekel}, {Fazio}, {Fontana}, {Grogin},
  {Huang}, {Kocevski}, {Rafelski}, {Weiner}, \& {Willner}}]{Finkelstein2013}
{Finkelstein}, S.~L., {Papovich}, C., {Dickinson}, M., {et~al.} 2013, \nat,
  502, 524

\bibitem[{{Finkelstein} {et~al.}(2015){Finkelstein}, {Ryan}, {Papovich},
  {Dickinson}, {Song}, {Somerville}, {Ferguson}, {Salmon}, {Giavalisco},
  {Koekemoer}, {Ashby}, {Behroozi}, {Castellano}, {Dunlop}, {Faber}, {Fazio},
  {Fontana}, {Grogin}, {Hathi}, {Jaacks}, {Kocevski}, {Livermore}, {McLure},
  {Merlin}, {Mobasher}, {Newman}, {Rafelski}, {Tilvi}, \&
  {Willner}}]{Finkelstein2015}
{Finkelstein}, S.~L., {Ryan}, Russell~E., J., {Papovich}, C., {et~al.} 2015,
  \apj, 810, 71

\bibitem[{{Gutkin} {et~al.}(2016){Gutkin}, {Charlot}, \&
  {Bruzual}}]{Gutkin2016}
{Gutkin}, J., {Charlot}, S., \& {Bruzual}, G. 2016, \mnras, 462, 1757

\bibitem[{{Hagen} {et~al.}(2016){Hagen}, {Zeimann}, {Behrens}, {Ciardullo},
  {Grasshorn Gebhardt}, {Gronwall}, {Bridge}, {Fox}, {Schneider}, {Trump},
  {Blanc}, {Chiang}, {Chonis}, {Finkelstein}, {Hill}, {Jogee}, \&
  {Gawiser}}]{Hagen2016}
{Hagen}, A., {Zeimann}, G.~R., {Behrens}, C., {et~al.} 2016, \apj, 817, 79

\bibitem[{{Harikane} {et~al.}(2018){Harikane}, {Ouchi}, {Shibuya}, {Kojima},
  {Zhang}, {Itoh}, {Ono}, {Higuchi}, {Inoue}, {Chevallard}, {Capak}, {Nagao},
  {Onodera}, {Faisst}, {Martin}, {Rauch}, {Bruzual}, {Charlot}, {Davidzon},
  {Fujimoto}, {Hilmi}, {Ilbert}, {Lee}, {Matsuoka}, {Silverman}, \&
  {Toft}}]{Harikane2018}
{Harikane}, Y., {Ouchi}, M., {Shibuya}, T., {et~al.} 2018, \apj, 859, 84

\bibitem[{{Hoag} {et~al.}(2019){Hoag}, {Brada{\v{c}}}, {Huang}, {Mason},
  {Treu}, {Schmidt}, {Trenti}, {Strait}, {Lemaux}, {Finney}, \&
  {Paddock}}]{Hoag2019}
{Hoag}, A., {Brada{\v{c}}}, M., {Huang}, K., {et~al.} 2019, \apj, 878, 12

\bibitem[{{Hutchison} {et~al.}(2019){Hutchison}, {Papovich}, {Finkelstein},
  {Dickinson}, {Jung}, {Zitrin}, {Ellis}, {Malhotra}, {Rhoads},
  {Roberts-Borsani}, {Song}, \& {Tilvi}}]{Hutchison2019}
{Hutchison}, T.~A., {Papovich}, C., {Finkelstein}, S.~L., {et~al.} 2019, \apj,
  879, 70

\bibitem[{{Itoh} {et~al.}(2018){Itoh}, {Ouchi}, {Zhang}, {Inoue}, {Mawatari},
  {Shibuya}, {Harikane}, {Ono}, {Kusakabe}, {Shimasaku}, {Fujimoto}, {Iwata},
  {Kajisawa}, {Kashikawa}, {Kawanomoto}, {Komiyama}, {Lee}, {Nagao}, \&
  {Taniguchi}}]{Itoh2018}
{Itoh}, R., {Ouchi}, M., {Zhang}, H., {et~al.} 2018, \apj, 867, 46

\bibitem[{{Jaskot} {et~al.}(2019){Jaskot}, {Dowd}, {Oey}, {Scarlata}, \&
  {McKinney}}]{Jaskot2019}
{Jaskot}, A.~E., {Dowd}, T., {Oey}, M.~S., {Scarlata}, C., \& {McKinney}, J.
  2019, arXiv e-prints, arXiv:1908.09763

\bibitem[{{Jaskot} \& {Oey}(2013)}]{Jaskot2013}
{Jaskot}, A.~E., \& {Oey}, M.~S. 2013, \apj, 766, 91

\bibitem[{{Jaskot} \& {Ravindranath}(2016)}]{Jaskot2016}
{Jaskot}, A.~E., \& {Ravindranath}, S. 2016, \apj, 833, 136

\bibitem[{{Jones} {et~al.}(2015){Jones}, {Martin}, \& {Cooper}}]{Jones2015}
{Jones}, T., {Martin}, C., \& {Cooper}, M.~C. 2015, The Astrophysical Journal,
  813, 126

\bibitem[{{Kimm} {et~al.}(2019){Kimm}, {Blaizot}, {Garel}, {Michel-Dansac},
  {Katz}, {Rosdahl}, {Verhamme}, \& {Haehnelt}}]{Kimm2019}
{Kimm}, T., {Blaizot}, J., {Garel}, T., {et~al.} 2019, \mnras, 486, 2215

\bibitem[{{Kornei} {et~al.}(2010){Kornei}, {Shapley}, {Erb}, {Steidel},
  {Reddy}, {Pettini}, \& {Bogosavljevi{\'c}}}]{Kornei2010}
{Kornei}, K.~A., {Shapley}, A.~E., {Erb}, D.~K., {et~al.} 2010, \apj, 711, 693

\bibitem[{{Kriek} {et~al.}(2015){Kriek}, {Shapley}, {Reddy}, {Siana}, {Coil},
  {Mobasher}, {Freeman}, {de Groot}, {Price}, {Sanders}, {Shivaei}, {Brammer},
  {Momcheva}, {Skelton}, {van Dokkum}, {Whitaker}, {Aird}, {Azadi}, {Kassis},
  {Bullock}, {Conroy}, {Dav{\'e}}, {Kere{\v s}}, \& {Krumholz}}]{Kriek2015}
{Kriek}, M., {Shapley}, A.~E., {Reddy}, N.~A., {et~al.} 2015, \apjs, 218, 15

\bibitem[{{Labb{\'e}} {et~al.}(2013){Labb{\'e}}, {Oesch}, {Bouwens},
  {Illingworth}, {Magee}, {Gonz{\'a}lez}, {Carollo}, {Franx}, {Trenti}, {van
  Dokkum}, \& {Stiavelli}}]{Labbe2013}
{Labb{\'e}}, I., {Oesch}, P.~A., {Bouwens}, R.~J., {et~al.} 2013, \apjl, 777,
  L19

\bibitem[{{Laporte} {et~al.}(2017){Laporte}, {Nakajima}, {Ellis}, {Zitrin},
  {Stark}, {Mainali}, \& {Roberts-Borsani}}]{Laporte2017}
{Laporte}, N., {Nakajima}, K., {Ellis}, R.~S., {et~al.} 2017, \apj, 851, 40

\bibitem[{{Le F{\`e}vre} {et~al.}(2019){Le F{\`e}vre}, {Lemaux}, {Nakajima},
  {Schaerer}, {Talia}, {Zamorani}, {Cassata}, {Garilli}, {Maccagni},
  {Pentericci}, {Tasca}, {Zucca}, {Amorin}, {Bardelli}, {Cimatti},
  {Giavalisco}, {Guaita}, {Hathi}, {Marchi}, {Vanzella}, {Vergani}, \&
  {Dunlop}}]{LeFe2017}
{Le F{\`e}vre}, O., {Lemaux}, B.~C., {Nakajima}, K., {et~al.} 2019, \aap, 625,
  A51

\bibitem[{{Leitherer} {et~al.}(2010){Leitherer}, {Ortiz Ot{\'a}lvaro},
  {Bresolin}, {Kudritzki}, {Lo Faro}, {Pauldrach}, {Pettini}, \&
  {Rix}}]{Leitherer2010}
{Leitherer}, C., {Ortiz Ot{\'a}lvaro}, P.~A., {Bresolin}, F., {et~al.} 2010,
  \apjs, 189, 309

\bibitem[{{Mainali} {et~al.}(2018){Mainali}, {Zitrin}, {Stark}, {Ellis},
  {Richard}, {Tang}, {Laporte}, {Oesch}, \& {McGreer}}]{Mainali2018}
{Mainali}, R., {Zitrin}, A., {Stark}, D.~P., {et~al.} 2018, Monthly Notices of
  the Royal Astronomical Society, 479, 1180

\bibitem[{{Mainali} {et~al.}(2019){Mainali}, {Stark}, {Tang}, {Chevallard},
  {Charlot}, {Sharon}, {Coe}, {Salmon}, {Bradley}, {Johnson}, {Frye}, {Avila},
  {Ogaz}, {Zitrin}, {Brada{\v{c}}}, {Lemaux}, {Mahler}, {Paterno-Mahler},
  {Strait}, \& {Andrade-Santos}}]{Mainali2019}
{Mainali}, R., {Stark}, D.~P., {Tang}, M., {et~al.} 2019, arXiv e-prints,
  arXiv:1909.09212

\bibitem[{{Maiolino} {et~al.}(2008){Maiolino}, {Nagao}, {Grazian}, {Cocchia},
  {Marconi}, {Mannucci}, {Cimatti}, {Pipino}, {Ballero}, {Calura}, {Chiappini},
  {Fontana}, {Granato}, {Matteucci}, {Pastorini}, {Pentericci}, {Risaliti},
  {Salvati}, \& {Silva}}]{Maiolino2008}
{Maiolino}, R., {Nagao}, T., {Grazian}, A., {et~al.} 2008, \aap, 488, 463

\bibitem[{{Markwardt}(2009)}]{Mark2009}
{Markwardt}, C.~B. 2009, in Astronomical Society of the Pacific Conference
  Series, Vol. 411, Astronomical Data Analysis Software and Systems XVIII, ed.
  D.~A. {Bohlender}, D.~{Durand}, \& P.~{Dowler}, 251

\bibitem[{{Maseda} {et~al.}(2017){Maseda}, {Brinchmann}, {Franx}, {Bacon},
  {Bouwens}, {Schmidt}, {Boogaard}, {Contini}, {Feltre}, {Inami},
  {Kollatschny}, {Marino}, {Richard}, {Verhamme}, \& {Wisotzki}}]{Maseda2017}
{Maseda}, M.~V., {Brinchmann}, J., {Franx}, M., {et~al.} 2017, \aap, 608, A4

\bibitem[{{Mason} {et~al.}(2018){Mason}, {Treu}, {de Barros}, {Dijkstra},
  {Fontana}, {Mesinger}, {Pentericci}, {Trenti}, \& {Vanzella}}]{Mason2018b}
{Mason}, C.~A., {Treu}, T., {de Barros}, S., {et~al.} 2018, \apjl, 857, L11

\bibitem[{{Mason} {et~al.}(2019){Mason}, {Fontana}, {Treu}, {Schmidt}, {Hoag},
  {Abramson}, {Amorin}, {Brada{\v{c}}}, {Guaita}, {Jones}, {Henry}, {Malkan},
  {Pentericci}, {Trenti}, \& {Vanzella}}]{Mason2019}
{Mason}, C.~A., {Fontana}, A., {Treu}, T., {et~al.} 2019, \mnras, 485, 3947

\bibitem[{{McLean} {et~al.}(2012){McLean}, {Steidel}, {Epps}, {Konidaris},
  {Matthews}, {Adkins}, {Aliado}, {Brims}, {Canfield}, {Cromer}, {Fucik},
  {Kulas}, {Mace}, {Magnone}, {Rodriguez}, {Rudie}, {Trainor}, {Wang}, {Weber},
  \& {Weiss}}]{McLean2012}
{McLean}, I.~S., {Steidel}, C.~C., {Epps}, H.~W., {et~al.} 2012, in Society of
  Photo-Optical Instrumentation Engineers (SPIE) Conference Series, Vol. 8446,
  \procspie, 84460J

\bibitem[{{McLeod} {et~al.}(2012){McLeod}, {Fabricant}, {Nystrom}, {McCracken},
  {Amato}, {Bergner}, {Brown}, {Burke}, {Chilingarian}, {Conroy}, {Curley},
  {Furesz}, {Geary}, {Hertz}, {Holwell}, {Matthews}, {Norton}, {Park}, {Roll},
  {Zajac}, {Epps}, \& {Martini}}]{McLeod2012}
{McLeod}, B., {Fabricant}, D., {Nystrom}, G., {et~al.} 2012, \pasp, 124, 1318

\bibitem[{{Momcheva} {et~al.}(2016){Momcheva}, {Brammer}, {van Dokkum},
  {Skelton}, {Whitaker}, {Nelson}, {Fumagalli}, {Maseda}, {Leja}, {Franx},
  {Rix}, {Bezanson}, {Da Cunha}, {Dickey}, {F{\"o}rster Schreiber},
  {Illingworth}, {Kriek}, {Labb{\'e}}, {Ulf Lange}, {Lundgren}, {Magee},
  {Marchesini}, {Oesch}, {Pacifici}, {Patel}, {Price}, {Tal}, {Wake}, {van der
  Wel}, \& {Wuyts}}]{Momcheva2016}
{Momcheva}, I.~G., {Brammer}, G.~B., {van Dokkum}, P.~G., {et~al.} 2016, \apjs,
  225, 27

\bibitem[{{Nakajima} {et~al.}(2018{\natexlab{a}}){Nakajima}, {Fletcher},
  {Ellis}, {Robertson}, \& {Iwata}}]{Nakajima2018mnras}
{Nakajima}, K., {Fletcher}, T., {Ellis}, R.~S., {Robertson}, B.~E., \& {Iwata},
  I. 2018{\natexlab{a}}, Monthly Notices of the Royal Astronomical Society,
  477, 2098

\bibitem[{{Nakajima} \& {Ouchi}(2014)}]{Nakajima2014}
{Nakajima}, K., \& {Ouchi}, M. 2014, \mnras, 442, 900

\bibitem[{{Nakajima} {et~al.}(2018{\natexlab{b}}){Nakajima}, {Schaerer}, {Le
  F{\`e}vre}, {Amor{\'\i}n}, {Talia}, {Lemaux}, {Tasca}, {Vanzella},
  {Zamorani}, {Bardelli}, {Grazian}, {Guaita}, {Hathi}, {Pentericci}, \&
  {Zucca}}]{Nakajima2018aa}
{Nakajima}, K., {Schaerer}, D., {Le F{\`e}vre}, O., {et~al.}
  2018{\natexlab{b}}, Astronomy and Astrophysics, 612, A94

\bibitem[{{Oesch} {et~al.}(2015){Oesch}, {van Dokkum}, {Illingworth},
  {Bouwens}, {Momcheva}, {Holden}, {Roberts-Borsani}, {Smit}, {Franx},
  {Labb{\'e}}, {Gonz{\'a}lez}, \& {Magee}}]{Oesch2015}
{Oesch}, P.~A., {van Dokkum}, P.~G., {Illingworth}, G.~D., {et~al.} 2015,
  \apjl, 804, L30

\bibitem[{{Oke} {et~al.}(1995){Oke}, {Cohen}, {Carr}, {Cromer}, {Dingizian},
  {Harris}, {Labrecque}, {Lucinio}, {Schaal}, {Epps}, \& {Miller}}]{Oke1995}
{Oke}, J.~B., {Cohen}, J.~G., {Carr}, M., {et~al.} 1995, \pasp, 107, 375

\bibitem[{{Ono} {et~al.}(2012){Ono}, {Ouchi}, {Mobasher}, {Dickinson},
  {Penner}, {Shimasaku}, {Weiner}, {Kartaltepe}, {Nakajima}, {Nayyeri},
  {Stern}, {Kashikawa}, \& {Spinrad}}]{Ono2012}
{Ono}, Y., {Ouchi}, M., {Mobasher}, B., {et~al.} 2012, \apj, 744, 83

\bibitem[{{Osterbrock} \& {Ferland}(2006)}]{Oster2006}
{Osterbrock}, D.~E., \& {Ferland}, G.~J. 2006, {Astrophysics of gaseous nebulae
  and active galactic nuclei}

\bibitem[{{Pentericci} {et~al.}(2014){Pentericci}, {Vanzella}, {Fontana},
  {Castellano}, {Treu}, {Mesinger}, {Dijkstra}, {Grazian}, {Brada{\v c}},
  {Conselice}, {Cristiani}, {Dunlop}, {Galametz}, {Giavalisco}, {Giallongo},
  {Koekemoer}, {McLure}, {Maiolino}, {Paris}, \& {Santini}}]{Pentericci2014}
{Pentericci}, L., {Vanzella}, E., {Fontana}, A., {et~al.} 2014, \apj, 793, 113

\bibitem[{{P{\'e}rez-Montero}(2014)}]{PM2014}
{P{\'e}rez-Montero}, E. 2014, \mnras, 441, 2663

\bibitem[{{Pettini} \& {Pagel}(2004)}]{Pettini2004}
{Pettini}, M., \& {Pagel}, B. E.~J. 2004, \mnras, 348, L59

\bibitem[{{Reddy} {et~al.}(2016){Reddy}, {Steidel}, {Pettini},
  {Bogosavljevi{\'c}}, \& {Shapley}}]{Reddy2016}
{Reddy}, N.~A., {Steidel}, C.~C., {Pettini}, M., {Bogosavljevi{\'c}}, M., \&
  {Shapley}, A.~E. 2016, \apj, 828, 108

\bibitem[{{Reddy} {et~al.}(2015){Reddy}, {Kriek}, {Shapley}, {Freeman},
  {Siana}, {Coil}, {Mobasher}, {Price}, {Sanders}, \& {Shivaei}}]{Reddy2015}
{Reddy}, N.~A., {Kriek}, M., {Shapley}, A.~E., {et~al.} 2015, The Astrophysical
  Journal, 806, 259

\bibitem[{{Reddy} {et~al.}(2018){Reddy}, {Shapley}, {Sanders}, {Kriek}, {Coil},
  {Shivaei}, {Freeman}, {Mobasher}, {Siana}, {Azadi}, {Fetherolf}, {Fornasini},
  {Leung}, {Price}, {Zick}, \& {Barro}}]{Reddy2018}
{Reddy}, N.~A., {Shapley}, A.~E., {Sanders}, R.~L., {et~al.} 2018, \apj, 869,
  92

\bibitem[{{Rigby} {et~al.}(2015){Rigby}, {Bayliss}, {Gladders}, {Sharon},
  {Wuyts}, {Dahle}, {Johnson}, \& {Pe{\~n}a-Guerrero}}]{Rigby2015}
{Rigby}, J.~R., {Bayliss}, M.~B., {Gladders}, M.~D., {et~al.} 2015, \apjl, 814,
  L6

\bibitem[{{Rix} {et~al.}(2004){Rix}, {Pettini}, {Leitherer}, {Bresolin},
  {Kudritzki}, \& {Steidel}}]{Rix2004}
{Rix}, S.~A., {Pettini}, M., {Leitherer}, C., {et~al.} 2004, \apj, 615, 98

\bibitem[{{Roberts-Borsani} {et~al.}(2016){Roberts-Borsani}, {Bouwens},
  {Oesch}, {Labbe}, {Smit}, {Illingworth}, {van Dokkum}, {Holden}, {Gonzalez},
  {Stefanon}, {Holwerda}, \& {Wilkins}}]{RB2016}
{Roberts-Borsani}, G.~W., {Bouwens}, R.~J., {Oesch}, P.~A., {et~al.} 2016,
  \apj, 823, 143

\bibitem[{{Rudie} {et~al.}(2012){Rudie}, {Steidel}, {Trainor}, {Rakic},
  {Bogosavljevi{\'c}}, {Pettini}, {Reddy}, {Shapley}, {Erb}, \&
  {Law}}]{Rudie2012}
{Rudie}, G.~C., {Steidel}, C.~C., {Trainor}, R.~F., {et~al.} 2012, \apj, 750,
  67

\bibitem[{{Salmon} {et~al.}(2015){Salmon}, {Papovich}, {Finkelstein}, {Tilvi},
  {Finlator}, {Behroozi}, {Dahlen}, {Dav{\'e}}, {Dekel}, {Dickinson},
  {Ferguson}, {Giavalisco}, {Long}, {Lu}, {Mobasher}, {Reddy}, {Somerville}, \&
  {Wechsler}}]{Salmon2015}
{Salmon}, B., {Papovich}, C., {Finkelstein}, S.~L., {et~al.} 2015, \apj, 799,
  183

\bibitem[{{Schenker} {et~al.}(2014){Schenker}, {Ellis}, {Konidaris}, \&
  {Stark}}]{Schenker2014}
{Schenker}, M.~A., {Ellis}, R.~S., {Konidaris}, N.~P., \& {Stark}, D.~P. 2014,
  \apj, 795, 20

\bibitem[{{Schmidt} {et~al.}(2017){Schmidt}, {Huang}, {Treu}, {Hoag},
  {Brada{\v{c}}}, {Henry}, {Jones}, {Mason}, {Malkan}, {Morishita},
  {Pentericci}, {Trenti}, {Vulcani}, \& {Wang}}]{Schmidt2017}
{Schmidt}, K.~B., {Huang}, K.~H., {Treu}, T., {et~al.} 2017, \apj, 839, 17

\bibitem[{{Senchyna} {et~al.}(2019){Senchyna}, {Stark}, {Chevallard},
  {Charlot}, {Jones}, \& {Vidal-Garc{\'\i}a}}]{Senchyna2019}
{Senchyna}, P., {Stark}, D.~P., {Chevallard}, J., {et~al.} 2019, Monthly
  Notices of the Royal Astronomical Society, 488, 3492

\bibitem[{{Senchyna} {et~al.}(2017){Senchyna}, {Stark}, {Vidal-Garc{\'{\i}}a},
  {Chevallard}, {Charlot}, {Mainali}, {Jones}, {Wofford}, {Feltre}, \&
  {Gutkin}}]{Senchyna2017}
{Senchyna}, P., {Stark}, D.~P., {Vidal-Garc{\'{\i}}a}, A., {et~al.} 2017,
  \mnras, 472, 2608

\bibitem[{{Shapley} {et~al.}(2003){Shapley}, {Steidel}, {Pettini}, \&
  {Adelberger}}]{Shapley2003}
{Shapley}, A.~E., {Steidel}, C.~C., {Pettini}, M., \& {Adelberger}, K.~L. 2003,
  \apj, 588, 65

\bibitem[{{Shapley} {et~al.}(2006){Shapley}, {Steidel}, {Pettini},
  {Adelberger}, \& {Erb}}]{Shapley2006}
{Shapley}, A.~E., {Steidel}, C.~C., {Pettini}, M., {Adelberger}, K.~L., \&
  {Erb}, D.~K. 2006, \apj, 651, 688

\bibitem[{{Shibuya} {et~al.}(2018){Shibuya}, {Ouchi}, {Harikane}, {Rauch},
  {Ono}, {Mukae}, {Higuchi}, {Kojima}, {Yuma}, {Lee}, {Furusawa}, {Konno},
  {Martin}, {Shimasaku}, {Taniguchi}, {Kobayashi}, {Kajisawa}, {Nagao}, {Goto},
  {Kashikawa}, {Komiyama}, {Kusakabe}, {Momose}, {Nakajima}, {Tanaka}, \&
  {Wang}}]{Shibuya2018}
{Shibuya}, T., {Ouchi}, M., {Harikane}, Y., {et~al.} 2018, \pasj, 70, S15

\bibitem[{{Shivaei} {et~al.}(2018){Shivaei}, {Reddy}, {Siana}, {Shapley},
  {Kriek}, {Mobasher}, {Freeman}, {Sanders}, {Coil}, {Price}, {Fetherolf},
  {Azadi}, {Leung}, \& {Zick}}]{Shivaei2018}
{Shivaei}, I., {Reddy}, N.~A., {Siana}, B., {et~al.} 2018, \apj, 855, 42

\bibitem[{{Skelton} {et~al.}(2014){Skelton}, {Whitaker}, {Momcheva}, {Brammer},
  {van Dokkum}, {Labb{\'e}}, {Franx}, {van der Wel}, {Bezanson}, {Da Cunha},
  {Fumagalli}, {F{\"o}rster Schreiber}, {Kriek}, {Leja}, {Lundgren}, {Magee},
  {Marchesini}, {Maseda}, {Nelson}, {Oesch}, {Pacifici}, {Patel}, {Price},
  {Rix}, {Tal}, {Wake}, \& {Wuyts}}]{Skelton2014}
{Skelton}, R.~E., {Whitaker}, K.~E., {Momcheva}, I.~G., {et~al.} 2014, \apjs,
  214, 24

\bibitem[{{Smit} {et~al.}(2014){Smit}, {Bouwens}, {Labb{\'e}}, {Zheng},
  {Bradley}, {Donahue}, {Lemze}, {Moustakas}, {Umetsu}, {Zitrin}, {Coe},
  {Postman}, {Gonzalez}, {Bartelmann}, {Ben{\'{\i}}tez}, {Broadhurst}, {Ford},
  {Grillo}, {Infante}, {Jimenez-Teja}, {Jouvel}, {Kelson}, {Lahav}, {Maoz},
  {Medezinski}, {Melchior}, {Meneghetti}, {Merten}, {Molino}, {Moustakas},
  {Nonino}, {Rosati}, \& {Seitz}}]{Smit2014}
{Smit}, R., {Bouwens}, R.~J., {Labb{\'e}}, I., {et~al.} 2014, \apj, 784, 58

\bibitem[{{Smit} {et~al.}(2015){Smit}, {Bouwens}, {Franx}, {Oesch}, {Ashby},
  {Willner}, {Labb{\'e}}, {Holwerda}, {Fazio}, \& {Huang}}]{Smit2015}
{Smit}, R., {Bouwens}, R.~J., {Franx}, M., {et~al.} 2015, \apj, 801, 122

\bibitem[{{Sobral} \& {Matthee}(2019)}]{Sobral2019}
{Sobral}, D., \& {Matthee}, J. 2019, \aap, 623, A157

\bibitem[{{Sobral} {et~al.}(2015){Sobral}, {Matthee}, {Darvish}, {Schaerer},
  {Mobasher}, {R{\"o}ttgering}, {Santos}, \& {Hemmati}}]{Sobral2015}
{Sobral}, D., {Matthee}, J., {Darvish}, B., {et~al.} 2015, \apj, 808, 139

\bibitem[{{Stark}(2016)}]{Stark2016}
{Stark}, D.~P. 2016, \araa, 54, 761

\bibitem[{{Stark} {et~al.}(2010){Stark}, {Ellis}, {Chiu}, {Ouchi}, \&
  {Bunker}}]{Stark2010}
{Stark}, D.~P., {Ellis}, R.~S., {Chiu}, K., {Ouchi}, M., \& {Bunker}, A. 2010,
  \mnras, 408, 1628

\bibitem[{{Stark} {et~al.}(2014){Stark}, {Richard}, {Siana}, {Charlot},
  {Freeman}, {Gutkin}, {Wofford}, {Robertson}, {Amanullah}, {Watson}, \&
  {Milvang-Jensen}}]{Stark2014}
{Stark}, D.~P., {Richard}, J., {Siana}, B., {et~al.} 2014, \mnras, 445, 3200

\bibitem[{{Stark} {et~al.}(2015{\natexlab{a}}){Stark}, {Walth}, {Charlot},
  {Cl{\'e}ment}, {Feltre}, {Gutkin}, {Richard}, {Mainali}, {Robertson},
  {Siana}, {Tang}, \& {Schenker}}]{Stark2015b}
{Stark}, D.~P., {Walth}, G., {Charlot}, S., {et~al.} 2015{\natexlab{a}},
  \mnras, 454, 1393

\bibitem[{{Stark} {et~al.}(2015{\natexlab{b}}){Stark}, {Richard}, {Charlot},
  {Cl{\'e}ment}, {Ellis}, {Siana}, {Robertson}, {Schenker}, {Gutkin}, \&
  {Wofford}}]{Stark2015}
{Stark}, D.~P., {Richard}, J., {Charlot}, S., {et~al.} 2015{\natexlab{b}},
  \mnras, 450, 1846

\bibitem[{{Stark} {et~al.}(2017){Stark}, {Ellis}, {Charlot}, {Chevallard},
  {Tang}, {Belli}, {Zitrin}, {Mainali}, {Gutkin}, {Vidal-Garc{\'\i}a},
  {Bouwens}, \& {Oesch}}]{Stark2017}
{Stark}, D.~P., {Ellis}, R.~S., {Charlot}, S., {et~al.} 2017, \mnras, 464, 469

\bibitem[{{Steidel} {et~al.}(2003){Steidel}, {Adelberger}, {Shapley},
  {Pettini}, {Dickinson}, \& {Giavalisco}}]{Steidel2003}
{Steidel}, C.~C., {Adelberger}, K.~L., {Shapley}, A.~E., {et~al.} 2003, \apj,
  592, 728

\bibitem[{{Steidel} {et~al.}(2010){Steidel}, {Erb}, {Shapley}, {Pettini},
  {Reddy}, {Bogosavljevi{\'c}}, {Rudie}, \& {Rakic}}]{Steidel2010}
{Steidel}, C.~C., {Erb}, D.~K., {Shapley}, A.~E., {et~al.} 2010, \apj, 717, 289

\bibitem[{{Steidel} {et~al.}(2004){Steidel}, {Shapley}, {Pettini},
  {Adelberger}, {Erb}, {Reddy}, \& {Hunt}}]{Steidel2004}
{Steidel}, C.~C., {Shapley}, A.~E., {Pettini}, M., {et~al.} 2004, \apj, 604,
  534

\bibitem[{{Steidel} {et~al.}(2014){Steidel}, {Rudie}, {Strom}, {Pettini},
  {Reddy}, {Shapley}, {Trainor}, {Erb}, {Turner}, {Konidaris}, {Kulas}, {Mace},
  {Matthews}, \& {McLean}}]{Steidel2014}
{Steidel}, C.~C., {Rudie}, G.~C., {Strom}, A.~L., {et~al.} 2014, \apj, 795, 165

\bibitem[{{Tang} {et~al.}(2019){Tang}, {Stark}, {Chevallard}, \&
  {Charlot}}]{Tang2018}
{Tang}, M., {Stark}, D.~P., {Chevallard}, J., \& {Charlot}, S. 2019, \mnras,
  489, 2572

\bibitem[{{Trainor} {et~al.}(2015){Trainor}, {Steidel}, {Strom}, \&
  {Rudie}}]{Trainor2015}
{Trainor}, R.~F., {Steidel}, C.~C., {Strom}, A.~L., \& {Rudie}, G.~C. 2015,
  \apj, 809, 89

\bibitem[{{Trainor} {et~al.}(2016){Trainor}, {Strom}, {Steidel}, \&
  {Rudie}}]{Trainor2016}
{Trainor}, R.~F., {Strom}, A.~L., {Steidel}, C.~C., \& {Rudie}, G.~C. 2016,
  \apj, 832, 171

\bibitem[{{Trainor} {et~al.}(2019){Trainor}, {Strom}, {Steidel}, {Rudie},
  {Chen}, \& {Theios}}]{Trainor2019}
{Trainor}, R.~F., {Strom}, A.~L., {Steidel}, C.~C., {et~al.} 2019, arXiv
  e-prints, arXiv:1908.04794

\bibitem[{{Treu} {et~al.}(2012){Treu}, {Trenti}, {Stiavelli}, {Auger}, \&
  {Bradley}}]{Treu2012}
{Treu}, T., {Trenti}, M., {Stiavelli}, M., {Auger}, M.~W., \& {Bradley}, L.~D.
  2012, \apj, 747, 27

\bibitem[{{Vanzella} {et~al.}(2011){Vanzella}, {Pentericci}, {Fontana},
  {Grazian}, {Castellano}, {Boutsia}, {Cristiani}, {Dickinson}, {Gallozzi},
  {Giallongo}, {Giavalisco}, {Maiolino}, {Moorwood}, {Paris}, \&
  {Santini}}]{Vanzella2011}
{Vanzella}, E., {Pentericci}, L., {Fontana}, A., {et~al.} 2011, \apjl, 730, L35

\bibitem[{{Williams} {et~al.}(2018){Williams}, {Curtis-Lake}, {Hainline},
  {Chevallard}, {Robertson}, {Charlot}, {Endsley}, {Stark}, {Willmer},
  {Alberts}, {Amorin}, {Arribas}, {Baum}, {Bunker}, {Carniani}, {Crand all},
  {Egami}, {Eisenstein}, {Ferruit}, {Husemann}, {Maseda}, {Maiolino}, {Rawle},
  {Rieke}, {Smit}, {Tacchella}, \& {Willott}}]{Williams2018}
{Williams}, C.~C., {Curtis-Lake}, E., {Hainline}, K.~N., {et~al.} 2018, \apjs,
  236, 33

\bibitem[{{Yang} {et~al.}(2017){Yang}, {Malhotra}, {Gronke}, {Rhoads},
  {Leitherer}, {Wofford}, {Jiang}, {Dijkstra}, {Tilvi}, \& {Wang}}]{Yang2017}
{Yang}, H., {Malhotra}, S., {Gronke}, M., {et~al.} 2017, \apj, 844, 171

\bibitem[{{Zheng} {et~al.}(2016){Zheng}, {Malhotra}, {Rhoads}, {Finkelstein},
  {Wang}, {Jiang}, \& {Cai}}]{Zheng2016}
{Zheng}, Z.-Y., {Malhotra}, S., {Rhoads}, J.~E., {et~al.} 2016, \apjs, 226, 23

\bibitem[{{Zitrin} {et~al.}(2015){Zitrin}, {Ellis}, {Belli}, \&
  {Stark}}]{Zitrin2015}
{Zitrin}, A., {Ellis}, R.~S., {Belli}, S., \& {Stark}, D.~P. 2015, \apjl, 805,
  L7

\end{thebibliography}
\end{CJK}
\end{document}